\documentclass[prl,10pt,aps,footinbib,floatfix,twocolumn,longbibliography,superscriptaddress]{revtex4-2}

\usepackage{amsmath,amssymb,bm}
\usepackage{graphicx}
\usepackage[dvipsnames]{xcolor}
\usepackage{tikz}
\usepackage{xr}%
\usepackage{hyperref}
\hypersetup{colorlinks=true,allcolors=blue!60!black}%
\usepackage{comment}
\usepackage{slashed,braket}%
\usepackage[normalem]{ulem}
\usetikzlibrary{decorations.markings,arrows.meta}
\tikzset{
  fd/.style={line width=0.8pt},
  arr/.style={postaction={decorate},decoration={markings,mark=at position 0.58 with {\arrow{Latex[length=3.6pt,width=3.4pt]}}}},
  nuc/.style={fd,arr},
  del/.style={fd,double distance=1.5pt},
  pio/.style={fd,dash pattern=on 2.6pt off 1.8pt},
  axi/.style={fd,dash pattern=on 2.6pt off 1.8pt},
  vlo/.style={circle,fill=black,inner sep=0pt,minimum size=4pt},
  vsq/.style={rectangle,fill=black,inner sep=0pt,minimum size=4.4pt},
  vosq/.style={rectangle,draw=black,fill=white,line width=0.7pt,inner sep=0pt,minimum size=4.6pt},
  slab/.style={font=\footnotesize,align=center},
}
\newcommand{\FDseg}[3]{\ifx N#1\draw[nuc] #2 -- #3;\else\draw[del] #2 -- #3; \path[arr] #2 -- #3;\fi}
\newcommand{\FDpole}[3]{%
  \ifx o#2
    \FDseg{N}{(0,1.5)}{(1.1,1.5)} \FDseg{#1}{(1.1,1.5)}{(2.2,1.5)} \FDseg{N}{(2.2,1.5)}{(3.3,1.5)}
    \FDseg{N}{(0,0)}{(1.1,0)} \FDseg{N}{(1.1,0)}{(3.3,0)}
    \draw[pio] (1.1,1.5) -- (1.1,0); \draw[axi] (2.2,1.5) -- (2.75,2.3);
    \node[vlo] at (1.1,1.5) {}; \node[vlo] at (1.1,0) {}; \node[vlo] at (2.2,1.5) {};
    \node[slab] at (0.82,0.75) {$\pi$};
  \else
    \FDseg{N}{(0,1.5)}{(1.1,1.5)} \FDseg{#1}{(1.1,1.5)}{(2.2,1.5)} \FDseg{N}{(2.2,1.5)}{(3.3,1.5)}
    \FDseg{N}{(0,0)}{(2.2,0)} \FDseg{N}{(2.2,0)}{(3.3,0)}
    \draw[pio] (2.2,1.5) -- (2.2,0); \draw[axi] (1.1,1.5) -- (1.65,2.3);
    \node[vlo] at (1.1,1.5) {}; \node[vlo] at (2.2,1.5) {}; \node[vlo] at (2.2,0) {};
    \node[slab] at (2.48,0.75) {$\pi$};
  \fi
  \node[slab] at (1.65,1.2) {\ifx N#1$N$\else$\Delta$\fi};
  \node[slab] at (\ifx o#2 2.95\else 1.85\fi,2.42) {$a$};
  \node[slab] at (1.65,-0.42) {#3};}
\newcommand{\FDcontact}[2]{%
  \FDseg{N}{(0,1.5)}{(1.65,1.5)} \FDseg{N}{(1.65,1.5)}{(3.3,1.5)}
  \FDseg{N}{(0,0)}{(1.65,0)} \FDseg{N}{(1.65,0)}{(3.3,0)}
  \draw[pio] (1.65,1.5) -- (1.65,0); \draw[axi] (1.65,1.5) -- (2.2,2.3);
  \node[#1] at (1.65,1.5) {}; \node[vlo] at (1.65,0) {};
  \node[slab] at (1.37,0.75) {$\pi$}; \node[slab] at (2.4,2.42) {$a$};
  \node[slab] at (1.65,-0.42) {#2};}
\newcommand{\FDabs}[3]{%
  \FDseg{N}{(0,0)}{(0.8,0)} \FDseg{#1}{(0.8,0)}{(1.8,0)} \FDseg{N}{(1.8,0)}{(2.6,0)}
  \ifx s#2
    \draw[axi] (0.25,1.1) -- (0.8,0); \draw[pio] (1.8,0) -- (2.35,1.1);
  \else
    \draw[axi] (0.25,1.1) -- (1.8,0); \draw[pio] (0.8,0) -- (2.35,1.1);
  \fi
  \node[vlo] at (0.8,0) {}; \node[vlo] at (1.8,0) {};
  \node[slab] at (0.25,1.38) {$a$}; \node[slab] at (2.35,1.38) {$\pi$};
  \node[slab] at (1.3,-0.4) {\ifx N#1$N$\else$\Delta$\fi};
  \node[slab] at (1.3,-0.9) {#3};}
\newcommand{\FDabscontact}[1]{%
  \FDseg{N}{(0,0)}{(1.3,0)} \FDseg{N}{(1.3,0)}{(2.6,0)}
  \draw[axi] (0.75,1.1) -- (1.3,0); \draw[pio] (1.3,0) -- (1.85,1.1);
  \node[vsq] at (1.3,0) {};
  \node[slab] at (0.75,1.38) {$a$}; \node[slab] at (1.85,1.38) {$\pi$};
  \node[slab] at (1.3,-0.9) {#1};}

\graphicspath{{figures/}}

\newcommand{\orcid}[1]{\begingroup
  \hypersetup{hidelinks}\href{https://orcid.org/#1}{\includegraphics[width=10pt]{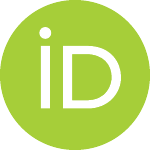}} \endgroup}

\newcommand{\MeV}{\,\mathrm{MeV}}
\newcommand{\GeV}{\,\mathrm{GeV}}
\newcommand{\kpc}{\,\mathrm{kpc}}

\newcommand{\dd}{\mathrm{d}}
\newcommand{\Gpi}{G_{\pi N\Delta}}
\newcommand{\Gax}{G_{aN\Delta}}
\newcommand{\cud}{c_{u-d}}
\newcommand{\vp}{\bm p}
\newcommand{\vl}{\bm l}
\newcommand{\vq}{\bm q}
\newcommand{\vs}{\bm\sigma}
\newcommand{\Amp}{\mathcal M}
\newcommand{\SMref}[1]{see SM Sec.~\ref{S-#1}}
\newcommand{\SMeq}[1]{SM Eq.~(\ref{S-#1})}

\newcommand{\gap}{g_{ap}}
\newcommand{\gag}{g_{a\gamma}}
\newcommand{\cpd}{c_{u+d}}
\newcommand{\fm}{\,\mathrm{fm}}
\newcommand{\vP}{\bm P}

\newcommand{\TSAboveTwoHundredRatio}{2.4^{+0.0}_{-0.8}}

\newcommand{\TSCDeltaFour}{1.58}
\newcommand{\TSCDeltaThree}{-3.15}

\newcommand{\TSCnUnc}{0.017(48)}
\newcommand{\TSContactAll}{0.0}
\newcommand{\TSContactConn}{-5.1}
\newcommand{\TSContactRes}{+5.3}

\newcommand{\TSCpCnCorr}{-0.28}
\newcommand{\TSCpUnc}{-0.438(48)}

\newcommand{\TSCvdUnc}{-0.178(30)}

\newcommand{\TSDraws}{500}

\newcommand{\TSFaN}{7.4^{+1.9}_{-0.9}\times10^{8}\,{\rm GeV}}
\newcommand{\TSFaT}{1.0^{+0.2}_{-0.2}\times10^{9}\,{\rm GeV}}

\newcommand{\TSFluMeanEN}{81}
\newcommand{\TSFluMeanET}{97}

\newcommand{\TSFracD}{38}

\newcommand{\TSHatCFour}{0.86}
\newcommand{\TSInBRNpi}{0.994}
\newcommand{\TSInCFour}{2.17(3)}
\newcommand{\TSInCThree}{-3.61(5)}

\newcommand{\TSInDeltaRun}{0.0237(52)}
\newcommand{\TSInDeltaS}{-0.035(13)}
\newcommand{\TSInFpi}{92.07(57)}
\newcommand{\TSInGA}{1.2754(13)}
\newcommand{\TSInGamma}{117}
\newcommand{\TSInGammaHi}{120}
\newcommand{\TSInGammaLo}{114}
\newcommand{\TSInGax}{1.445(86)}
\newcommand{\TSInGpi}{1.437(27)}
\newcommand{\TSInGpiStatic}{1.04}
\newcommand{\TSInGzero}{0.440(60)}
\newcommand{\TSInHatCEight}{-0.60}
\newcommand{\TSInHatCFour}{2.44(3)}
\newcommand{\TSInHatCTwo}{1.59(3)}
\newcommand{\TSInMDelta}{1232}
\newcommand{\TSInMDeltaHi}{1234}
\newcommand{\TSInMDeltaLo}{1230}
\newcommand{\TSInMN}{938.9}
\newcommand{\TSInMpi}{139.6}
\newcommand{\TSInZ}{0.474(65)}

\newcommand{\TSLOFracD}{41}

\newcommand{\TSLumT}{5.62^{+2.74}_{-1.76}\times10^{76}}

\newcommand{\TSMaT}{5.6^{+1.2}_{-1.0}\,{\rm meV}}

\newcommand{\TSNtotT}{1.11\times10^{81}}

\newcommand{\TSOrderRatioD}{0.701}

\newcommand{\TSOrderRatioT}{0.749^{+0.128}_{+0.010}}
\newcommand{\TSPeakDensity}{1.23}

\newcommand{\TSPeakLapse}{0.76}
\newcommand{\TSPeakRadius}{8.0}
\newcommand{\TSPeakT}{45.1}

\newcommand{\TSRatioTN}{1.87^{+0.12}_{-0.38}}

\newcommand{\TSResCThree}{-0.46}

\newcommand{\TSSpecPeakD}{155}

\newcommand{\TSHotAboveTwoHundredRatio}{2.3^{-0.1}_{-0.7}}

\newcommand{\TSHotDCorrZero}{350}

\newcommand{\TSHotFluMeanEN}{97}
\newcommand{\TSHotFluMeanET}{113}

\newcommand{\TSHotFracD}{49}
\newcommand{\TSHotFracDUnc}{49^{+2}_{-14}}

\newcommand{\TSHotFracIUnc}{4^{+1}_{-3}}

\newcommand{\TSHotFracNUnc}{47^{+15}_{-2}}

\newcommand{\TSHotLumT}{2.51^{+1.42}_{-0.96}\times10^{77}}

\newcommand{\TSHotNtotT}{4.91\times10^{81}}

\newcommand{\TSHotOrderRatioT}{0.734^{+0.163}_{+0.016}}

\newcommand{\TSHotPeakT}{57.3}

\newcommand{\TSHotRatioTN}{2.14^{+0.08}_{-0.51}}

\newcommand{\TSHotSrcDeltaDominates}{175}
\newcommand{\TSHotSrcGainDelta}{2.3}

\newcommand{\TSHotSrcMeanEN}{148}
\newcommand{\TSHotSrcMeanET}{171}

\newcommand{\TSLOHotFracD}{50}

\newcommand{\TSColdAboveTwoHundredRatio}{2.5^{+0.0}_{-0.8}}

\newcommand{\TSColdDCorrZero}{460}

\newcommand{\TSColdFluMeanEN}{73}
\newcommand{\TSColdFluMeanET}{87}

\newcommand{\TSColdFracD}{31}

\newcommand{\TSColdLumT}{2.41^{+1.09}_{-0.66}\times10^{76}}

\newcommand{\TSColdNtotT}{4.83\times10^{80}}

\newcommand{\TSColdOrderRatioT}{0.775^{+0.116}_{+0.009}}

\newcommand{\TSColdPeakT}{38.9}

\newcommand{\TSColdRatioTN}{1.69^{+0.11}_{-0.29}}

\newcommand{\TSLOColdFracD}{34}

\newcommand{\SEBDCold}{371}
\newcommand{\SEBDHot}{101}

\newcommand{\SEBDLOTUnc}{38^{+44}_{-13}}

\newcommand{\SEBDNLONUnc}{68^{+14}_{-16}}

\newcommand{\SEBDNLOTUnc}{46^{+14}_{-9}}

\newcommand{\SEBDParam}{212}

\newcommand{\SEDeltaGainLightNLOUnc}{1.77^{+0.11}_{-0.33}}

\newcommand{\SEDeltaGainNLOUnc}{2.18^{+0.07}_{-0.55}}

\newcommand{\SEModColdBDLOTUnc}{66^{+68}_{-22}}

\newcommand{\SEModColdBDNLONUnc}{112^{+19}_{-24}}

\newcommand{\SEModColdBDNLOTUnc}{78^{+19}_{-15}}
\newcommand{\SEModColdBDParam}{342}

\newcommand{\SEModHotBDLOTUnc}{16^{+21}_{-6}}

\newcommand{\SEModHotBDNLONUnc}{28^{+8}_{-8}}

\newcommand{\SEModHotBDNLOTUnc}{19^{+7}_{-4}}
\newcommand{\SEModHotBDParam}{91}

\newcommand{\SEOZColdBDDelta}{284}
\newcommand{\SEOZColdBDDeltaFull}{131}
\newcommand{\SEOZColdBDDeltaFullUnc}{131^{+40}_{-21}}

\newcommand{\SEOZColdBDDeltaUnc}{284^{+36}_{-14}}

\newcommand{\SEOZHotBDDelta}{69}

\newcommand{\SEOZHotBDDeltaFullUnc}{29^{+13}_{-6}}

\newcommand{\SEOZHotBDDeltaUnc}{69^{+13}_{-5}}

\newcommand{\SEOZHotGainDelta}{2.1}
\newcommand{\SEOZHotGainDeltaFull}{12.1}

\newcommand{\SEOZHotLightShapeGainDelta}{1.11}
\newcommand{\SEOZHotLightShapeGainDeltaFull}{1.32}

\newcommand{\SEPFColdBDDelta}{251}
\newcommand{\SEPFColdBDDeltaFull}{108}
\newcommand{\SEPFColdBDDeltaFullUnc}{108^{+39}_{-19}}

\newcommand{\SEPFColdBDDeltaUnc}{251^{+38}_{-15}}

\newcommand{\SEPFHotBDDelta}{62}
\newcommand{\SEPFHotBDDeltaFull}{23}
\newcommand{\SEPFHotBDDeltaFullUnc}{23^{+11}_{-5}}

\newcommand{\SEPFHotBDDeltaUnc}{62^{+12}_{-5}}

\newcommand{\SEDFBDDeltaFullUnc}{40^{+16}_{-7}}
\newcommand{\SEDFBDDeltaUnc}{111^{+19}_{-10}}
\newcommand{\SEDFBDParam}{190}

\newcommand{\SEDFCnUnc}{0.279(43)}

\newcommand{\SEDFCpUnc}{-0.593(43)}

\newcommand{\SEDFCvdUnc}{-0.342(30)}
\newcommand{\SEDFFaT}{1.7^{+0.3}_{-0.3}\times10^{9}\,{\rm GeV}}
\newcommand{\SEDFGainDelta}{2.9}
\newcommand{\SEDFGainDeltaFull}{22.6}
\newcommand{\SEDFKappaMaxUnc}{2.1^{+0.5}_{-0.4}}
\newcommand{\SEDFLumRatio}{2.81}
\newcommand{\SEDFMaT}{3.4^{+0.8}_{-0.5}\,{\rm meV}}

\newcommand{\SEDFRatioDelta}{2.00}
\newcommand{\SEDFRatioN}{1.21}
\newcommand{\SEDFTanBeta}{10}

\newcommand{\CKDeltaThree}{7.0}

\newcommand{\CKDtwoColdNLOT}{0.11}

\newcommand{\CKDtwoHotNLOT}{0.64}

\newcommand{\CKDtwoNLOTFullUnc}{0.22^{+0.11}_{-0.10}}

\newcommand{\CKHatCTwoRes}{-1.55}

\newcommand{\CKLamNLONBorn}{0.26}

\newcommand{\CKLamNLOTFull}{2.4}
\newcommand{\CKLamNLOTFullUnc}{2.4^{+3.1}_{-1.7}}

\newcommand{\CKMdet}{187}

\newcommand{\CKOursOverPCRed}{4.9}

\newcommand{\CKPCThirtyKSVZFull}{1.6}

\newcommand{\CKPCThirtyKSVZRedFull}{0.49}

\newcommand{\CWAbsPole}{128}

\newcommand{\CWOSPole}{106}

\newcommand{\CWSpPole}{91}
\newcommand{\CWSpPoleHi}{111}
\newcommand{\CWSpPoleLo}{68}

\newcommand{\RBDChangeHK}{7}

\newcommand{\RBDChangeHard}{7}
\newcommand{\RBDChangeSE}{22}

\newcommand{\RBHKrefMid}{9.1}

\newcommand{\RBHKvdMid}{11.1}

\newcommand{\RBLateDensMidHi}{2.9}
\newcommand{\RBLateDensMidLo}{2.1}

\newcommand{\RBOneSDmthirtyDMax}{1.73}

\newcommand{\RBOneSDmthirtyDMin}{1.41}

\newcommand{\RBOneSDmthirtyTMax}{1.24}

\newcommand{\RBOneSDmthirtyTMin}{1.17}

\newcommand{\RBOneSDptwentyfiveDMax}{0.92}

\newcommand{\RBOneSDptwentyfiveDMin}{0.91}

\newcommand{\RBOneSDptwentyfiveTMax}{0.97}

\newcommand{\RBOneSDptwentyfiveTMin}{0.97}

\newcommand{\RBVDLumDFiveMid}{1.8}

\newcommand{\RBVDLumDNineMid}{1.2}

\newcommand{\SIEvFlexND}{17.7}

\newcommand{\SIEvFlexN}{-0.5}

\newcommand{\SIFigNNA}{920}
\newcommand{\SIFigNNB}{550}
\newcommand{\SIFigPiA}{1700}
\newcommand{\SIFigPiB}{23}

\newcommand{\SIFigRatioB}{0.04}

\newcommand{\SIGtAgree}{1.9}

\newcommand{\SIGtNNA}{1500}
\newcommand{\SIGtNNB}{3800}

\newcommand{\SIGtPiA}{4600}
\newcommand{\SIGtPiB}{3300}

\newcommand{\SIGtPiFacB}{143}

\newcommand{\SILellaPiB}{2.0\times10^{4}}

\newcommand{\SINoDetRatioB}{0.64}

\newcommand{\SIdTMzeroABand}{14.5^{+1.3}_{-5.8}}
\newcommand{\SIdTMzeroB}{12.1}
\newcommand{\SIdTMzeroBBand}{12.1^{+1.0}_{-5.8}}

\newcommand{\SIdTMzeroCBand}{10.7^{+0.9}_{-5.7}}

\newcommand{\SIdTMzeroDBand}{10.3^{+0.9}_{-5.1}}

\newcommand{\SIdTMzeroEBand}{9.4^{+1.1}_{-3.5}}
\newcommand{\SIdTMzeroF}{8.2}
\newcommand{\SIdTMzeroFBand}{8.2^{+1.1}_{-2.5}}

\newcommand{\SICalMzeroHi}{5}
\newcommand{\SICalMzeroLo}{2}

\newcommand{\LACountsLB}{6500}

\newcommand{\LACountsRB}{1.1\times10^{4}}

\newcommand{\LACountsUB}{2.3\times10^{4}}

\newcommand{\LAEvFixedLB}{-7.4}

\newcommand{\LAEvFixedLPiB}{20509}

\newcommand{\LAEvFixedRB}{-7.0}

\newcommand{\LAEvFixedRPiB}{13497}

\newcommand{\LAEvFixedUB}{-6.3}

\newcommand{\LAEvFixedUPiB}{7205}

\newcommand{\LAEvFlexLB}{-0.9}

\newcommand{\LAEvFlexLPiB}{71.0}

\newcommand{\LAEvFlexRB}{1.7}

\newcommand{\LAEvFlexRPiB}{36.5}

\newcommand{\LAEvFlexUB}{13.1}

\newcommand{\LAEvFlexUPiB}{63.1}

\newcommand{\LAMeanTB}{34.6}
\newcommand{\LAMeanTC}{33.5}
\newcommand{\LAMeanTD}{30.1}
\newcommand{\LAMeanTE}{24.7}
\newcommand{\LAMeanTF}{19.2}
\newcommand{\LANucRatioMax}{6.6}
\newcommand{\LANucRatioMin}{3.5}

\newcommand{\LAShareRB}{42}

\newcommand{\LAShareRD}{45}

\newcommand{\LAShareUB}{71}

\newcommand{\LAShareUD}{77}

\newcommand{\LATLB}{35.6}
\newcommand{\LATLC}{35.1}
\newcommand{\LATLD}{29.3}
\newcommand{\LATLE}{22.2}
\newcommand{\LATLF}{15.5}

\newcommand{\LATempCalAbs}{3.3}
\newcommand{\LATempCalHi}{2.5}
\newcommand{\LATempCalLo}{-3.3}

\newcommand{\LATempNtotR}{2.5\times10^{4}}
\newcommand{\LATempNtotU}{3100}

\newcommand{\LATempWin}{0.5}

\newcommand{\LATempZthreeAllR}{7 of 10}
\newcommand{\LATempZthreeAllU}{7 of 10}

\newcommand{\LAdTRBBand}{13.2^{+1.0}_{-5.9}}

\newcommand{\LAdTRCBand}{12.3^{+1.2}_{-5.4}}

\newcommand{\LAdTRDBand}{12.0^{+0.7}_{-4.4}}

\newcommand{\LAdTREBand}{11.4^{+0.3}_{-3.8}}

\newcommand{\LAdTRFBand}{8.7^{+1.2}_{-2.7}}
\newcommand{\LAdTRMax}{13.2}
\newcommand{\LAdTRMin}{8.7}

\newcommand{\LAdTUBBand}{21.7^{+4.6}_{-6.9}}

\newcommand{\LAdTUCBand}{18.7^{+3.9}_{-5.8}}

\newcommand{\LAdTUDBand}{19.3^{+3.2}_{-4.9}}

\newcommand{\LAdTUEBand}{18.7^{+2.2}_{-3.5}}

\newcommand{\LAdTUFBand}{16.5^{+1.4}_{-2.2}}
\newcommand{\LAdTUMax}{21.7}
\newcommand{\LAdTUMin}{16.5}

\newcommand{\LAfpiLD}{0}

\newcommand{\LAfpiRDBand}{5^{+4}_{-5}}

\newcommand{\LAfpiREBand}{7^{+0}_{-3}}

\newcommand{\LAfpiUCHi}{7}

\newcommand{\LAfpiUDBand}{22^{+5}_{-13}}

\newcommand{\LAfpiUEBand}{19^{+2}_{-5}}

\newcommand{\PRLBWenh}{25}

\newcommand{\PRLFpi}{92.1}
\newcommand{\PRLGamma}{117}

\newcommand{\PRLHardCut}{220}
\newcommand{\PRLHardGainCut}{5}

\newcommand{\PRLMN}{938.9}

\newcommand{\PRLPeakDApproxThirtyFive}{270}

\newcommand{\PRLPeakDFortyFive}{281}

\newcommand{\PRLQGainMax}{2}

\newcommand{\PRLSlopeThirtyFive}{5.9}

\newcommand{\PRLTwoDelta}{586}
\newcommand{\PRLdelta}{293}
\newcommand{\PRLkappa}{0.28}

\newcommand{\PRLuPeakN}{2.32}

\newcommand{\PRLCkGainDetector}{3.8}

\newcommand{\PRLCkGainSource}{2.4}
\newcommand{\PRLCkGainUnc}{9.1^{+2.7}_{-5.0}}
\newcommand{\PRLSMGammaColl}{106}
\newcommand{\PRLSMGammaMes}{95}
\newcommand{\PRLSMGammaTot}{201}
\newcommand{\PRLSMGammaVacPole}{116}

\newcommand{\PRLSMPeakDens}{1.18}
\newcommand{\PRLSMPeakR}{8.2}
\newcommand{\PRLSMPeakT}{45.4}
\newcommand{\PRLSMPeakYp}{0.20}
\newcommand{\PRLSMPoleW}{1069}
\newcommand{\PRLSMShift}{12}
\newcommand{\PRLSMShiftp}{26}
\newcommand{\PRLSMSigmaS}{-245}
\newcommand{\PRLSMSigmaV}{209}
\newcommand{\PRLSMU}{-37}
\newcommand{\PRLSMUp}{-78}

\newcommand{\DXcFiveHat}{-0.091}
\newcommand{\ERNewASignedA}{+0.7}
\newcommand{\ERNewASignedB}{+0.6}
\newcommand{\ERNewASignedC}{+0.7}
\newcommand{\ERSpecTTwoMin}{+1}
\newcommand{\ERSpecTTwoMax}{+3}
\newcommand{\ERSpecNFourMin}{+32}
\newcommand{\ERSpecNFourMax}{+45}
\newcommand{\ERSpecTFourMin}{-12}
\newcommand{\ERSpecTFourMax}{-9}

\newcommand{\EDHotT}{45}

\makeatletter\let\inputrows\@@input\makeatother

\makeatletter
\renewcommand\onecolumngrid{\do@columngrid{one}{\@ne}\def\set@footnotewidth{\onecolumngrid}%
  \def\footnoterule{\kern-6pt\hrule width 1.5in\kern6pt}}
\renewcommand\twocolumngrid{\def\footnoterule{\dimen@\skip\footins\divide\dimen@\thr@@
  \kern-\dimen@\hrule width.5in\kern\dimen@}\do@columngrid{mlt}{\tw@}}
\makeatother

\makeatletter
\newcommand*{\addFileDependency}[1]{\typeout{(#1)}\@addtofilelist{#1}\IfFileExists{#1}{}{\typeout{No file #1.}}}
\makeatother

\begin{document}

\title{Harder, Better, Faster, Stronger: The \texorpdfstring{$\Delta(1232)$}{Delta(1232)} Resonance\texorpdfstring{\\}{} in Supernova QCD Axion Emission}

\author{{Paolo Arcangeli}\,\orcid{0009-0009-7263-898X}}
\affiliation{Technical University of Munich, TUM School of Natural Sciences, Physics Department, James-Franck-Str.~1, 85748 Garching, Germany}
\author{Konstantin Springmann\,\orcid{0000-0002-9617-6136}}
\affiliation{Department of Particle Physics and Astrophysics, Weizmann Institute of Science, Rehovot, Israel 7610001}
\author{{Michael Stadlbauer}\,\orcid{0000-0003-1711-0647}}
\affiliation{Technical University of Munich, TUM School of Natural Sciences, Physics Department, James-Franck-Str.~1, 85748 Garching, Germany}
\affiliation{Max Planck Institute for Physics, Boltzmannstr.~8, 85748 Garching, Germany}
\affiliation{Theoretical Particle Physics Laboratory (LPTP), Institute of Physics, EPFL, Lausanne, Switzerland}
\author{{Stefan Stelzl}\,\orcid{0000-0001-5964-1054}}
\affiliation{Institut de F\'isica d'Altes Energies (IFAE) and Barcelona Institute of Science and Technology (BIST), Campus UAB, 08193 Bellaterra (Barcelona), Spain}
\author{Andreas Weiler\,\orcid{0000-0001-9253-0831}}
\affiliation{Technical University of Munich, TUM School of Natural Sciences, Physics Department, James-Franck-Str.~1, 85748 Garching, Germany}

\date{\today}

\begin{abstract}
The high-energy tail of supernova axion emission controls searches whose response grows with axion energy.
We identify a novel production channel, in which the axion is emitted through an intermediate $\Delta(1232)$ resonance, which adds a hard component to the spectrum.
We calculate the axion production rate from nucleon-nucleon Bremsstrahlung in chiral effective field theory through next-to-leading order. 
Including the $\Delta(1232)$ roughly doubles the luminosity at one second after bounce and increases the number of escaping axions above $200\MeV$ by a factor of about $3$ over the first nine seconds, relative to the nucleon-only calculation.
Adding this new production channel roughly doubles $\gamma$-ray counts from conversion in the progenitor's magnetic field and increases water-Cherenkov counts by about an order of magnitude through its combined effects on emission and absorption.
The same hard component can mimic a higher core temperature or thermal-pion emission in $\gamma$-ray spectral fits.
The $\Delta(1232)$ resonance is therefore crucial for predicting supernova axion signals and interpreting their spectra.

\end{abstract}

\maketitle

\textbf{\textit{Introduction}}---
The QCD axion solves the strong CP problem~\cite{Peccei:1977hh,Peccei:1977ur, Weinberg:1977ma,Wilczek:1977pj,DiLuzio:2020wdo} and may constitute dark matter~\cite{Preskill:1982cy,Abbott:1982af,Dine:1982ah}.
Supernova (SN)~1987A~\cite{Kamiokande-II:1987idp,Bionta:1987qt} provides one of the strongest constraints on axion couplings to nucleons: axion emission from the hot proto-neutron star would have shortened the neutrino burst~\cite{Raffelt:1987yt,Turner:1987by, Burrows:1988ah,Keil:1996ju,Raffelt:2006cw,Carenza:2019pxu,Caputo:2024oqc,Springmann:2024mjp,Springmann:2024ret}.
Axions from a future Galactic SN could be detected as $\gamma$ rays after conversion in the magnetic field of the progenitor~\cite{Manzari:2024jns, Candon:2025sdm} or of the Galaxy~\cite{Meyer:2016wrm,Fiorillo:2026rgo}, or through absorption in large water Cherenkov detectors~\cite{Engel:1990zd,Vonk:2022tho,Carenza:2023wsm,Cavan-Piton:2025nsj,DiLuzio:2026sxp}.
Whereas the SN~1987A bound depends on the total emission, these signals depend on the \textit{axion spectrum}: for meV-mass QCD axions the probability of conversion in the progenitor field grows with the axion energy, and absorption with pion production requires an energy above the pion threshold.
A fit to the $\gamma$-ray spectrum can be used to infer the supernova core temperature and the pion content \cite{Lella:2024hfk}.
Including the $\Delta(1232)$ as an explicit degree of freedom significantly enhances the high energy tail of the spectrum.

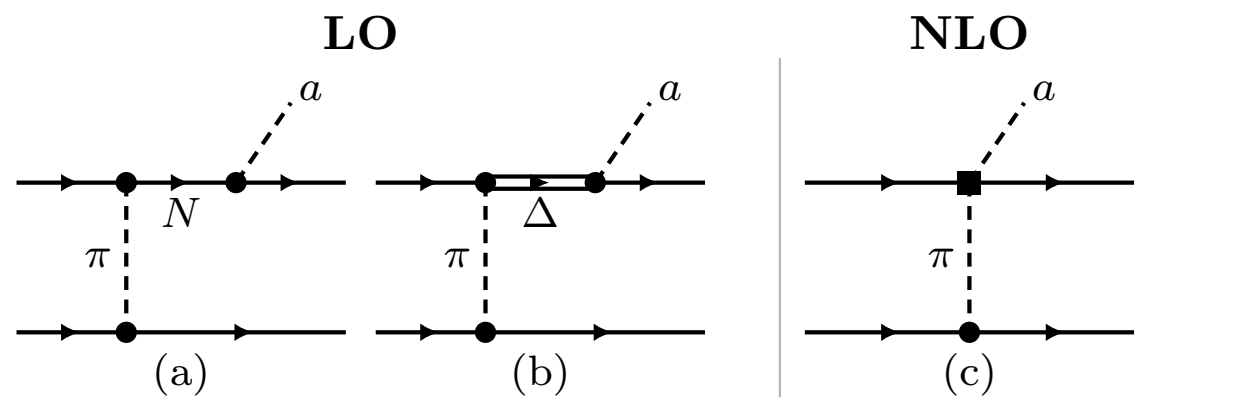
\begin{figure}[t]
\centering
\begin{tikzpicture}[scale=0.66]
\node[font=\small\bfseries] at (3.45,3.0) {LO};
\node[font=\small\bfseries] at (9.55,3.0) {NLO};
\draw[gray!60,line width=0.4pt] (7.65,-0.65) -- (7.65,2.75);
\begin{scope}[shift={(0,0)}]\FDpole{N}{o}{(a)}\end{scope}
\begin{scope}[shift={(3.6,0)}]\FDpole{D}{o}{(b)}\end{scope}
\begin{scope}[shift={(7.9,0)}]\FDcontact{vsq}{(c)}\end{scope}
\end{tikzpicture}
\caption{Example diagrams of $NN\to NNa$ at leading and next-to-leading order (LO, NLO), with the constants at their vertices.
LO: intermediate nucleon (a; $G_A$, $C_p$, $C_n$) and $\Delta$ (b; $\Gpi$, $\Gax$).
NLO: the $a\pi NN$ contact term (c) of the chiral connection, fixed by chiral symmetry, and of the LECs $\hat c_3^{\rm res}$, $\hat c_4^{\rm res}$ and $\hat c_9$.}
\label{fig:diagrams}
\end{figure}

The dominant production process during a SN is nucleon--nucleon Bremsstrahlung, $NN\to NNa$, in which a nucleon radiates the axion while it exchanges a pion with another nucleon~\cite{Iwamoto:1984ir,Brinkmann:1988vi}, see Fig.~\ref{fig:diagrams}.
The nuclear side of this process has been refined phenomenologically~\cite{Hanhart:2000ae,Carenza:2019pxu,Raffelt:1991pw,Chang:2018rso,Ericson:1988wr,Janka:1995ir}.
Vertex corrections in vacuum and at finite density have been systematically computed in chiral perturbation theory (ChPT)~\cite{Balkin:2020dsr,Vonk:2020zfh,Vonk:2021sit,Springmann:2024mjp,Springmann:2024ret,DiLuzio:2024vzg}.
Throughout, the baryon that absorbs the pion and radiates the axion is a nucleon.

At sufficiently high energies, the intermediate baryon can instead be a $\Delta(1232)$.
It has spin and isospin $3/2$, lies $\delta=M_\Delta-M_N\simeq\PRLdelta\MeV$ above the nucleon, and decays in vacuum into a nucleon and a pion with width $\Gamma\simeq\PRLGamma\MeV$~\cite{ParticleDataGroup:2024cfk}.
Pions and QCD axions can both induce the spin and isospin transition from a nucleon to a $\Delta$.
Consider an axion emitted by an outgoing nucleon.
The intermediate nucleon is off shell by $\sim E_a$, suppressing its propagator at large $E_a$.
An intermediate $\Delta$, however, can be close to its mass shell for $E_a\sim \delta$.
At typical SN temperatures, $T\simeq30$--$50\MeV$, such axions lie in the Boltzmann tail. %
We show that the $\Delta$-resonance compensates for a large part of this suppression due to its small width. 
The $\Delta$ has been included in pion-induced production, $\pi N\to Na$~\cite{Ho:2022oaw,Vonk:2022tho,Li:2023thv}, which relies on a debated abundance of thermal pions~\cite{Carenza:2020cis,Fore:2023gwv,Fiorillo:2025gnd}, and in photon Bremsstrahlung, $pp\to pp\gamma$~\cite{Tiator:1978sj,deJong:1994krc,Jetter:1994zza,Martinus:1998zz}, but not in axion Bremsstrahlung, the dominant channel, which needs no thermal pions.

We compute $NN\to NNa$ in ChPT with nucleon and explicit $\Delta$ intermediate states, integrate it over SN simulations, and derive its consequences for the signals from a future Galactic SN.
We provide details in the Supplemental Material (SM)~\cite{SM}.

\textbf{\textit{Analytic estimate}}---%
We first derive an analytic estimate of the axion spectrum in a simplified setting to isolate the role of the $\Delta$.
The estimate shows that the $\Delta$ produces a peak at $E_a\simeq\delta$, fixed by the mass splitting far above the nucleon peak at $E_a\simeq\PRLuPeakN\,T$, and that the height of this peak relative to the nucleon peak grows steeply with the temperature.
Since the $\Delta$ is heavier than the nucleon, an axion with $E_a\simeq\delta$ must be supplied by nucleon pairs from the Boltzmann tail, whose phase space is suppressed by $e^{-\delta/T}$.
The width $\Gamma$ of the $\Delta$ compensates for this because it is small compared to $2\delta\simeq\PRLTwoDelta\MeV$, and the squared amplitude is enhanced by $(2\delta/\Gamma)^2\simeq\PRLBWenh$ relative to the off-shell nucleon, see Eq.~\eqref{eq:FGD}.

In $NN\to NNa$ Bremsstrahlung, two nucleons exchange a pion, and one emits an axion through an intermediate nucleon, Fig.~\ref{fig:diagrams}\hyperref[fig:diagrams]{(a)}~\cite{Brinkmann:1988vi} or $\Delta$, Fig.~\ref{fig:diagrams}\hyperref[fig:diagrams]{(b)}.
Both the pion and the axion couple the nucleon to the $\Delta$ through the isovector axial current.
Isospin forbids the axion isoscalar source to couple a nucleon to a Delta.
For a systematic derivation of the Lagrangian and the vertices, \SMref{sec:lagrangian}.
For this analytic estimate, we assume axion momenta small compared to the nucleon momenta (soft-axion approximation), and a $\Delta$ with constant vacuum width, and neglect the pion mass.
In this approximation, the $\Delta$ can go on shell only when pion absorption precedes axion emission.
The reverse ordering remains off-shell.

\begin{figure}[t]
\centering
\includegraphics[width=\columnwidth]{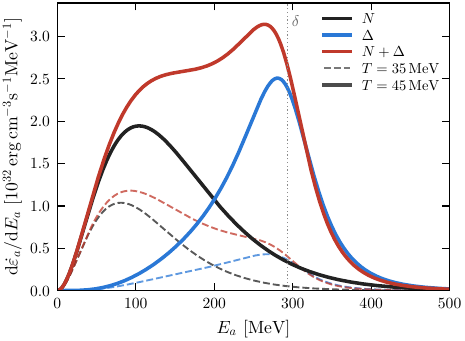}
\caption{Estimate of the axion emissivity per unit energy in a Fermi gas for the KSVZ axion with $f_a=10^9\GeV$: nucleon contribution (black), $\Delta$ contribution (blue) and their sum including the interference (red), at $T=35\MeV$ (dashed) and $45\MeV$ (solid).
The dotted line marks $E_a=\delta$.}
\label{fig:analytic}
\end{figure}

We approximate the supernova background as non-degenerate, non-relativistic nucleons in symmetric matter at nuclear saturation density, $n_0=0.16\fm^{-3}$.
The emissivity $\dot\varepsilon_a$ is the energy radiated in axions per unit volume and time (see \hyperref[em:rate]{Rate}).
The spectrum including only nucleons, denoted as $\dot\varepsilon_{aN}$ and shown in black in Fig.~\ref{fig:analytic}, is
\begin{equation}
 \frac{\dd\dot\varepsilon_{aN}}{\dd E_a}=\frac{n_0^2G_A^4M_N^{3/2}C_{pn}}{1536\pi^{7/2}F_\pi^4f_a^2}\,\frac{E_a^3}{T^{1/2}}\,
 e^{-E_a/2T}K_1\Big(\frac{E_a}{2T}\Big),
 \label{eq:FGN}
\end{equation}
with the nucleon axial coupling $G_A$, the pion decay constant $F_\pi=\PRLFpi\MeV$, the nucleon mass $M_N=\PRLMN\MeV$, the modified Bessel function $K_1$ and $C_{pn}=7(C_p^2+C_n^2)+4C_pC_n$, where $C_p$ and $C_n$ are the dimensionless axion couplings to protons and neutrons.
The $\Delta$ diagrams differ in the couplings, the spin and isospin factors and the propagator.
Near the resonance, the spectrum including only the $\Delta$ diagrams, $\dot\varepsilon_{a\Delta}$ (blue line in Fig.~\ref{fig:analytic}), is
\begin{equation} \frac{\dd\dot\varepsilon_{a\Delta}}{\dd E_a}\simeq\kappa\,\frac{E_a^2}{(E_a-\delta)^2+\Gamma^2/4}\,\frac{\dd\dot\varepsilon_{aN}}{\dd E_a},
 \label{eq:FGD}
\end{equation}
where $E_a^2/|E_a-\delta+\tfrac i2\Gamma|^2$ is the ratio of the $\Delta$ to the nucleon propagator.
The intermediate nucleon is off shell by $E_a$, and the intermediate $\Delta$ by $E_a-\delta$.
At the pole this ratio is $(2\delta/\Gamma)^2$.
The constant $\kappa$ collects the couplings and the spin--isospin factors and for the KSVZ axion~\cite{Kim:1979if,Shifman:1979if} is given by  $\kappa=\tfrac{352}{81}\Gpi^2\Gax^2\cud^2/(G_A^2C_{pn})\simeq \PRLkappa$.
The peak of the $\Delta$ spectrum is dictated by the interplay of the resonant enhancement and the Boltzmann suppression,
\begin{equation}
 E_a^{*}\simeq\delta-\frac{\Gamma^2}{8T}\Big(1-\frac{9T}{2\delta}\Big),
 \label{eq:FGpeak}
\end{equation}
$\PRLPeakDApproxThirtyFive\MeV$ at $T=35\MeV$.
Its position is set by $\delta$, the temperature enters only through the correction of order $\Gamma^2/T$ (Fig.~\ref{fig:analytic}).
Its height relative to the nucleon peak, $R(T)\propto(2\delta/\Gamma)^2(\delta/T)^{5/2}e^{-\delta/T}$, grows steeply with temperature, $\dd\ln R/\dd\ln T=\delta/T-5/2\simeq\PRLSlopeThirtyFive$ at $35\MeV$.

The ratio in Eq.~\eqref{eq:FGD} is small near the nucleon peak and reaches $\kappa(2\delta/\Gamma)^2$ at $E_a=\delta$.
Integrated over the spectrum at $T=35$--$45\MeV$, the $\Delta$ raises the emissivity by a factor of up to $\PRLQGainMax$, but the number of axions above $\PRLHardCut\MeV$, to which the $\gamma$-ray and Cherenkov signals are most sensitive, by a factor of $\sim\PRLHardGainCut$.

While the analytic estimate gives physical intuition, it is not quantitatively correct.
Pions are massive, SN matter is neutron-rich and partially degenerate, and nucleons interact with the medium.
Collisions broaden the $\Delta$ width to $1.5$--$2$ times its vacuum width.
Recoil also reshapes the tail of the spectrum: forward axion emission is Doppler-hardened by the relative momentum of the $\Delta$ to the fluid, while the relativistic transition coupling depends on the axion momentum in the $\Delta$ rest-frame.
At next-to-leading order (NLO) in the chiral expansion, vertices start to account for this distinction (\SMref{sec:results}).
We therefore retain the axion momentum, NLO vertices and in-medium width in the full calculation.
The central result of the estimate carries over: the $\Delta$ enhances the hard axion tail more than the total emission, since the ratio of the $\Delta$ to the nucleon propagator grows with $E_a$ up to the resonance.

\begin{samepage}
\textbf{\textit{ChPT with an explicit $\Delta$}}---
We investigate the axion emission in the effective theory of the QCD axion, pions, and non-relativistic baryons.
When the $\Delta$ is integrated out, ChPT counts $\delta$ as a hard scale and replaces the $\Delta$ exchange by a series of contact terms in powers of $E_a/\delta$, whose leading terms contribute to the LECs $\hat c_3$ and $\hat c_4$, defined in Eq.~\eqref{eq:L}.
We keep the $\Delta$ dynamical and count $\delta$ as a soft scale, known as the small-scale expansion~\cite{Jenkins:1991es,Hemmert:1996xg,Hemmert:1997ye,Fettes:2000bb}: each order adds one power of $E_a/\Lambda_\chi$, $\delta/\Lambda_\chi$ or $|\vp|/M_N$ to the amplitude, with $\Lambda_\chi\simeq4\pi F_\pi\simeq1.2\GeV$ and the nucleon momenta $|\vp|$, which are at least of the order of the Fermi momentum $k_f$.
The LO and NLO heavy-baryon (HB) Lagrangians are
\begin{align}
 \mathcal L_{\rm LO}={}&\bar N\big(iv\cdot D+g_A\,S\cdot u+g_0\,S\cdot\hat u\big)N\nonumber\\
 &-\bar T^\mu\big(iv\cdot D-\delta\big)T_\mu
 +\big(g_{\pi N\Delta}\,\bar T^\mu_iw^i_\mu N+{\rm h.c.}\big),\nonumber\\
 \mathcal L_{\rm NLO}={}&\bar N\big(\hat c_3^{\rm res}\,u\cdot u+\hat c_4^{\rm res}[S^\mu,S^\nu]u_\mu u_\nu+\hat c_9\,u\cdot\hat u\big)N
 \nonumber\\
 &+\mathcal L_{1/M_N}+\dots,
 \label{eq:L}
\end{align}
\end{samepage}
Here $N$, $T_\mu$ are the nucleon and $\Delta$ fields in the HB limit, $v$ and $S_\mu$ their velocity and spin, and $g_A$, $g_0$ the isovector and isoscalar nucleon axial couplings.
The axion and pions are captured in $u_\mu=\tau^iw^i_\mu=-\tau^i\partial_\mu\pi^i/f_\pi+\cud\,\tau^3\partial_\mu a/f_a$ and $\hat u_\mu=\cpd\,\partial_\mu a/f_a$, with $c_{u\pm d}=\tfrac12(c_u\pm c_d)$ and the axion couplings $c_u$, $c_d$ to the up and down quarks.
At NLO, the recoil Lagrangian $\mathcal L_{1/M_N}$ contains nucleon and $\Delta$ relativistic vertex corrections of order $1/M_N$, that arise when integrating out the heavy field components of the baryons (\SMref{sec:lagrangian}).
Equation~\eqref{eq:L} shows only the dominant operators.
The full Lagrangians are given in SM Sec.~\ref{S-sec:lagrangian}.

Furthermore, vertices of the form $a\pi NN$ start contributing, see Fig.~\ref{fig:diagrams} (c).
They include a term from the chiral connection in $D_\mu$~\cite{Carena:1988kr,Choi:2021ign,Springmann:2024mjp}, and contributions from $\hat c_3^{\rm res}$, $\hat c_4^{\rm res}$.
Note that $\hat{c}_i^{\rm res}$ only includes the residual values $\hat c_i^{\rm res}=\hat c_i-c_i^\Delta$ that remain when the $\Delta$ is included explicitly~\cite{Krebs:2007rh,Hoferichter:2015tha,Siemens:2016jwj}~(\SMref{sec:lagrangian}).
Due to its isospin structure, the constant $\hat c_9$ is not affected when integrating out the $\Delta$.
Its value is unknown and we draw it around $\pm 0.5\,\GeV^{-1}$~\cite{Springmann:2024mjp}.
Loops enter at the next order, as do $NNN\Delta$ contact terms, since through NLO these cancel against their exchange diagrams (\SMref{sec:nnndelta}).
Once SN simulations including $\Delta$s in the EOS become available, one should also add $\Delta$ external states.

We follow this power counting except for internal propagators, where the width and the kinetic recoil terms are resummed to preserve the resonance pole.
When calculating the emissivity, we use the same dispersion relations for external-state normalization and phase space (see \hyperref[em:amp]{Amplitude}).

We consistently match each constant to data at the kinematic point where it is measured.
Following Ref.~\cite{Springmann:2024mjp}, lower-case symbols denote parameters of the Lagrangian and capitals denote physical constants; see \hyperref[em:constants]{Constants} and SM Sec.~\ref{S-sec:matching} for details.

\begin{figure}[t]
\centering
\includegraphics[width=\columnwidth]{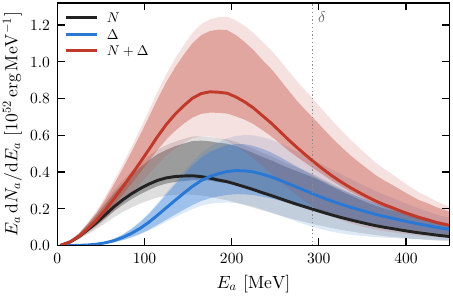}
\caption{Energy spectrum $E_a\,\dd N_a/\dd E_a$ of the axions at the source, the emissivity per unit energy integrated over the star and over the first nine seconds of the Garching model SFHo-20.0, against the local axion energy, for the KSVZ axion with $f_a=10^9\GeV$ at NLO: nucleon contribution (black), $\Delta$ contribution (blue) and the sum including interference (red), with uncertainty from the constants (dark bands) and the truncation (light bands).
The dotted line marks $E_a=\delta$.}
\label{fig:spectrum}
\end{figure}

\begin{figure*}[t]
\centering
\includegraphics[width=\textwidth]{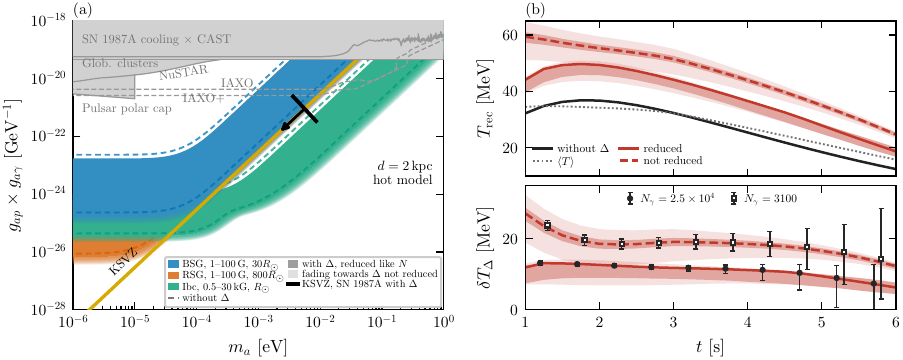}
\caption{(a) \emph{Fermi}-LAT sensitivity to $\gap\times\gag$ for a SN at $2\kpc$ with a blue or red supergiant (radius $(30$--$800)\,R_\odot$, surface field $(1$--$100)\,$G) or a Type Ibc progenitor ($R_\odot$, $(0.5$--$30)\,$kG), with the KSVZ axion (yellow, for the DFSZ axion see SM Fig.~\ref{S-fig:dfszreach}).
Grey: relevant bounds and projections~\cite{Springmann:2024mjp,CAST:2024eil,Dolan:2022kul,Ruz:2024gkl, Noordhuis:2022ljw,IAXO:2019mpb}.
Dashed: SFHo-20.0 as one-zone source of Ref.~\cite{Candon:2025sdm}; shaded: this source with our NLO $\Delta$ contribution in the same cell, reduced like the nucleon part and fading out towards an unreduced $\Delta$ contribution.
Black bar: SN~1987A bound on the KSVZ axion with the $\Delta$, i.e. $m_a\lesssim\TSMaT$.
(b) Temperature reconstructed from the \emph{Fermi}-LAT counts of SFHo-18.8 with the three-parameter Bremsstrahlung fit of Ref.~\cite{Lella:2024hfk}, for its Bremsstrahlung spectrum alone (black) and with our $\Delta$ part reduced like it (red solid) or not reduced (red dashed); the dotted line shows the emission-weighted temperature $\langle T\rangle$ of the profile.
Below, the difference $\delta T_\Delta$ between the reconstructed temperatures with the statistical $16$--$84\%$ ranges of $\LATempWin\,$s windows (filled points reduced, open points not reduced) for the largest photon numbers $N_\gamma$ in $1$--$6\,$s at which the template is not rejected.
Bands as in Fig.~\ref{fig:spectrum}.}
\label{fig:detection}
\end{figure*}

\textbf{\textit{Supernova axion emission}}---We evaluate the full NLO axion emission rate in simulated supernova cores.
In the SN core, the density background shifts the baryon self-energies~\cite{Hempel:2014ssa}, while Pauli blocking and collisions modify the $\Delta$ width.
We use the value of the $\Delta$ potential given by the GIBUU package~\cite{Buss:2011mx}, $U_\Delta =\tfrac23 U_N$.
In the mean-field models adopted in~\cite{Hempel:2014ssa}, these potentials come from a scalar and a vector contribution, $U_N=\Sigma_S^N+\Sigma_V^N$ and similar for the $\Delta$, for the impact of alternative choices, \SMref{sec:delta-width}.
We adopt the same splitting, $\Sigma^\Delta_{S,V}=\tfrac23\Sigma^N_{S,V}$. %
Varying $U_\Delta-U_N$ within $[-30,25]\MeV$~\cite{Bodek:2020wbk,Cozma:2021tfu} changes the luminosity $1\,$s post-bounce by $\fpeval{round(100*(\RBOneSDptwentyfiveTMin-1))}\%$ to $+\fpeval{round(100*(\RBOneSDmthirtyTMax-1))}\%$, but the number of axions above $200\MeV$ by at most $\RBDChangeHard\%$ (\SMref{sec:medium-profiles}).
Varying the splitting into scalar and vector components has little influence.
The in-medium $\Delta$ width has two terms, $\Gamma_\Delta=\Gamma_{\Delta\to N\pi}^{\rm blocked}+\Gamma_{\rm coll}$: the decay $\Delta\to N\pi$, reduced by Pauli blocking of the final nucleon, and collisions between nucleons and the $\Delta$.
We adopt the zero-temperature collision width from~\cite{Oset:1987re,Buss:2011mx} (for finite temperature, see~\cite{Ghosh:2016hln}).
Near the emission peak the total width is up to twice its vacuum value.
A similar result can also be found using the leading ChPT calculation of the collision width from $N \Delta \to N N$.
An independent spreading-width prescription gives a comparable luminosity at $1\,$s~\cite{VanDaele:2001cz} (\SMref{sec:width} for a detailed discussion).
Furthermore, we use one-pion exchange in the Born approximation.
Iterating the tensor force between nucleons can reduce the soft-radiation rate from nucleons in neutron matter to roughly a quarter~\cite{Hanhart:2000ae}.
The corresponding correction to the $\Delta$-contribution however is unknown (\SMref{sec:integration}).
A systematic study of the potential for Bremsstrahlung beyond the one-pion-exchange (including an intermediate $\Delta$) is left for a future publication~\cite{Arcangeli:2026nuc}.

We evaluate the emissivity with exact kinematics in three spherical Garching simulations with the SFHo equation of state and muons, for progenitors of $18.8$, $18.6$ and $20.0\,M_\odot$~\cite{Bollig:2020xdr,Garching,Steiner:2012rk}.
Their cores range from cold to hot: at $1\,$s the emission peaks at $T\simeq\fpeval{round(\TSColdPeakT)}$, $\fpeval{round(\TSPeakT)}$ and $\fpeval{round(\TSHotPeakT)}\MeV$, respectively, at one to a few times $n_0$ (SM Fig.~\ref{S-fig:width-profiles}).
We integrate the emissivity per unit energy over the star and the first nine seconds, assuming free streaming, and obtain the spectrum at the source, shown in Fig.~\ref{fig:spectrum} and SM Figs.~\ref{S-fig:spectrum-s188} and~\ref{S-fig:spectrum-s186}.
Its uncertainty bands contain the matched constants and the chiral truncation~\cite{Springmann:2024mjp}, but none of the effects discussed above (\SMref{sec:uncertainty}).
For SFHo-20.0 the $\Delta$ contribution overtakes the nucleon contribution above $\TSHotSrcDeltaDominates\MeV$ and raises the spectrum at $E_a=\delta$ by a factor $\TSHotSrcGainDelta$, so that the mean axion energy grows from $\TSHotSrcMeanEN$ to $\TSHotSrcMeanET\MeV$.

\textbf{\textit{Implications for detection}}---
The intermediate $\Delta$ enhances the high-energy tail more strongly than the total emission.
Searches that weight high-energy axions therefore benefit most from this contribution.

\emph{Gamma rays.}
Axions convert into photons in the dipole field of the progenitor star~\cite{Manzari:2024jns, Candon:2025sdm}.
The coherence length between relativistic massive axions and the photon with negligible plasma mass is set by their momentum difference, $\ell  \sim 1/\Delta p \sim 2E_a/m_a^2$.
For QCD axions with meV masses, axion and photon dephase over a distance smaller than the coherence length of the magnetic field, and the conversion probability grows with the energy, $P=\gag^2B_0^2E_a^2/m_a^4$ for a surface field $B_0$.
Since $\gap\gag\propto m_a^2$ for the QCD axion, the photon count then does not depend on the axion mass.
Using the effective area $A_{\rm eff}$ of \emph{Fermi}-LAT~\cite{Fermi-LAT:2009ihh} the spectrum is weighted with $\sim E_a^2A_{\rm eff}$, and the $\Delta$ raises the photon count by $\SEDeltaGainNLOUnc$ compared to the nucleon-only contribution (\SMref{sec:gamma}).
For lighter axions the dephasing length exceeds the stellar radius and, as long as the vacuum birefringence is negligible, the probability does not depend on the energy. The count then weights the spectrum with $\sim A_{\rm eff}$, and the $\Delta$ raises it by $\SEDeltaGainLightNLOUnc$.
The hot and the cold source of Ref.~\cite{Candon:2025sdm}, i.e. one-zone models of SFHo-20.0 and SFHo-18.8, use a nucleon rate calibrated to the measured neutron--proton cross section, several times below one-pion exchange~\cite{Fiorillo:2025gnd}.
This calibration contains the corrections beyond the Born approximation that our rates neglect~\cite{Hanhart:2000ae}. For comparison we take them into account, although their effect on the $\Delta$ part is not known.
Adding our $\Delta$ part, evaluated in the same one-zone cells, to these sources lowers the ratio of field strength to distance, $B/d$, above which the KSVZ axion is detectable, from $\SEBDHot$ and $\SEBDCold\,{\rm G}/{\rm kpc}$ to $\SEOZHotBDDeltaUnc$ and $\SEOZColdBDDeltaUnc\,{\rm G}/{\rm kpc}$ assuming it is reduced like the nucleon part, and to $\SEOZHotBDDeltaFullUnc$ and $\SEOZColdBDDeltaFullUnc\,{\rm G}/{\rm kpc}$ assuming it is not reduced (Fig.~\ref{fig:detection}(a), \SMref{sec:gamma}).
In the latter case, a hot core makes the KSVZ axion detectable for red and blue supergiants with fields of $100\,$G at $2\kpc$; for a cold core they must lie within about $\fpeval{round(100/\SEOZColdBDDeltaFull,1)}\kpc$.

For sufficiently light axion-like particles converting in the Galactic magnetic field~\cite{Brockway:1996yr,Grifols:1996id,Payez:2014xsa,Hoof:2022xbe}, the $\gamma$-ray spectrum might reveal information about the SN core.
Ref.~\cite{Lella:2024hfk} proposed to measure the core temperature and the abundance of thermal pions with the \emph{Fermi}-LAT spectrum, and thereby test the EOS. 
The inference fits two templates. 
The nucleon Bremsstrahlung template is a Gamma distribution in energy with fitted normalization, mean energy and shape parameter, whose mean energy gives the temperature $T_{\rm rec}$ through a relation calibrated on one-zone models. 
The other template is associated to the $\pi^-p\to na$ process~\cite{Carenza:2020cis} and has a normalization that measures the pion abundance~\footnote{Note that thermal pions require a $\pi^-$ chemical potential near the pion mass, which softens the equation of state in tension with observed heavy pulsars~\cite{Fore:2019wib,Fore:2023gwv,Vijayan:2023qrt,Balkin:2023xtr,Antoniadis:2013pzd,Fonseca:2021wxt}.}.
Our calculation tests how the omitted $\Delta$ resonance affects this inference.
The calibration and the templates match to the Bremsstrahlung spectrum of Ref.~\cite{Lella:2024hfk}, which corrects one-pion exchange for multiple nucleon scattering~\cite{Raffelt:1991pw,Janka:1995ir} and for two-pion exchange through a $\rho$-meson term~\cite{Ericson:1988wr}, and is reduced by a factor $\LANucRatioMin$--$\LANucRatioMax$ compared to one-pion exchange.
To isolate the effect of the $\Delta$ we take this spectrum as the nucleon part and add our $\Delta$ part, reduced like the nucleon part or not reduced, as in Fig.~\ref{fig:detection}(a) (\SMref{sec:inference}).
Without the $\Delta$, the template then returns the emission-weighted temperature of the pion-free SFHo-18.8 profile to within $\LATempCalAbs\MeV$, see Fig.~\ref{fig:detection}(b).
Including the $\Delta$ raises $T_{\rm rec}$ by $\delta T_\Delta=\LAdTRMin$--$\LAdTRMax\MeV$ between $1$ and $6\,$s, or by $\LAdTUMin$--$\LAdTUMax\MeV$ if it is not reduced.
Fitted in $\LATempWin\,$s windows, the template describes the data up to $N_\gamma=\LATempNtotR$ photons in $1$--$6\,$s ($\LATempNtotU$ without reduction), the largest number with a median $p$-value of at least $0.05$, and for these photons $\delta T_\Delta$ exceeds three standard deviations of $T_{\rm rec}$ in \LATempZthreeAllR{} windows (\LATempZthreeAllU{}). 
A fit that omits the $\Delta$ but allows pion conversion assigns $\LAfpiRDBand\%$ ($\LAfpiUDBand\%$) of the photons in $t\in[3,4]\,$s to that channel, although the spectrum contains no thermal pions.
Including the $\Delta$ in the templates is therefore necessary to infer the core temperature and pion content from an axion spectrum and use them to constrain the equation of state.

\begin{figure}[t]
\centering
\includegraphics[width=\columnwidth]{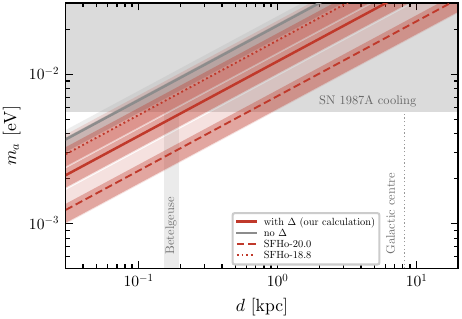}
\caption{Smallest KSVZ mass that gives two $\pi^0$ events in one Hyper-Kamiokande tank ($\CKMdet\,$kt) against the distance, for SFHo-18.6 with the $\Delta$ in emission and absorption (red), and without the $\Delta$ (grey), and with the $\Delta$ for SFHo-20.0 (dashed) and SFHo-18.8 (dotted); shaded: excluded by SN~1987A cooling with the $\Delta$.
Bands as in Fig.~\ref{fig:spectrum}.}
\label{fig:cherenkov}
\end{figure}

\emph{Water Cherenkov detectors.}
An axion with $E_a>145\MeV$ can be absorbed by a hydrogen nucleus in water and produce a neutral pion, $ap\to p\pi^0$~\cite{Vonk:2022tho,Li:2023thv,Cavan-Piton:2025nsj}.
We compute its cross section including the $\Delta$ diagrams of process $a N\to \pi N$, which is the inverse process considered in~\cite{Choi:2021ign,Ho:2022oaw} (\SMref{sec:cherenkov}).
The excitation of the $\Delta$ dominates: at $E_a=300\MeV$ it raises the cross section of the nucleon diagrams by a factor $\CKDeltaThree$. %
In one Hyper-Kamiokande tank~\cite{Hyper-Kamiokande:2018ofw}, a KSVZ axion with $f_a=10^9\GeV$ from a SN at $0.2\kpc$ gives an expected yield of $N_{\pi^0}=\CKLamNLOTFullUnc$ events, compared with $\CKLamNLONBorn$ when the $\Delta$ is omitted from both emission and absorption.
The $\Delta$ thus raises the count by a factor $\PRLCkGainUnc$, by $\PRLCkGainSource$ in the source and by $\PRLCkGainDetector$ in the detector.
Hyper-Kamiokande would see at least two events from a SN up to a distance of $\CKDtwoNLOTFullUnc\kpc\,(10^9\GeV/f_a)^2$ (SFHo-18.6), $\CKDtwoColdNLOT\kpc\,(10^9\GeV/f_a)^2$ (SFHo-18.8) and $\CKDtwoHotNLOT\kpc\,(10^9\GeV/f_a)^2$ (SFHo-20.0), see Fig~\ref{fig:cherenkov}.
Note that for this analysis, the axion production is calculated in the one-pion-exchange approximation.
With the SN~1987A bound, only such a nearby SN is within reach.
References~\cite{Vonk:2022tho,Cavan-Piton:2025nsj} predicted events in this channel from pion-induced axion production~\cite{Carenza:2020cis}, which requires a large population of thermal pions whose abundance is debated~\cite{Fore:2023gwv,Fiorillo:2025gnd}.
The $N_{\pi^0}=\CKLamNLOTFull$ events from Bremsstrahlung with the $\Delta$ already exceed the $\CKPCThirtyKSVZFull$ events of the pion source at $T=30\MeV$, quoted in Ref.~\cite{Cavan-Piton:2025nsj} without the gravitational redshift, which that work lists among the neglected effects; redshifted, the pion source gives $\CKPCThirtyKSVZRedFull$ events, $\CKOursOverPCRed$ times fewer.

\textbf{\textit{Conclusions}}---An intermediate $\Delta(1232)$ adds a hard component to supernova axion Bremsstrahlung and contributes to the luminosity at a level comparable to nucleon Bremsstrahlung in the models considered.
We computed $NN\to NNa$ with an explicit $\Delta$ at NLO in the chiral small-scale expansion.
Near $E_a\simeq\delta$, the resonant propagator partly compensates the Boltzmann suppression of hard axions.
In the simulated KSVZ benchmarks, the $\Delta$ roughly doubles the luminosity at one second after bounce, which strengthens the SN cooling bound, and approximately triples the number of escaping axions above $200\MeV$ over the first nine seconds.
Its effects on emission and absorption roughly double progenitor-conversion $\gamma$-ray counts and increase water-Cherenkov counts by about an order of magnitude in the benchmarks considered.
For light axion-like particles converting in the Galactic field, omitting this component can bias the reconstructed core temperature beyond its statistical uncertainty and mimic thermal-pion emission.
The $\Delta$ contribution thus affects both detection prospects and the interpretation of supernova axion spectra.
Related resonance effects can arise when nucleons radiate other light particles through an isovector axial coupling.

\textbf{\textit{Acknowledgments}}---We thank Hans-Thomas Janka and his group, especially Malte Heinlein, for providing the SN profiles and the authors of the GiBUU transport code for making their in-medium $\Delta$ width publicly available.
We acknowledge the usage of OpenAI's Codex and Anthropic's Claude.
PA and AW have been supported by the Collaborative Research Center SFB1258, the Munich Institute for Astro-, Particle and BioPhysics (MIAPbP), and by the Excellence Cluster ORIGINS, which is funded by the Deutsche Forschungsgemeinschaft (DFG, German Research Foundation) under Germany’s Excellence Strategy – EXC 2094 – 39078331.
MS is supported by the Swiss National Science Foundation under contract 200020-213104 and through the National Center of Competence in Research SwissMAP.
S.S. acknowledges financial support from the Spanish Ministry of Science and Innovation (MICINN) through the Spanish State Research Agency, under Severo Ochoa Centres of Excellence Programme 2025-2029 (CEX2024001442-S).
This work is also part of the R\&D\&i project PID2023-146686NB-C31, funded by MICIU/AEI/10.13039/501100011033/ and by ERDF/EU.
IFAE is partially funded by the CERCA program of the Generalitat de Catalunya.
KS is supported by a research grant from Mr. and Mrs. George Zbeda and by the Minerva foundation.
\nocite{Raffelt:1996wa,Raffelt:1987im,Balkin:2022qer,Chen:2013kt,FlavourLatticeAveragingGroupFLAG:2021npn,ParticleDataGroup:2022pth,Arcangeli:2026long,FlavourLatticeAveragingGroup:2019iem,Fettes:2000gb,Pascalutsa:1999zz,Pascalutsa:2000kd,
Tang:1996sq,Krebs:2009bf,Bernard:1995dp,Long:2009wq,Aeppli:1993rs,Denner:2006ic,Mosel:2020zdw,Kolomeitsev:2016ptu,
Gegelia:2016pjm,Radecky:1981fn,Kitagaki:1986ct,Wilkinson:2014yfa,Rodrigues:2016xjj,Hernandez:2010bx,Low:1958sn,
Epelbaum:2007sq,Gorghetto:2018ocs,Stoks:1993tb,NNOnline,Feldman:1997qc,Jansson:2012pc,Planck:2016gdp,Calore:2023srn,Bruel:2018lac,Cowan:2010js,
Cavan-Piton:2024ayu,Fogli:2004ff,Gomez-Banon:2024oux,Banerjee:2025dlo,
Weinberg:1990rz,Weinberg:1991um,Pascalutsa:2002pi,BMW:2014pzb,Baroni:2018fdn,Gnech:2023mvb,Rampp:2002bq,Raffelt:1993ix,Epelbaum:2008ga}

\bibliography{bibliography}

\onecolumngrid
\newpage
\begin{center}
\textbf{\large End Matter}
\end{center}
\twocolumngrid

\phantomsection\label{em:constants}\textbf{\textit{Constants}}---Table~\ref{tab:constants} lists the constants used for the computation of the rate.
Each is fixed by an observable at its kinematic point, with the vertices, the normalization and the exact kinematics of the rate (\SMref{sec:matching}).
At every matching point the $1/M_N$ terms of the vertices vanish, so LO and NLO use the same constants.
At $q^2=0$ and $E_a=0$ the recoil vanishes, so the leading-order values of $G_A$, $C_p$ and $C_n$ apply.
$\Gpi$ is fixed by the width of a $\Delta$ at rest, $\Gamma(\Delta\to N\pi)=\Gpi^2q^3(\epsilon_N+M_N)/12\pi F_\pi^2M_\Delta$ with the decay momentum $q$ and the nucleon energy $\epsilon_N$; the static formula would give $\TSInGpiStatic$.
$\Gax$ is fixed by the axial $N\to\Delta$ transition measured in pion production by neutrinos~\cite{Hernandez:2016yfb}.
The two agree within errors, as the off-diagonal Goldberger--Treiman relation predicts~\cite{Procura:2008ze}.
The residual LECs subtract the static $\Delta$ exchange, $c_3^\Delta=\TSCDeltaThree$ and $c_4^\Delta=\TSCDeltaFour\GeV^{-1}$, which is the $\Delta$ exchange of the heavy-baryon amplitude at $E_a\to0$.
For simplicity assume Gaussian errors for all known constants.
For the unknown $\hat{c}_9$, we use a superposition of of two normal distributions with mean $\pm 0.5$.
\begin{table}[h]
\caption{Constants of the rate; $c_i$ in ${\rm GeV}^{-1}$.}
\label{tab:constants}
\begin{ruledtabular}
\begin{tabular}{lll}
constant & value & fixed by\\
\hline
$G_A$ & $\TSInGA$ & neutron $\beta$ decay\\
$C_p$, $C_n$ & $\TSCpUnc$, $\TSCnUnc$ & $G_A$, $\Delta u+\Delta d$, $\Delta s$, $z$~\cite{GrillidiCortona:2015jxo}\\
$\cud$ & $\TSCvdUnc$ & $z=m_u/m_d=\TSInZ$\\
$M_\Delta$, $\Gamma_\Delta$ & $\TSInMDelta$, $\TSInGamma(\fpeval{(\TSInGammaHi-\TSInGammaLo)/2})\MeV$ & PDG Breit--Wigner~\cite{ParticleDataGroup:2024cfk}\\
$\Gpi$ & $\TSInGpi$ & $\Gamma(\Delta\to N\pi)$ at $W=M_\Delta$\\
$\Gax$ & $\TSInGax$ & axial $N\to\Delta$ transition~\cite{Hernandez:2016yfb}\\
$c_3$, $c_4$ & $\TSInCThree$, $\TSInCFour$ & Roy--Steiner $\pi N$~\cite{Hoferichter:2015tha}\\
$\hat c_3^{\rm res}$, $\hat c_4^{\rm res}$ & $\TSResCThree$, $\TSHatCFour$ & $\hat c_i-c^\Delta_i$\\
$\hat c_9$ & $0$, $\pm0.5(5)$ & unknown\\
\end{tabular}
\end{ruledtabular}
\end{table}

\phantomsection\label{em:amp}\textbf{\textit{Amplitude}}---We give the amplitude of proton--proton collisions, $p(p_1)+p(p_2)\to p(p_3)+p(p_4)+a(p_a)$, with the axion four-momentum $p_a=(E_a,\vp_a)$.
The other channels differ only in their isospin factors (\SMref{sec:fermigas}).
In each diagram of Fig.~\ref{fig:diagrams} the two protons exchange a pion, and each proton runs along a line labeled by its initial and final momenta.
On the line that radiates the axion, a nucleon or a $\Delta$ of four-momentum $P$ propagates between the vertices of the pion and the axion, with the propagators
\begin{align}
 S_N(P)&=\frac{P^0+M_N}{P^2-M_N^2},\nonumber\\
 S_\Delta(P)\,P_{3/2}^{ab}\xi_{3/2}^{ij}&=\frac{P^0+M_\Delta}{P^2-M_\Delta^2+i\sqrt{P^2}\,\Gamma}\nonumber\\
 &\quad\times\Bigl(-\delta^{ab}+\frac{\sigma^a\sigma^b}{3}\Bigr)\xi_{3/2}^{ij}.
 \label{eq:props}
\end{align}
The $\Delta$ propagator carries the spin-$\tfrac32$ projector, whose spatial components are $P_{3/2}^{ab}=-(\delta^{ab}-\sigma^a\sigma^b/3)$, the isospin-$\tfrac32$ projector $\xi_{3/2}^{ij}=\delta^{ij}-\frac13\tau^i\tau^j$, and the width $\Gamma$ of the $\Delta$ (\hyperref[em:medium]{Medium}).
These propagators go beyond LO HB, which would just be the static propagators $1/l^0$ for the nucleon and $1/(l^0-\delta)$ for the $\Delta$ of the residual momentum $l$.
Near $E_a\simeq\delta$ the latter diverges and one has to resum the self-energies.
The real part of the self-energy is absorbed into the physical mass $M_\Delta$, and its imaginary part is the width $\Gamma$, which by unitarity is the decay width of Eq.~\eqref{eq:width} with the same coupling $\Gpi$.
The kinetic energy $\vl^{\,2}/2M_N$ of the internal baryon, which shifts the pole by amounts comparable to $\Gamma/2$, is included in the same way, since Eq.~\eqref{eq:props} is the upper $2\times2$ block of the relativistic propagator, with its pole at the exact position.

Consider the diagrams in which the line $1\to3$, from $P_1$ to $P_3$, radiates the axion and absorbs the pion of four-momentum $q=(q^0,\vq)=P_2-P_4$ from the line $2\to4$.
The axion is emitted after ($s=+$) or before ($s=-$) the pion is absorbed, so that the internal baryon carries the four-momentum $P=P_3+P_a$ or $P=P_1-P_a$.
With the large-component of the proton Dirac spinor $\xi_i = \chi_i\sqrt{P_i^0 + M_N}$ where $P^0_i=(M_N^2+\vp_i^{\,2})^{1/2}$ and $\chi_i$ the unit-normalised spinor, $(\chi_i^s)^\dagger \chi_i^s = \delta_{ss'}$ (\SMref{sec:NLOamplitudes} and \cite{Scherer:2002tk}), the LO diagrams (a) and (b) are
\begin{align}
 \Amp_N^s={}&-\frac{G_A^2C_p}{8F_\pi^2f_a}\,\frac{S_N(P)\;\xi_4^\dagger\,\vs\cdot\vq\,\xi_2}{\vq^{\,2}+M_\pi^2-q_0^2}\nonumber\\
 &\times\xi_3^\dagger\big[\vp_a\cdot\vq+is\,\vs\cdot(\vp_a\times\vq)\big]\xi_1,\nonumber\\
 \Amp_\Delta^s={}&-\frac23\,\frac{G_A\Gpi\Gax\cud}{2F_\pi^2f_a}\,
 \frac{S_\Delta(P)\;\xi_4^\dagger\,\vs\cdot\vq\,\xi_2}{\vq^{\,2}+M_\pi^2-q_0^2}\nonumber\\
 &\times\xi_3^\dagger\big[\tfrac23\vp_a\cdot\vq-\tfrac{is}3\vs\cdot(\vp_a\times\vq)\big]\xi_1.
 \label{eq:LOamp}
\end{align}
The brackets follow from contracting the spin indices of the vertices and of the propagators, which gives $(\vs\cdot\vp_a)(\vs\cdot\vq)=\vp_a\cdot\vq+i\vs\cdot(\vp_a\times\vq)$ on a nucleon line and $-p_a^aP_{3/2}^{ab}q^b=\tfrac23\vp_a\cdot\vq-\tfrac i3\vs\cdot(\vp_a\times\vq)$ on a $\Delta$ line for $s=+$, and the reverse order for $s=-$.
Only the neutral pion connects two protons, and its isospin factor at each proton is 1, so $\Amp_N^s$ carries none.
In the $\Delta$ diagrams the axion couples through its isovector coupling $\cud$, which has the isospin structure of a $\pi^0$, so both vertices carry the isospin Clebsch--Gordan coefficient $\langle\Delta^+|p\,\pi^0\rangle=\sqrt{2/3}$, which gives the factor $\tfrac23$ in $\Amp_\Delta^s$.
For nucleon momenta $|\vp|\ll M_N$ and $E_a\ll M_N$, the two orderings give $S_N(P)\simeq\pm1/E_a$, so the spin-independent term $\vp_a\cdot\vq$ cancels in the sum over $s$.
With $S_\Delta(P)\simeq1/(\pm E_a-\delta)$ it does not cancel, and the $\Delta$ also radiates through $\vp_a\cdot\vq$.

The axion can leave any of the four external protons.
We therefore add the diagrams with $s=\pm$ on the line $1\to3$ and the same diagrams with the line $2\to4$ radiating, and subtract those with the two final protons interchanged.
This gives the nucleon part $\Amp_N$ and the $\Delta$ part $\Amp_\Delta$ of the amplitude $\Amp=\Amp_N+\Amp_\Delta$.
Each contains the factor $\prod_{i=1}^4\sqrt{P_i^0+M_N}$ arising from the normalization of the spinors $\xi_i$.

At NLO the vertices receive their first corrections in $|\vp|/M_N$ and $E_a/M_N$, which arise when the heavy components of the baryon fields are integrated out (\SMref{sec:lagrangian}).
On a nucleon line with the momenta $\vp$ and $\vp^{\,\prime}$ the axion vertex $\vs\cdot\vp_a$ becomes $\vs\cdot[\vp_a-E_a(\vp+\vp^{\,\prime})/2M_N]$, and the transition vertex $\vp_a$ becomes $\vp_a-E_a\vl/M_N$ with the $\Delta$ momentum $\vl$.
To first order in the baryon velocity both are the Doppler-shifted axion momentum in the baryon rest frame (\SMref{sec:doppler}).
The pion vertices change in the same way with $q^0$ and $\vq$.
Diagram (c) adds the contact vertices of Eq.~\eqref{eq:L}, at which the axion and the pion attach to the same point of a nucleon line.
For $pp$ the contact diagram of the line $1\to3$ is
\begin{equation}
 \Amp_c=-\frac{G_A\big(\cud\hat c_3^{\rm res}+\tfrac12\cpd\hat c_9\big)\,\vp_a\cdot\vq}{F_\pi^2f_a\,(\vq^{\,2}+M_\pi^2-q_0^2)}\,
 \xi_3^\dagger\xi_1\;\xi_4^\dagger\,\vs\cdot\vq\,\xi_2,
 \label{eq:contact}
\end{equation}
while the chiral connection and $\hat c_4^{\rm res}$ contribute only through charged pions (\SMref{sec:NLOamplitudes}), and $\Amp_c$, which contains no $\Delta$, belongs to the nucleon part $\Amp_N$.
We sum $|\Amp|^2$ over the spins numerically and split it into the nucleon part $|\Amp_N|^2$, the $\Delta$ part $|\Amp_\Delta|^2$ and the interference $2\,{\rm Re}\,\Amp_N^*\Amp_\Delta$.

\phantomsection\label{em:medium}\textbf{\textit{Medium}}---In the supernova core the nucleons have the scalar and vector self-energies $\Sigma_S^N$ and $\Sigma_V^N$ of the equation of state, which differ for protons and neutrons~\cite{Hempel:2014ssa}.
A nucleon then has the effective mass $m_N^*=M_N+\Sigma_S^N$ and the energy $\Sigma_V^N+\epsilon$ with $\epsilon=(m_N^{*2}+\vp^{\,2})^{1/2}$.
A $\Delta^+$ has two thirds of the proton self-energies, $m_\Delta^*=M_\Delta+\tfrac23\Sigma_S^p$ and $\tfrac23\Sigma_V^p$~\cite{Buss:2011mx}.
The amplitude keeps its form, with the effective masses in the propagators of Eq.~\eqref{eq:props} and in the normalization $(\epsilon_i+m_N^*)^{1/2}$, and with the kinetic four-momentum $(P^0-\Sigma_V,\vP)$ of the internal baryon in place of $P$, so the self-energies, like the kinetic energy, enter the propagator to all orders.
For a final proton at rest the resonance then lies at $E_a\simeq\delta-U_N/3+\delta^2/2m_N^*$, with the nucleon potential $U_N=\Sigma_S^N+\Sigma_V^N$.

The width of the $\Delta$ has a mesonic and a collisional part.
The mesonic part is the decay into a nucleon and a pion,
\begin{equation}
 \Gamma_N=\frac{\Gpi^2q_N^3\,(\epsilon_N+m_N^*)}{12\pi F_\pi^2W_N}\,B_N,
 \label{eq:width}
\end{equation}
the imaginary part of the $N\pi$ loop of the $\Delta$ self-energy.
Here $W_N$ is the invariant mass of the nucleon and the pion, $W_N^2=(P^0-\Sigma_V^N)^2-\vP^{\,2}$, with the four-momentum $P$ of the $\Delta$ and the vector self-energy $\Sigma_V^N$ of the final nucleon.
Two-body kinematics then fixes the momentum $q_N=\lambda^{1/2}(W_N^2,m_N^{*2},M_\pi^2)/2W_N$ and the energy $\epsilon_N=(m_N^{*2}+q_N^2)^{1/2}$ of the nucleon in the $N\pi$ rest frame, which is $\epsilon$ at the momentum $q_N$, with the Källén function $\lambda$ of \SMeq{eq:width-vacuum-kernel}.
The factor $B_N=\langle1-f_N\rangle$ is the probability that the final nucleon state is empty, averaged over the decay directions.
A $\Delta^+$ decays into $p\pi^0$ and $n\pi^+$ with the squared Clebsch--Gordan coefficients $\tfrac23$ and $\tfrac13$, so its mesonic width is $\tfrac23\Gamma_p+\tfrac13\Gamma_n$.
In vacuum, with $m_N^*=M_N$ and $B_N=1$, this is the width that fixes $\Gpi$.
The collisional part, from $\Delta N\to\Delta N$, $\Delta N\to NN$ and $\Delta NN\to NNN$ in the medium, is taken from Ref.~\cite{Oset:1987re} as implemented in GiBUU~\cite{Buss:2011mx,GiBUU}, adapted to the mean fields of the rate (\SMref{sec:width}).

\phantomsection\label{em:rate}\textbf{\textit{Rate}}---The emissivity is
\begin{equation}
 \dot\varepsilon_a=\sum_{\rm ch}S\int\dd\Phi\;E_a\,f_1f_2(1-f_3)(1-f_4)\sum_{\rm spins}|\Amp|^2,
 \label{eq:emis}
\end{equation}
with the phase space of the four nucleons and the axion
\begin{equation}
 \dd\Phi=\prod_{i=1}^4\frac{\dd^3p_i}{(2\pi)^32\epsilon_i}\,\frac{\dd^3k}{(2\pi)^32E_a}\,
 (2\pi)^4\delta^{(4)}\Big(\sum_{\rm in}P-\sum_{\rm out}P\Big),
 \label{eq:dPhi}
\end{equation}
$S=\tfrac14$ for $nn$, $pp$ and $1$ for $np$, Fermi--Dirac occupations $f_i$ and Pauli blocking of the final nucleons.
The factors $1/(2\epsilon_i)$ follow from the relativistic normalization of the one-nucleon external states.
We integrate the eight remaining dimensions with exact kinematics, including the axion momentum in momentum conservation and in the internal propagators, by quasi-Monte Carlo integration with randomized Sobol points (\SMref{sec:integration}).

\textbf{\textit{Uncertainties}}---We estimate the uncertainties as in Ref.~\cite{Springmann:2024mjp}.
The Lagrangian parameter uncertainty band is the 16--84\% interval of $\TSDraws$ draws of all inputs and of $\hat c_9$, each with its rematched residual contact term, evaluated on the same phase-space points; ratios are formed draw by draw.
The truncation error is the last computed order times one further power of
\begin{equation}
 Q=\max\big(E_a/4\pi F_\pi,\ \delta/4\pi F_\pi,\ |\vp|/M_N\big),
 \label{eq:Qtrunc}
\end{equation}
evaluated for each cell and axion energy with the nucleon momentum $|\vp|=\max(k_f,\sqrt{M_NE_a})$, the larger of the Fermi momentum and the smallest momentum of the colliding nucleons, in their center-of-mass frame, that supplies the axion energy.
At LO the amplitude changes by a factor $1\pm Q$, and at NLO the emissivity per unit energy of each part by $Q$ times the larger of its NLO correction, the difference of its values at NLO and at LO, and $Q$ times its LO value, $Q\max\big(|\dd\dot\varepsilon_a^{\rm NLO}/\dd E_a-\dd\dot\varepsilon_a^{\rm LO}/\dd E_a|,\, Q\,|\dd\dot\varepsilon_a^{\rm LO}/\dd E_a|\big)$.
    The second term keeps the estimate finite where the NLO correction changes sign, as that of the $\Delta$ part does at $E_\infty\simeq\TSHotDCorrZero$--$\TSColdDCorrZero\MeV$ (\SMref{sec:uncertainty}).

\onecolumngrid
\newpage
\begin{center}
\textbf{\large Supplemental Material for ``Harder, Better, Faster, Stronger: The $\Delta(1232)$ Resonance in Supernova QCD Axion Emission''}
\end{center}
\setcounter{section}{0}\setcounter{equation}{0}\setcounter{figure}{0}\setcounter{table}{0}\setcounter{footnote}{0}
\setcounter{secnumdepth}{3}%
\makeatletter
\newif\if@SMtoc
\let\SM@contentsline\contentsline
\def\SM@skipline#1#2#3#4{}
\renewcommand*{\contentsline}[1]{\if@SMtoc\expandafter\SM@contentsline\else\expandafter\SM@skipline\fi{#1}}
\addtocontents{toc}{\protect\@SMtoctrue}
\makeatother
\renewcommand{\thesection}{S\arabic{section}}\renewcommand{\theHsection}{S\arabic{section}}
\renewcommand{\theequation}{S\arabic{equation}}\renewcommand{\theHequation}{S\arabic{equation}}
\renewcommand{\thefigure}{S\arabic{figure}}\renewcommand{\theHfigure}{S\arabic{figure}}
\renewcommand{\thetable}{S\Roman{table}}\renewcommand{\theHtable}{S\Roman{table}}
\renewcommand{\topfraction}{0.9}\renewcommand{\bottomfraction}{0.8}\renewcommand{\textfraction}{0.07}
\renewcommand{\floatpagefraction}{0.75}
\makeatletter\ltx@footnote@pop\makeatother%
\tableofcontents
\newpage

\section{\texorpdfstring{$\Delta$}{Delta}-full Lagrangian, Feynman rules and propagators}\label{sec:lagrangian}

\paragraph*{\textbf{Lagrangian and Feynman rules.}}
We follow the conventions of Refs.~\cite{Springmann:2024mjp,Hemmert:1996xg}.
The heavy-baryon (HB) theory is the nonrelativistic reduction of the relativistic chiral Lagrangian, following the chiral power counting.
As in the End Matter, the full four-momentum of a baryon is decomposed as $P^\mu=m_Nv^\mu+l^\mu$, where $v^\mu$ is a four-velocity, $v^2=1$, $m_N$ is a large subtraction scale which we choose to be the nucleon mass and $l^\mu$ is the residual momentum.
Removing the common mass dependence and integrating out the heavy components of the Dirac doublet $\Psi$ gives the HB nucleon doublet
\begin{equation}
    N=e^{im_Nv\cdot x}P_{v+}\Psi\,,
\end{equation}
with the projector $P_{v\pm}=(1\pm\slashed v)/2$.
The spin of the nucleon is defined as $S^\mu=\tfrac{i}{2}\gamma_5\sigma^{\mu\nu}v_\nu=(0,\boldsymbol\sigma/2)$ for $v^\mu=(1,\bm0)$, which is the choice we adopt throughout this work.

As discussed in the main text, the power-counting scheme is the small-scale expansion (SSE), which treats the splitting $\delta=m_\Delta-m_N$ as a soft scale, in the same way as the pion mass and external momenta~\cite{Jenkins:1991es,Hemmert:1996xg,Hemmert:1997ye,Fettes:2000bb}. 
The LO and NLO Lagrangians are given in Eq.~\eqref{M-eq:L}.
The derivative axion interactions couple to the axial quark current,
\begin{equation}
    \mathcal L_{aq}=\frac{\partial_\mu a}{2f_a}
\sum_qc_q\bar q\gamma^\mu\gamma_5q\,,
\label{eq:derivaive coupling axion and quarks}
\end{equation}
after the chiral rotation on the quark fields removing the axion--gluon coupling \cite{GrillidiCortona:2015jxo,Springmann:2024mjp}. 
They enter the chiral building blocks through the external sources $r_\mu=-l_\mu= c_{u-d}\tau^3\partial_\mu a/(2f_a)$. 
Collecting the pions in $U=u^2=\exp(i\tau^i\pi^i/f_\pi)$, the Lagrangian depends on
\begin{equation}
u_\mu =i\big[u^\dagger(\partial_\mu-ir_\mu)u -u(\partial_\mu-il_\mu)u^\dagger\big] =\tau^iw^i_\mu= -\frac{\partial_\mu\pi^i}{f_\pi}\tau^i +c_{u-d}\frac{\partial_\mu a}{f_a}\tau^3+\cdots, \qquad \hat u_\mu=c_{u+d}\frac{\partial_\mu a}{f_a}\,,
\label{eq:hb-building-blocks}
\end{equation}
where $c_{u\pm d}=(c_u\pm c_d)/2$, and on the chiral covariant derivative of the nucleon
\begin{equation}
D_\mu=\partial_\mu+\Gamma_\mu,
\qquad
\Gamma_\mu=\frac12\big[u^\dagger(\partial_\mu-ir_\mu)u
                   +u(\partial_\mu-il_\mu)u^\dagger\big]=\tau^k\Gamma^k_\mu .
\label{eq:hb-connection}
\end{equation}
The $\Delta$ field $T^\mu_i$ of Eq.~\eqref{M-eq:L} is an isospinor that carries in addition an isovector index $i$, with $\tau^iT^\mu_i=0$, and its covariant derivative also carries isospin indices~\cite{Hemmert:1997ye},
\begin{equation}
D^{ij}_\mu=\delta^{ij}D_\mu-2i\,\epsilon^{ijk}\,\Gamma^k_\mu .
\label{eq:hb-delta-derivative}
\end{equation}
$\Gamma_\mu$ is the chiral connection and gives an $a\pi NN$ interaction, the Weinberg--Tomozawa (WT) term
\begin{equation}
\mathcal L^{(1)}_{\pi N} \supset \mathcal L_{a\pi NN}^{\rm WT}
 =-\frac{c_{u-d}}{2f_\pi f_a}\epsilon^{3bc}
 (\partial_\mu a)\pi^b
 \bar{ N}v^\mu\tau^c N.
\label{eq:hb-wt}
\end{equation}
From the LO Lagrangian of Eq.~\eqref{M-eq:L}, with pion and axion momenta outgoing, the Feynman rules involving a single boson field are
\begin{equation}
\begin{alignedat}{2}
NN\pi^i(q):&\quad -g_A\tau^i\frac{\boldsymbol\sigma\cdot\bm q}{2f_\pi},
&\qquad NNa(p_a):&\quad c_N\frac{\boldsymbol\sigma\cdot\bm p_a}{2f_a},\\[4pt]
N T^{\mu,n}\pi^i(q):&\quad
 g_{\pi N \Delta}\frac{q_\mu}{f_\pi}\delta^{in},
&\qquad NT^{\mu,n}a(p_a):&\quad
 -c_{u-d}g_{a N \Delta}\frac{p_{a\mu}}{f_a}\delta^{n3}
\end{alignedat}
\label{eq:hb-lo-vertices}
\end{equation}
and including two boson fields,
\begin{equation}
 NNa(p_a)\pi^i(q)\big|_{\rm WT}:\qquad
 \frac{c_{u-d}}{2f_\pi f_a}\,
 (v\cdot p_a)\,\epsilon^{3ij}\tau^j .
\label{eq:hb-wt-vertex}
\end{equation}
where $c_N=g_Ac_{u-d}\tau^3+g_0c_{u+d}$.
Notice that the $N$-$\Delta$ transition vertices share at LO one common coefficient, as clear from the LO Lagrangians \eqref{M-eq:L}, and therefore $g_{a N \Delta} = g_{\pi N \Delta}$.
The N and $\Delta$ propagators extracted from the LO Lagrangians \eqref{M-eq:L} read
\begin{equation}
    iS_{N}(l) = \frac{i}{(l^0+i\epsilon)}P_{v+}\,,\quad iS_{\Delta}^{\mu\nu,ij}(l) =-iS_\Delta(l) P_{3/2}^{\mu\nu}(v)
       \xi_{3/2}^{ij}P_{v+}\,,
\label{eq:hb-lo-propagators}
\end{equation}
with $S_\Delta(l) = 1/(v\cdot l-\delta+i0)$, and
\begin{equation}
P_{3/2}^{\mu\nu}(v)
 =\eta^{\mu\nu}-v^\mu v^\nu+
   \frac{4}{d-1}S^\mu S^\nu,
\qquad
\xi_{3/2}^{ij}=\delta^{ij}-\frac13\tau^i\tau^j,
\label{eq:hb-projectors}
\end{equation}
are the spin-$3/2$ and isospin-$3/2$ projectors, respectively.
$d$ is the spacetime dimension, such that for $d=4$, $P_{3/2}^{ab}=-(\delta^{ab}-\sigma^a\sigma^b/3)$.

At NLO the HB reduction generates fixed $1/m_N$ or recoil interactions, whose local two-source contributions are absorbed into the residual hatted coefficients $\hat c_i^{\rm res}$ of the $\Delta$-full theory. 
With $ D_\perp^\mu= D^\mu-v^\mu v\cdot D$, the nucleon operators relevant to $NN\to NNa$ and $aN\to\pi N$ are\footnote{The operator $\sim \hat c_1$ comes quadratic in the axion field \cite{Balkin:2020dsr,Balkin:2022qer,Balkin:2023xtr,Gomez-Banon:2024oux,Banerjee:2025dlo}.}
\begin{equation}
\begin{aligned}
\mathcal L_{\pi N}^{(2)}
={}&
 \bar{N}\Big[
 -\frac{i}{2m_N}\left(\{S\cdot D,v^\mu(g_Au_\mu+g_0\hat u_\mu)\}-iD_\perp^2\right)
 +\hat c_2^{\rm res}(v\cdot u)^2
 +\hat c_3^{\rm res}u\cdot u\\
&\hspace{15mm}
 +\hat c_4^{\rm res}[S^\mu,S^\nu]u_\mu u_\nu
 +\hat c_5\,\widetilde\chi_+
 +\frac{\hat c_8}{4}(v\cdot u)(v\cdot\hat u)
 +\hat c_9u\cdot\hat u
 \Big] N,
\end{aligned}
\label{eq:hb-nlo-nucleon}
\end{equation}
where $\widetilde\chi_+$ is the traceless part of the quark-mass term, which contains the product of a pion and the axion,
\begin{equation}
 \widetilde\chi_+\supset-\frac{4z}{(1+z)^2}\,M_\pi^2\,\frac{\tau^i\pi^i\,a}{f_\pi f_a},\qquad z=\frac{m_u}{m_d},
 \label{eq:hb-chi}
\end{equation}
and does not vanish for equal quark masses~\cite{Springmann:2024mjp}.
The term $\sim D_\perp^2$ contains the NLO contribution to the nucleon self energy, which we account for by making use of the resummed nucleon propagator, see the next paragraph.
Through the connection in $D_\mu$ the same term corrects the WT vertex \eqref{eq:hb-wt-vertex} by the recoil of the nucleon, $v\cdot p_a\to v\cdot p_a-\vp_a\cdot(\vp+\vp')/2m_N$, with $\vp$ and $\vp'$ the residual nucleon momenta adjacent to the vertex.
The spin-dependent correction of the same order is the $1/4M_N$ contained in $\hat c_4^{\rm res}$ below.
The $\hat c_i$ coefficients are related to those of the relativistic nucleon chiral Lagrangian (unhatted $c_i$), by relativistic corrections and contributions from integrating out the $\Delta$\footnote{Ref.~\cite{Springmann:2024mjp} writes $\tfrac12\hat c_2(v\cdot u)^2$, with $\hat c_2$ twice ours.
Here $\hat c_2$ has the normalization of Refs.~\cite{Fettes:2000gb,Hoferichter:2015tha}, in which the Roy--Steiner value $c_2=1.81\,\mathrm{GeV}^{-1}$ is quoted.}
\begin{equation}
    \begin{aligned}
 \hat c_2^{\rm res}&=\hat c_2-c_2^\Delta=c_2-\frac{G_A^2}{8M_N}-\frac{4\Gpi\Gax}{9\delta},\\
 \hat c_3^{\rm res}&=\hat c_3-c_3^\Delta=c_3+\frac{4\Gpi\Gax}{9\delta},\\
 \hat c_4^{\rm res}&=\hat c_4-c_4^\Delta=c_4+\frac{1}{4M_N}-\frac{2\Gpi\Gax}{9\delta},\\
 \hat c_5&=c_5,\\
 \hat c_8&=c_8-\frac{G_AG_0}{M_N},\\
 \hat c_9&=c_9,
\end{aligned}
\label{eq: hat ci coeff and residual}
\end{equation}
where $\hat c_5$, $\hat c_8$ and $\hat c_9$ are not affected by the $\Delta$.
The dependence of the $\Delta$-mediated contributions, and hence of $c_i^\Delta$, on redundant off-shell parameters in the $\Delta$ Lagrangian is absorbed into $\hat c_i^{\rm res}$ ~\cite{Tang:1996sq, Pascalutsa:2000kd,Krebs:2009bf}.

In the $\Delta$ sector, the only NLO Lagrangian relevant for $NN\to NNa$ is
\begin{equation}
\mathcal L^{(2)}_{\pi N\Delta}=-\frac{g_{\pi N\Delta}}{m_N}\,\bar T^\mu_i\,iD^{ij}_\mu\big[(v\cdot w^j)\,N\big]+{\rm h.c.}
\label{eq:hb-nlo-transition}
\end{equation}
The relativistic corrections that arise at NLO effectively provide momentum-dependent shifts of the LO Feynman rules \eqref{eq:hb-lo-vertices}, and in particular
\begin{equation}
\boldsymbol\sigma\cdot\bm p_a\ \longrightarrow\ \boldsymbol\sigma\cdot\left[\bm p_a-E_a\frac{\bm p+\bm p'}{2m_N}\right],\qquad
\boldsymbol\sigma\cdot\bm q\ \longrightarrow\ \boldsymbol\sigma\cdot\left[\bm q-q^0\frac{\bm p+\bm p'}{2m_N}\right],
\label{eq:hb-nn-recoil-rule}
\end{equation}
in $NNa$ and $NN\pi$, with $\bm p,\bm p'$ the three-momenta of the nucleons adjacent to the vertex, and
\begin{equation}
\bm p_a\ \longrightarrow\ \bm p_a-\frac{E_a}{m_N}\bm l,\qquad
\bm q\ \longrightarrow\ \bm q-\frac{q^0}{m_N}\bm l,
\label{eq:hb-delta-recoil-rule}
\end{equation}
in the $N\Delta a$ and $N\Delta\pi$ vertices, with the three-momentum $\bm l$ of the $\Delta$, the spatial part of its residual momentum $l=P_\Delta-m_Nv$.

We give the NLO Feynman rules for $a\pi NN$ vertices, separated into a pion energy independent contribution
\begin{equation}
\begin{aligned}
NNa(p_a)\pi^i(q):\quad
&\frac{i}{f_\pi f_a}
\Big[
  -2c_{u-d}\hat c_3^{\rm res}\delta^{i3}\left(\bm p_a\cdot\bm q\right)
  +c_{u-d}\hat c_4^{\rm res}\epsilon^{3ij}\tau^j
   \boldsymbol\sigma\cdot(\bm q\times\bm p_a) -c_{u+d}\tau^i\hat c_9\left(\bm p_a\cdot\bm q\right)
  -\frac{4z}{(1+z)^2}M_\pi^2\,\hat c_5\,\tau^i
\Big]
\end{aligned}
\label{eq: ci NLO Feynman rules - contributions independent on pion energy}
\end{equation}
and an energy dependent contribution
\begin{equation}
NNa(p_a)\pi^i(q):\quad
\frac{i}{f_\pi f_a}\Bigg[2c_{u-d}\delta^{i3}
 \left(\hat c_2^{\rm res}+\hat c_3^{\rm res}\right) +c_{u+d}\tau^i
 \left(\frac{\hat c_8}{4}
      +\hat c_9\right)
\Bigg]E_aq^0
\label{eq: ci NLO Feynman rules - contributions dependent on pion energy}
\end{equation}
This distinction is useful because in Bremsstrahlung the energy $q^0$ of the exchanged pion is a recoil energy, suppressed by $|\bm q|/m_N$ relative to its momentum, while in absorption the pion is real and its energy is not suppressed, see Sec.~\ref{sec:NLOamplitudes}.

The contact interactions of two nucleons without derivatives are~\cite{Weinberg:1990rz,Weinberg:1991um}
\begin{equation}
 \mathcal L_{NN}=-\tfrac12C_S\,(\bar NN)(\bar NN)+2C_T\,(\bar NS^\mu N)(\bar NS_\mu N) ,
 \label{eq:hb-nn-contact}
\end{equation}
and are of the same order as one-pion exchange.
With one axial source they enter at NLO~\cite{Epelbaum:2007sq,Springmann:2024mjp},
\begin{equation}
 \mathcal L_{NNa}=-\frac{c_D}{2f_\pi^2\Lambda_\chi}\,(\bar NN)(\bar NS\cdot u\,N)
 -\frac{\tilde c_D}{2f_\pi^2\Lambda_\chi}\,(\bar NN)(\bar NS\cdot\hat u\,N) ,
 \label{eq:hb-cd}
\end{equation}
where $\Lambda_\chi\simeq700\MeV$ is the scale with which the literature on three-nucleon forces normalizes the first term~\cite{Epelbaum:2007sq}, so that $c_D$ is of order one.
The first term is the contact term of the two-nucleon axial current, and the second is its partner with the isoscalar source~\cite{Springmann:2024mjp}.
The corresponding Feynman rules are
\begin{equation}
\begin{alignedat}{2}
NN\to NN:&\quad C_S+C_T\,\boldsymbol\sigma_1\cdot\boldsymbol\sigma_2,
&\qquad NN\to NNa(p_a):&\quad
-\frac{1}{4f_\pi^2\Lambda_\chi f_a}\sum_{i=1,2}\big(c_Dc_{u-d}\,\tau^3_i+\tilde c_Dc_{u+d}\big)\,\boldsymbol\sigma_i\cdot\bm p_a ,
\end{alignedat}
\label{eq:hb-nn-contact-rules}
\end{equation}
where $\boldsymbol\sigma_i$ and $\tau^3_i$ act on nucleon $i$.
A contact term without derivatives acts only in $S$ waves~\cite{Epelbaum:2008ga}.
The constants $C_S$ and $C_T$ are matched in Sec.~\ref{sec:matching}.

\paragraph*{\textbf{Resummed propagators.}}
The two LO propagators of Eq.~\eqref{eq:hb-lo-propagators} are static.
They depend only on the residual momentum, without the rest mass, which HB removes, and without the kinetic energy.
The processes discussed in the main text depend on the physical width of the $\Delta$, so the propagators must have their poles at the physical positions, both in vacuum and in the medium.
The static propagators have their poles at $v\cdot l=0$ and $v\cdot l=\delta$, and an expansion in the self-energy and the kinetic energy does not move them.
The $\Delta$-propagator needs to be resummed close to its on-shell point, at distances smaller than $\Gamma/2$ away from the pole~\cite{Pascalutsa:2002pi}, leading to
\begin{equation}
S_N(l)=
\frac{1+(v\cdot l)/(2m_N)}
 {v\cdot l+l^2/(2m_N)+i0}\,,\quad
S_\Delta(l)
 =\frac{2m_N+v\cdot l+\delta}
 {2m_N(v\cdot l-\delta)+(l^2-\delta^2)
       +iW(P)\Gamma(P)}\,,
\label{eq:hb-resummed-propagators}
\end{equation}
with $W(P) =\sqrt{P^2}= \sqrt{\left(m_N v_\mu +l_\mu\right)^{2}}$.
For the $\Delta$ propagator, we include the Breit--Wigner width $\Gamma$, consistently with the matching to its decay width.
With $m_\Delta=m_N+\delta$, the dressed denominator is
\begin{equation}
P^2-m_\Delta^2+iW(P)\Gamma(P).
\label{eq:hb-bw-denominator}
\end{equation}
In terms of the full four-momentum $P=m_Nv+l$ these are the propagators $S_N(P)$ and $S_\Delta(P)$ of Eq.~\eqref{M-eq:props} of the main text, the upper $2\times2$ blocks of the relativistic propagators.
The real part of the $\Delta$ self-energy is absorbed into the physical mass, and its imaginary part is the width.

\section{Analytic estimate in a Fermi gas}
\label{sec:fermigas}
In this section we provide details about the derivation of the analytical results in Eqs.~\eqref{M-eq:FGN}--\eqref{M-eq:FGpeak} of the main text.
We evaluate the axion emissivity integral in Eq.~\eqref{M-eq:emis} using the following approximations: \textit{(i)} the non-degenerate limit, for which the Fermi-Dirac distributions $f_i = f(E_i, \mu_i)$ of the external nucleons with energies $E_i$ and chemical potentials $\mu_i$ are approximated as $f_i\to e^{-(E_i-\mu_i)/T}$ and $1-f_{j}\to 1$, with $E_i \simeq M_N + \bm p_i^2/(2M_N)$; \textit{(ii)} the soft-axion approximation (SAA) $\bm p_a\ll\bm p_i$ in the matrix element and in the three-momentum-conserving Dirac delta function; \textit{(iii)} the non-relativistic limit in the phase-space measure, $E_i\simeq M_N$, and in the normalization of the external nucleons in the spin sum; \textit{(iv)} the massless pion limit; \textit{(v)} the LO (also called \textit{static}) nucleon and the $\Delta$ propagators, instead of the recoil-resummed version in \eqref{eq:hb-resummed-propagators}, and the constant $\Delta$ width in vacuum, $\Gamma=\PRLGamma\,\mathrm{MeV}$; and \textit{(vi)} symmetric nucleon matter at nuclear saturation density $n_n=n_p=n_0/2$, and $M_n = M_p$, treated as a free Fermi gas at temperature $T$.
These approximations are often seen in the literature~\cite{Brinkmann:1988vi,Raffelt:1996wa,Raffelt:2006cw,Carenza:2019pxu}.

We relax all these approximations in Sec.~\ref{sec:integration}, which brings Eq.~\eqref{M-eq:emis} in a form that can be handled by a numerical Monte Carlo integrator.

\paragraph*{\textbf{Approximated matrix elements.}}
We first evaluate the LO nucleon and $\Delta$ diagrams, Fig.~\ref{M-fig:diagrams}(a) and \ref{M-fig:diagrams}(b), in the soft-axion kinematics (ii), with the notation of Eq.~\eqref{M-eq:LOamp} of the End Matter extended to all channels.
Each nucleon $i$ has a two-component spinor $\xi_i$ (more details in Sec.~\ref{sec:NLOamplitudes}) and an isospin state $|t_i\rangle$, with $|p\rangle=(1,0)^T$ and $|n\rangle=(0,1)^T$, on which the Pauli matrices $\tau^j$ act.
For the line $1\to3$ radiating the axion from its outgoing nucleon ($s=+$), with the pion momentum $\bm q=\bm p_2-\bm p_4$, the diagrams are
\begin{equation}
\begin{aligned}
\mathcal M_N^+&=-\frac{G_A^2}{8F_\pi^2f_a}\,\frac{1}{\bm q^2+M_\pi^2}\,\frac{1}{E_a}\,
 \xi_3^\dagger(\boldsymbol\sigma\cdot\bm p_a)(\boldsymbol\sigma\cdot\bm q)\xi_1\;\xi_4^\dagger\boldsymbol\sigma\cdot\bm q\,\xi_2\;
 \langle t_3|\,C\,\tau^j|t_1\rangle\langle t_4|\tau^j|t_2\rangle,\\
\mathcal M_\Delta^+&=-\frac{G_A G_{\pi N\Delta}G_{aN\Delta}c_{u-d}}{2F_\pi^2f_a}\,\frac{1}{\bm q^2+M_\pi^2}\,
 \frac{1}{E_a-\delta+i\Gamma/2}\\
&\quad\times\xi_3^\dagger\Big(\frac23\,\bm p_a\cdot\bm q-\frac{i}{3}\boldsymbol\sigma\cdot(\bm p_a\times\bm q)\Big)\xi_1\;
 \xi_4^\dagger\boldsymbol\sigma\cdot\bm q\,\xi_2\;
 \Big\langle t_3\Big|\,\delta^{3j}-\frac13\tau^3\tau^j\,\Big|t_1\Big\rangle\langle t_4|\tau^j|t_2\rangle ,
\end{aligned}
\label{eq:matrix elements}
\end{equation}
with the axion coupling $C=C_p(1+\tau^3)/2+C_n(1-\tau^3)/2$ of the emitting nucleon and a sum over the isospin index $j$ of the pion.
For $pp$ only the neutral pion, $j=3$, contributes, and the isospin factors are $1$ and $\tfrac23$ as in the End Matter.
For $np$ and $nn$ the charged pions, $j=1,2$, also contribute.

Approximation (v) gives the static baryon propagator factors $1/E_a$ and $1/(E_a-\delta+i\Gamma/2)$ for the nucleon and the $\Delta$ respectively.
With approximation (iv) we set $M_\pi=0$ in the pion propagator.
The spin structure of $\mathcal M_\Delta^+$ is the contraction $-p_a^aP_{3/2}^{ab}q^b=p_a^a(\delta^{ab}-\sigma^a\sigma^b/3)q^b$ with the spatial components $P_{3/2}^{ab}$ of the spin-$\tfrac32$ projector of Eq.~\eqref{eq:hb-projectors}.

Adding the diagrams with $s=-$, those in which the line $2\to4$ radiates, and the exchange diagrams, in which the final nucleons $3$ and $4$ are interchanged and which enter with a minus sign, squaring and summing over the spins, we are left with three channels, $nn$ ($nn\rightarrow nna$), $pp$ ($pp\rightarrow ppa$), and $np$ ($np \rightarrow npa$).
Each can be decomposed in its $N$ part $|\mathcal M_N|^2$, $\Delta$ part $|\mathcal M_\Delta|^2$ and interference $|\mathcal M_{\rm int}|^2=2\,{\rm Re}\,\mathcal M_N^*\mathcal M_\Delta$.
With the static $\Delta$ propagators $D^\pm=1/(\pm E_a-\delta+i\Gamma/2)$ for an axion radiated from an outgoing ($+$) or an incoming ($-$) nucleon, and the momentum transfers $\bm q=\bm p_2-\bm p_4$ and $\bm q^{\,\prime}=\bm p_2-\bm p_3$ of the direct and the exchange diagrams, the three parts, summed over the spins and averaged over the axion direction, are
\begin{align}
 \sum_{\rm spins}|\mathcal M_N|^2&=\frac{(2M_N)^4G_A^4}{F_\pi^4f_a^2}\times
 \begin{cases}
 C_n^2\big[1-(\hat{\bm q}\cdot\hat{\bm q}^{\,\prime})^2\big] & (nn),\\
 C_p^2\big[1-(\hat{\bm q}\cdot\hat{\bm q}^{\,\prime})^2\big] & (pp),\\
 \tfrac13\big(C_n^2+C_p^2+C_nC_p\big)\big[1+2(\hat{\bm q}\cdot\hat{\bm q}^{\,\prime})^2\big] & (np),
 \end{cases}
 \label{eq:FGampN}\\
 \sum_{\rm spins}|\mathcal M_\Delta|^2&=\frac{16\,(2M_N)^4G_A^2\Gpi^2\Gax^2\cud^2}{243\,F_\pi^4f_a^2}\,E_a^2\nonumber\\
 &\quad\times\begin{cases}
 \big[7\big(|D^+|^2+|D^-|^2\big)+2\,{\rm Re}\big(D^+D^{-*}\big)\big]\big[1-(\hat{\bm q}\cdot\hat{\bm q}^{\,\prime})^2\big] & (nn,\,pp),\\
 2\big(|D^+|^2+|D^-|^2\big)\big[1+2(\hat{\bm q}\cdot\hat{\bm q}^{\,\prime})^2\big] & (np),
 \end{cases}
 \label{eq:FGampD}\\
 \sum_{\rm spins}|\mathcal M_{\rm int}|^2&=\frac{(2M_N)^4G_A^3\Gpi\Gax\cud}{F_\pi^4f_a^2}\,E_a\,{\rm Re}\big(D^+-D^-\big)\times
 \begin{cases}
 \tfrac89\,C_n\big[1-(\hat{\bm q}\cdot\hat{\bm q}^{\,\prime})^2\big] & (nn),\\
 -\tfrac89\,C_p\big[1-(\hat{\bm q}\cdot\hat{\bm q}^{\,\prime})^2\big] & (pp),\\
 -\tfrac4{27}\,(C_p-C_n)\big[1+2(\hat{\bm q}\cdot\hat{\bm q}^{\,\prime})^2\big] & (np).
 \end{cases}
 \label{eq:FGampI}
\end{align}
The terms in $(\hat{\bm q}\cdot\hat{\bm q}^{\,\prime})^2$ come from the interference of the direct and exchange diagrams; integrated over the phase space they give the function $J$ in the emissivities below.

From \eqref{eq:matrix elements} one can see that for intermediate nucleons, the derivative axion vertex and the nucleon propagator supply compensating factors of $E_a$ and $1/E_a$.
With a massless pion, the retained spin-summed matrix element is independent of the nucleon momentum magnitudes.
In the $\Delta$ diagram, only the second observation remains true, and the mass splitting $\delta$ in the $\Delta$ propagator is at the heart of the enhancement discussed in the main text.

\paragraph*{\textbf{Approximation of the phase-space.}}
We now evaluate the phase space of Eq.~\eqref{M-eq:emis} using the center of mass frame.
We define total incoming momentum, relative incoming and relative outgoing momenta as $\bm P=\bm p_1+\bm p_2$, $\bm p=(\bm p_1-\bm p_2)/2$ and $\bm p'=(\bm p_3-\bm p_4)/2$.

Neglecting the axion momentum in three-momentum conservation, as prescribed by (ii), makes $\bm P$ common to the incoming and outgoing pairs.
Using the Boltzmann limit (i) with nonrelativistic nucleon energies and equal nucleon masses (vi), the distributions factorize as
\begin{equation}
 f_1f_2\propto
 e^{-\bm P^2/(4M_NT)}
 e^{-\bm p^2/(M_NT)},
\end{equation}
so the integral over $\bm P$ is Gaussian.
Energy conservation fixes $\bm p'^2=\bm p^2-M_NE_a$.
With
\begin{equation}
 u=\frac{E_a}{T},\qquad
 x=\frac{\bm p^2}{M_NT},\qquad
 \mu=\widehat{\bm p}\cdot\widehat{\bm p'},
\end{equation}
the axion emissivity therefore reads
\begin{equation}
 \frac{\dd\dot\varepsilon_a}{\dd E_a}
 =\frac{T^{1/2}E_a^2}
 {512\pi^{7/2}M_N^{5/2}}
 \sum_{\rm ch}S_{\rm ch}n_1n_2
 \int_u^\infty \dd x\,e^{-x}\sqrt{x(x-u)}\,\frac1{4\pi}\int \dd\Omega_a
 \int_{-1}^{1}\dd\mu\,
 \sum_{\rm spins}|\mathcal M|^2 .
\label{eq:fermigas-phase-space}
\end{equation}
Here $\Omega_a$ specifies the axion direction, $\sum_{\rm ch}$ is the sum over the isospin channels, and $S_{nn}=S_{pp}=1/4$ and $S_{np}=1$.
The same expression applies separately to the nucleon, $\Delta$, and interference contributions.

\paragraph*{\textbf{Approximate emissivity.}}
Using the modified Bessel function $K_1$ and the function $J$ defined as \footnote{The reason why two functions are needed lies in the approximated matrix elements.
In the massless-pion limit, the momentum-independent part of the squared matrix element gives $K_1$.
Interference between the direct and exchange diagrams also contains an angular dependence that changes with the energy transferred to the axion.
Integrating this additional term gives $J$.
This term vanishes when the incoming and outgoing relative momenta are taken to have equal magnitudes in the strict soft-energy limit.}
\begin{equation}
\begin{aligned}
 K_1(z)=\int_0^\infty \dd t\,e^{-z\cosh t}\cosh t\,,\quad J(u)=\frac{u^2}{2}e^{-u/2}
   \int_0^\infty \dd t\,t\tanh t\,e^{-(u/2)\cosh t},
\end{aligned}
\label{eq:fermigas-functions}
\end{equation}
applying the massless-pion limit (iv) and specializing to symmetric matter (vi), the integrals in Eq.~\eqref{eq:fermigas-phase-space} give:
\begin{equation}
\begin{aligned}
 \frac{\dd\dot\varepsilon_{aN}}{\dd E_a}
 =\frac{n_0^2G_A^4M_N^{3/2}T^{1/2}E_a^2}
 {1536\pi^{7/2}F_\pi^4f_a^2}\,\Big\{
 [7(C_p^2+C_n^2)+4C_pC_n]\,
 u e^{-u/2}K_1(u/2)
 +[5(C_p^2+C_n^2)+8C_pC_n]\,J(u)
 \Big\},
\end{aligned}
\label{eq:fermigas-nucleon-full}
\end{equation}
and
\begin{equation}
\begin{aligned}
 \frac{\dd\dot\varepsilon_{a\Delta}}{\dd E_a}
 =&
 \frac{n_0^2G_A^2G_{\pi N\Delta}^2G_{aN\Delta}^2
       c_{u-d}^2M_N^{3/2}T^{1/2}}
 {1944\pi^{7/2}F_\pi^4f_a^2}\,
 \frac{E_a^4}
 {\left((E_a-\delta)^2+\Gamma^2/4\right)
  \left((E_a+\delta)^2+\Gamma^2/4\right)}\\
 &\times\Big\{
 (10E_a^2+12\delta^2+3\Gamma^2)\,
 u e^{-u/2}K_1(u/2)
 +2E_a^2J(u)
 \Big\}.
\end{aligned}
\label{eq:fermigas-delta-full}
\end{equation}
Their interference is
\begin{equation}
\begin{aligned}
 \frac{\dd\dot\varepsilon_{a,\mathrm{int}}}{\dd E_a}
 ={}&
 \frac{n_0^2G_A^3G_{\pi N\Delta}G_{aN\Delta}
       (C_p-C_n)c_{u-d}M_N^{3/2}T^{1/2}}
 {864\pi^{7/2}F_\pi^4f_a^2}\,
 \frac{E_a^4(\delta^2-E_a^2-\Gamma^2/4)}
 {\left((E_a-\delta)^2+\Gamma^2/4\right)
  \left((E_a+\delta)^2+\Gamma^2/4\right)}\\
 &\times
 \Big\{5u e^{-u/2}K_1(u/2)+J(u)\Big\}.
\end{aligned}
\label{eq:fermigas-interference-full}
\end{equation}
Since $(C_p-C_n)c_{u-d} =2G_Ac_{u-d}^2 \geq 0$ at the retained order, the interference is constructive below $E_a=\sqrt{\delta^2-\Gamma^2/4}$ and destructive above it.
Without the term $J$, Eq.~\eqref{eq:fermigas-nucleon-full} is Eq.~\eqref{M-eq:FGN} of the main text.

\paragraph*{\textbf{The simplified resonance estimate.}}
To simplify the approximate treatment even further, we further neglect the terms proportional to $J$, adopting a stronger version of approximation (ii): besides $|\bm p_a|\ll|\bm p|$, we require $E_a\ll\bm p^2/M_N$.

Near $E_a=\delta$, the remaining term in \eqref{eq:fermigas-delta-full} is Eq.~\eqref{M-eq:FGD} of the main text,
\begin{equation}
 \frac{\dd\dot\varepsilon_{a\Delta}}{\dd E_a}
 \simeq
 \kappa\,
 \frac{E_a^2}{(E_a-\delta)^2+\Gamma^2/4}\,
 \frac{\dd\dot\varepsilon_{aN}}{\dd E_a},
\end{equation}
where also the nucleon spectrum in this expression is evaluated without its $J$ terms, and
\begin{equation}
 \kappa
 =\frac{64\,G_{\pi N\Delta}^2G_{aN\Delta}^2c_{u-d}^2}{81G_A^2\left(7(C_p^2+C_n^2)+4C_pC_n\right)}\,\frac{22\delta^2+3\Gamma^2}{4\delta^2+\Gamma^2/4}
\label{eq:fermigas-kappa}
\end{equation}
contains the ratio of the numerical coefficients and contributions from the non-resonant factor in the $\Delta$ squared propagators.
The numerical value quoted in the main text, $\kappa = \PRLkappa$, is obtained by neglecting $\Gamma^2/\delta^2$.

\paragraph*{\textbf{Peak energies and heights.}}
In the same approximation without $J$, the nucleon and $\Delta$ spectra peak respectively at
\begin{equation}
 E_a^{*(N)}\simeq\PRLuPeakN\,T\,,\quad  E_a^{*(\Delta)}
 \simeq
 \delta-\frac{\Gamma^2}{8}
 \left(\frac1T-\frac{9}{2\delta}\right),
\end{equation}
as reported in the main text.
The nucleon peak tracks $T$, whereas the $\Delta$ peak remains at an energy set primarily by the mass splitting.

To leading order in the expansion about the resonance, the ratio of the peak heights is
\begin{equation}
\begin{aligned}
 R(T) \equiv
 \frac{ \left.\dd\dot\varepsilon_{a\Delta}/\dd E_a \right|_{E_a=E_a^{*(\Delta)}}} { \left.\dd\dot\varepsilon_{aN}/\dd E_a \right|_{E_a=E_a^{*(N)}}} \simeq \frac{\sqrt{\pi}\,\kappa}{1.812} \left(\frac{2\delta}{\Gamma}\right)^2 \left(\frac{\delta}{T}\right)^{5/2} e^{-\delta/T},
\end{aligned}
\label{eq:fermigas-peak-ratio}
\end{equation}
which is quoted in main text.
Here $1.812$ is the maximum of $u^3e^{-u/2}K_1(u/2)$.
The expression shows the competition between the resonant enhancement $(2\delta/\Gamma)^2$ and the Boltzmann suppression $e^{-\delta/T}$ for producing a hard axion.
Increasing the temperature rapidly reduces this suppression, making the $\Delta$ component more prominent.

\begin{figure}[t]
    \centering
    \includegraphics[width=0.5\linewidth]{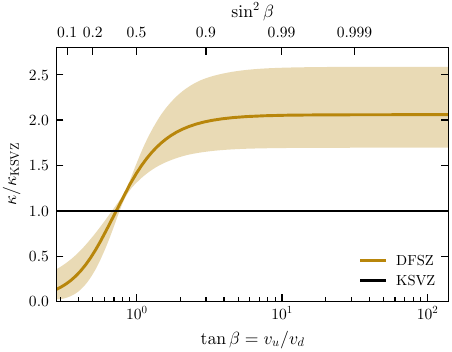}
    \caption{Ratio $\kappa_{\rm DFSZ}/\kappa_{\rm KSVZ}$ of Eq.~\eqref{eq:kappadfsz} over the range of $\tan\beta=v_u/v_d$ allowed by perturbative unitarity, with the constants of Table~\ref{M-tab:constants}.
    The band shows the $16$--$84\%$ range from the uncertainties of $G_A$, $\Delta u+\Delta d$, $\Delta s$, $z$ and the running.}
    \label{fig:dfsz}
\end{figure}

\paragraph*{\textbf{DFSZ versus KSVZ.}}
For a DFSZ axion~\cite{Dine:1981rt,Zhitnitsky:1980tq} the couplings $C_p$, $C_n$ and $c_{u-d}$ depend on the angle $\beta$, with $\tan\beta=v_u/v_d$ the ratio of the vacuum expectation values of the two Higgs doublets.
At $f_a$ this axion couples to the up-type quarks with $\cos^2\beta/3$ and to the down-type quarks with $\sin^2\beta/3$.
These add to the couplings $-1/(1+z)$ and $-z/(1+z)$ of the $u$ and $d$ quarks, with $z=m_u/m_d$, that the quark rotation removing the axion--gluon coupling induces in every model, and that are the only ones of the KSVZ axion~\cite{GrillidiCortona:2015jxo}.
Within the approximations above, and normalizing to the peak of the $N$-mediated spectrum, the only model-dependent quantity is $\kappa$.
Writing $c_\pm=\tfrac{1}{2}(C_p\pm C_n)$, so that $c_-=G_Ac_{u-d}$ at leading order, one has $A_N=18c_+^2+10c_-^2$ and $C_\Delta\propto c_-^2$, hence $\kappa\propto\big[10+18\,(c_+/c_-)^2\big]^{-1}$ and
\begin{equation}
    \frac{\kappa_{\rm DFSZ}}{\kappa_{\rm KSVZ}}=\frac{10+18\,(c_+^2/c_-^2)_{\rm KSVZ}}{10+18\,(c_+^2/c_-^2)_{\rm DFSZ}}.
    \label{eq:kappadfsz}
\end{equation}
With the constants of Table~\ref{M-tab:constants}, the running of Ref.~\cite{GrillidiCortona:2015jxo} and the perturbative-unitarity range $0.28<\tan\beta<140$~\cite{ParticleDataGroup:2022pth,Chen:2013kt}, we obtain Fig.~\ref{fig:dfsz}.
Over most of the allowed range the $\Delta$ contribution relative to the $N$ one is larger than in the KSVZ model, by up to a factor $\SEDFKappaMaxUnc$.
It is suppressed only where $c_-$, and hence the isovector coupling $c_{u-d}$, becomes small.

\section{Power counting and NLO amplitudes}
\label{sec:NLOamplitudes}
In this section we give the contributions through NLO to $NN\to NNa$ that supplement the amplitudes in the End Matter, using the same notation and conventions introduced there.
We also discuss diagrams where the interaction between the two initial nucleons is provided by contact terms instead of the OPE, of the type $NNNN$ or $NNN\Delta$.
We show that the former provide small contributions to the total axion emissivity rate, justifying why the corresponding Lagrangian operators are not explicitly shown in the main text, while the latter do not contribute at NLO because of a cancellation.
All the numerical results of this work contain the $NNNN$ contact interactions as well, although small.

\paragraph*{\textbf{Power counting in Bremsstrahlung and absorption.}}
Let $Q$ denote a soft hadronic scale.
The exchanged pion in Bremsstrahlung carries $q^0=\mathcal O(Q^2/M_N)$, whereas the on-shell pion in absorption has $q^0\geq M_\pi=\mathcal O(Q)$.

Due to the absence of an internal baryonic propagator, diagrams in which the axion is emitted or absorbed at an NNa$\pi$ vertex, namely diagrams (c) in Fig.~\ref{M-fig:diagrams} of the main text and in Fig.~\ref{fig:absorption}, are one order higher in SSE power counting compared to those in which the axion interacts with two baryons only.
As a consequence, the contact interaction diagrams proportional to $E_aq^0$ enter at NLO in absorption (see Eq.~\eqref{eq:nlo-abs-neutral-contact}) but only at NNLO in Bremsstrahlung.
We nevertheless keep them in the Bremsstrahlung rate, since in the hard tail the nucleon momenta are large and $q^0=\mathcal O(Q^2/M_N)$ is no longer small compared with $E_a$.

\paragraph*{\textbf{Normalization of the external nucleon spinors.}}
For an external nucleon with on-shell four-momentum $P_i^\mu=(P_i^0,\vp_i)$ and spin $s$, the relativistic Dirac spinor in the Dirac representation is \cite{Scherer:2002tk}
\begin{equation}
    u_i^s(P_i)
    =
    \sqrt{P_i^0+M_N}
    \begin{pmatrix}
        \chi_i^s\\[1mm]
        \dfrac{\boldsymbol{\sigma}\cdot\vp_i}
              {P_i^0+M_N}\chi_i^s
    \end{pmatrix},
    \qquad
    (\chi_i^s)^\dagger\chi_i^{s'}=\delta_{ss'}.
\end{equation}
We denote its large component by the two-component spinor
\begin{equation}
    \xi_i^s(P_i)=\sqrt{P_i^0+M_N}\,\chi_i^s.
    \label{eq:large component of spinor}
\end{equation}
Therefore, each external nucleon supplies a factor $\sqrt{P_i^0+M_N}$ to the amplitude, which produces the factor
\begin{equation}
    \prod_{i\in N_{\rm ext}}(P_i^0+M_N)
\end{equation}
in the spin-summed squared amplitudes, with four external nucleons for Bremsstrahlung (Eqs.~\eqref{M-eq:LOamp} and \eqref{M-eq:contact} for LO and NLO, respectively) and two for absorption (Eqs.~\eqref{eq:cherenkov-LO} and \eqref{eq:nlo-abs-neutral-contact}).
In the medium the spinor depends on the kinetic four-momentum $(P_i^0-\Sigma_V^N,\vp_i)$, so that $P_i^0+M_N$ becomes $\epsilon_i+m_N^*$, with the effective mass $m_N^*$ and $\epsilon_i=P_i^0-\Sigma_V^N=(m_N^{*2}+\vp_i^{\,2})^{1/2}$ (Sec.~\ref{sec:medium-profiles}).
The phase space contains the factors $1/2\epsilon_i$ of relativistic states, Eq.~\eqref{M-eq:dPhi}.

In the analytic estimate of Sec.~\ref{sec:fermigas}, we expand the external normalization factors to leading nonrelativistic order, $P_i^0+M_N\simeq 2M_N$, consistently with the static HB propagators.
The four external nucleons then give the factor $(2M_N)^4$ in the squared matrix elements \eqref{eq:FGampN}--\eqref{eq:FGampI}.

\paragraph*{\textbf{NLO Bremsstrahlung amplitudes.}}
At NLO, two new sets of Bremsstrahlung diagrams have to be included.
The first is simply obtained by promoting the $NNa$ and $N \Delta a$ vertices of Eq.~\eqref{M-eq:LOamp} to NLO, by making use of \eqref{eq:hb-nn-recoil-rule} and \eqref{eq:hb-delta-recoil-rule}, and diagrammatically corresponds to NLO corrections in diagrams (a) and (b) of Fig.~\ref{M-fig:diagrams} of the main text.

In the second set of diagrams the axion is emitted from the vertex, see Fig.~\ref{M-fig:diagrams} (c).
We obtain NN$\pi$a vertices both from the LO Lagrangian (the WT interaction \eqref{eq:hb-wt}) and from the NLO Lagrangian (the $\hat c_i$ terms).
We count both as NLO because these diagrams do not contain an internal nucleon propagator, and therefore, unlike the first group of diagrams, there is no small scale suppressing the $\mathcal O\left(p_a\right)$ scaling of the vertex function.
For the line $1\to3$ radiating, as in Eq.~\eqref{M-eq:LOamp}, the WT diagram is
\begin{equation}
 \mathcal M_{\rm WT}
 =\frac{G_Ac_{u-d}\,\eta}{2F_\pi^2f_a}\,
 \Big[E_a-\frac{\vp_a\cdot(\vp_1+\vp_3)}{2M_N}\Big]\,
 \frac{\xi_3^\dagger\xi_1\;\xi_4^\dagger\boldsymbol\sigma\cdot\bm q\,\xi_2}{\bm q^2+M_\pi^2-(q^0)^2},
\label{eq:nlo-brem-wt}
\end{equation}
with the isospin factor $\eta=+1$ if the radiating line turns a neutron into a proton, $\eta=-1$ if it turns a proton into a neutron, and $\eta=0$ if it keeps its charge.
The second term in the bracket is the recoil correction of the WT vertex from the operator $D_\perp^2$ of Eq.~\eqref{eq:hb-nlo-nucleon}.
The $\hat c_i$-mediated diagrams follow from the vertices \eqref{eq: ci NLO Feynman rules - contributions independent on pion energy} and \eqref{eq: ci NLO Feynman rules - contributions dependent on pion energy}.
The contributions that are independent of the pion energy $q^0$ are
\begin{equation}
\begin{aligned}
 \mathcal M_c
 ={}&-\frac{G_A}{F_\pi^2f_a\big[\bm q^2+M_\pi^2-(q^0)^2\big]}
 \Big\{
  \left(c_{u-d}\hat c_3^{\rm res}I_3
       +\frac{c_{u+d}}{2}\hat c_9 I_\pi\right)
  (\bm q\cdot\bm p_a)\,\xi_3^\dagger\xi_1\\
 &\hspace{24mm}
  +i c_{u-d}\hat c_4^{\rm res}\eta\,
       \xi_3^\dagger\boldsymbol\sigma\cdot(\bm q\times\bm p_a)\xi_1
 \Big\}
 \xi_4^\dagger\boldsymbol\sigma\cdot\bm q\,\xi_2 ,
\end{aligned}
\label{eq:nlo-brem-spatial-contact}
\end{equation}
with the isospin factors of the two lines
\begin{equation}
 I_\pi=\langle t_3|\tau^b|t_1\rangle\langle t_4|\tau^b|t_2\rangle,\qquad
 I_3=\langle t_3|t_1\rangle\langle t_4|\tau^3|t_2\rangle .
\label{eq:nlo-brem-isospin}
\end{equation}
For $pp$, $I_\pi=I_3=1$ and $\eta=0$, and $\mathcal M_c$ is Eq.~\eqref{M-eq:contact} of the End Matter.
The contributions depending on the pion energy $q^0$ and on $\hat c_5$ are
\begin{equation}
 \mathcal M_c^{\prime}
 =\frac{G_A}{F_\pi^2f_a\big[\bm q^2+M_\pi^2-(q^0)^2\big]}
 \Big\{
  \Big[c_{u-d}\big(\hat c_2^{\rm res}+\hat c_3^{\rm res}\big)I_3
       +\frac{c_{u+d}}{2}\Big(\frac{\hat c_8}{4}+\hat c_9\Big)I_\pi\Big]E_aq^0
  +\frac{2z}{(1+z)^2}M_\pi^2\,\hat c_5\,I_\pi\Big\}\,
 \xi_3^\dagger\xi_1\;\xi_4^\dagger\boldsymbol\sigma\cdot\bm q\,\xi_2 .
\label{eq:nlo-brem-energy-contact}
\end{equation}
The constant $\hat c_5$ is the isospin-breaking counterpart of $\hat c_1$.

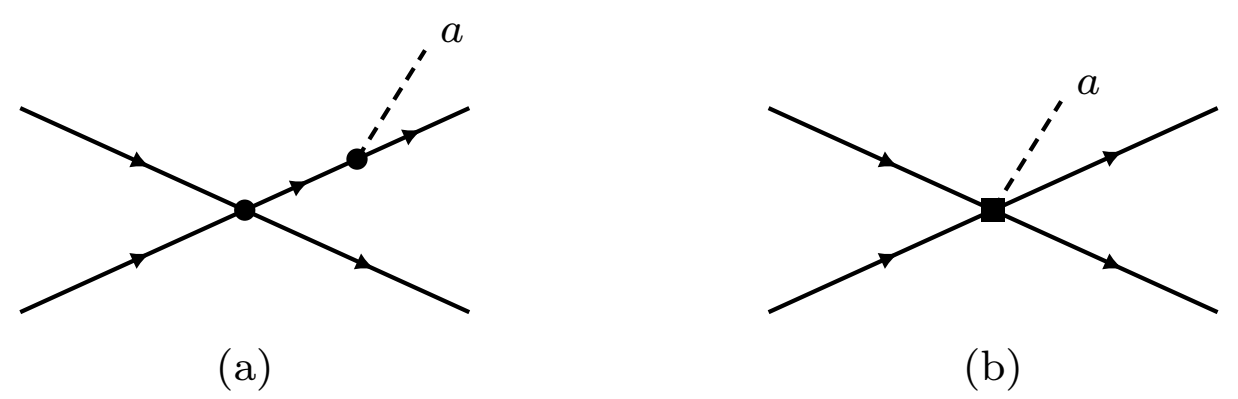
\begin{figure}[tbp]
\centering
\begin{tikzpicture}[scale=0.9]
\begin{scope}[shift={(0,0)}]
  \FDseg{N}{(0,1.5)}{(1.65,0.75)} \FDseg{N}{(0,0)}{(1.65,0.75)}
  \FDseg{N}{(1.65,0.75)}{(2.475,1.125)} \FDseg{N}{(2.475,1.125)}{(3.3,1.5)} \FDseg{N}{(1.65,0.75)}{(3.3,0)}
  \draw[axi] (2.475,1.125) -- (2.975,1.925);
  \node[vlo] at (1.65,0.75) {}; \node[vlo] at (2.475,1.125) {};
  \node[slab] at (3.175,2.05) {$a$};
  \node[slab] at (1.65,-0.42) {(a)};
\end{scope}
\begin{scope}[shift={(5.5,0)}]
  \FDseg{N}{(0,1.5)}{(1.65,0.75)} \FDseg{N}{(1.65,0.75)}{(3.3,1.5)}
  \FDseg{N}{(0,0)}{(1.65,0.75)} \FDseg{N}{(1.65,0.75)}{(3.3,0)}
  \draw[axi] (1.65,0.75) -- (2.15,1.55);
  \node[vsq] at (1.65,0.75) {};
  \node[slab] at (2.35,1.67) {$a$};
  \node[slab] at (1.65,-0.42) {(b)};
\end{scope}
\end{tikzpicture}
\caption{Diagrams of $NN\to NNa$ with a contact term of two nucleons.
Solid lines are nucleons, the dashed line is the axion.
Filled circles are vertices of LO, the filled square is a vertex of NLO, as in Ref.~\cite{Springmann:2024mjp}.
(a) A nucleon of the contact term $C_S$, $C_T$ radiates the axion.
(b) The two nucleons emit the axion at the contact term $c_D$, $\tilde c_D$.
The terms proportional to $\hat c_2^{\rm res}$, $\hat c_5$ and $\hat c_8$ enter at the $a\pi NN$ vertex of diagram (c) of Fig.~\ref{M-fig:diagrams} of the main text.}
\label{fig:nn-contact}
\end{figure}

\paragraph*{\textbf{$NNNN$ and $NNNNa$ contact terms.}}
Figure~\ref{fig:nn-contact} shows the diagrams in which the two incoming nucleons interact through a contact term instead of a pion.

In diagram (a) they scatter through $C_S$, $C_T$ of Eq.~\eqref{eq:hb-nn-contact}, and one of them radiates the axion after or before the collision.
Nucleon $i$ has the four-momentum $P_i=(P_i^0,\vp_i)$, the axion coupling $C_i$ and the two-component spinor $\xi_i$ before the collision and $P_i'$, $C_i'$, $\xi_i'$ after it.
As in the diagrams with pion exchange, the intermediate nucleon has the resummed propagator $S_N$ of Eq.~\eqref{M-eq:props}.
Including the NLO recoil corrections to the axion vertex, Eq.~\eqref{eq:hb-nn-recoil-rule}, the amplitudes of diagram (a) for radiation after (out) and before (in) the collision are
\begin{align}
 \mathcal M_{\rm out}&=\sum_{i=1,2}\frac{C_i'}{2f_a}\,S_N(P_i'+P_a)\;\xi_1^{\prime\dagger}\xi_2^{\prime\dagger}\,\vs_i\cdot\Big[\vp_a-E_a\,\frac{2\vp_i'+\vp_a}{2M_N}\Big]\,\big(C_S+C_T\,\vs_1\cdot\vs_2\big)\,\xi_1\xi_2,\label{eq:nn-contact-out}\\
 \mathcal M_{\rm in}&=\sum_{i=1,2}\frac{C_i}{2f_a}\,S_N(P_i-P_a)\;\xi_1^{\prime\dagger}\xi_2^{\prime\dagger}\big(C_S+C_T\,\vs_1\cdot\vs_2\big)\,\vs_i\cdot\Big[\vp_a-E_a\,\frac{2\vp_i-\vp_a}{2M_N}\Big]\,\xi_1\xi_2 .\label{eq:nn-contact-in}
\end{align}

Notice that, in the small momentum limit $|\bm p|\ll M_N$ and $E_a\ll M_N$, both the recoil of the vertex and the relativistic corrections of the propagator can be dropped, and the sum of the two terms is the commutator of the axion vertex with the contact term,
\begin{equation}
 \mathcal M_{\rm out}+\mathcal M_{\rm in}\;\simeq\;\frac{i\,C_T\,(C_p-C_n)}{2f_aE_a}\,\big(\tau_1^3-\tau_2^3\big)\,
 \xi_1^{\prime\dagger}\xi_2^{\prime\dagger}\,(\vs_2\times\vs_1)\cdot\vp_a\,\xi_1\xi_2 ,
 \label{eq:nn-contact-lo}
\end{equation}
where $\tau_i^3$ is $+1$ for a proton and $-1$ for a neutron, since the contact term keeps the charge of each nucleon.
In particular, the term $C_S$ is a number and commutes with the axion vertex, so it does not contribute in this limit.
The term $C_T$ contributes only in the $np$ channel.

In diagram (b) the two nucleons emit the axion at the contact term of Eq.~\eqref{eq:hb-cd},
\begin{equation}
 \mathcal M_{\rm (b)}=-\frac{1}{4F_\pi^2\Lambda_\chi f_a}\sum_{i=1,2}\big(c_Dc_{u-d}\,\tau^3_i+\tilde c_Dc_{u+d}\big)\,\xi_1^{\prime\dagger}\xi_2^{\prime\dagger}\,\vs_i\cdot\vp_a\,\xi_1\xi_2 ,
 \label{eq:nn-contact-cd}
\end{equation}
which is of NLO and also contributes only in the $np$ channel.

\paragraph*{\textbf{Size of the additional contact terms.}}
We now comment on the size of the contact terms for Bremsstrahlung discussed in this section, namely $\textit{(i)}$ the $NN\pi a$, which only include two nucleons, and \textit{(ii)} the $NNNN$ and $NNNNa$ contact terms, which include four nucleons.

\textit{(i)} The terms proportional to $\hat c_2^{\rm res}$, $\hat c_5$ and $\hat c_8$ in Eq.~\eqref{eq:nlo-brem-energy-contact} and the diagrams of Fig.~\ref{fig:nn-contact} change the luminosity at one second by $\ERNewASignedA$, $\ERNewASignedB$ and $\ERNewASignedC\%$ in SFHo-18.6, SFHo-18.8 and SFHo-20.0.
Most of the luminosity comes from axions below $200\MeV$, and at $E_\infty=200\MeV$ the spectrum of the luminosity at one second, $dL/dE_\infty$, changes by only $\ERSpecTTwoMin$ to $\ERSpecTTwoMax\%$ in total.
This agrees with the power counting above in this section, since the terms proportional to $\hat c_2^{\rm res}$ and $\hat c_8$ carry the energy $q^0$ of the exchanged pion, which is small in most collisions.
The $\hat c_5$ term is instead numerically smaller than the other $\hat c_i$ parameters (see Table~\ref{tab:matched-constants}) because it is an isospin-breaking interaction.

\textit{(ii)} The $NNNN$ contact terms act only in the $np$ channel at leading order.
In the hard tail the contact terms of \textit{(i)} and \textit{(ii)} are not small, because $q^0$ and the nucleon momenta grow with the axion energy.
Together they change $dL/dE_\infty$ at $E_\infty=400\MeV$ by $\ERSpecNFourMin$ to $\ERSpecNFourMax\%$ in the nucleon part and, through their interference with the $\Delta$ part, by $\ERSpecTFourMax$ to $\ERSpecTFourMin\%$ in the total.

\paragraph*{\textbf{$NNN\Delta$ contact terms.}}\label{sec:nnndelta}
Two nucleons can also excite the $\Delta$ of diagram (b) of Fig.~\ref{M-fig:diagrams} at a single vertex.
Without derivatives the only such contact term is~\cite{Epelbaum:2007sq}
\begin{equation}
 \mathcal L_{NNN\Delta}\propto\big(\bar T^\mu_iN\big)\big(\bar NS_\mu\tau^iN\big)+{\rm h.c.}.
 \label{eq:contact-ndelta}
\end{equation}
Spin and isospin conservation imply that this vertex does not contribute to $NN \to N \Delta$. 
A term with one time derivative reduces by the equations of motion to $\delta$ times Eq.~\eqref{eq:contact-ndelta} or to terms that do not enter $NN\to NNa$ at NLO.

\section{Matching of the Lagrangian parameters}
\label{sec:matching}
We match the constants to vacuum observables using the HB vertices, external Dirac normalization and exact recoil kinematics.
At the tree order retained here, $g_A=G_A$, $g_0=G_0$, $f_\pi=F_\pi$, $m_N=M_N$, $m_\pi=M_\pi$ and $m_\Delta=M_\Delta$.
Likewise, $g_{\pi N\Delta}=G_{\pi N\Delta}$ and $g_{aN\Delta}=G_{aN\Delta}$ at their matching points.
The matching points below receive no contribution from the retained vertex recoil terms, so the same physical inputs are used at LO and NLO.
The residual hatted contact coefficients enter only at NLO.
Table~\ref{tab:matched-constants} collects the inputs and their sources, and in the following paragraph we provide details on the matching.

\begin{table}[tbp]
 \centering
 \small
 \setlength{\tabcolsep}{4pt}
 \renewcommand{\arraystretch}{1.08}
 \begin{ruledtabular}
 \begin{tabular}{p{0.13\linewidth}p{0.21\linewidth}p{0.17\linewidth}p{0.41\linewidth}}
 Constant & Value & Kinematic point & Observable or matching, and source\\
 \hline
 $G_A$ & $\TSInGA$ & $Q^2=0$
 & Neutron $\beta$ decay~\cite{ParticleDataGroup:2024cfk,Springmann:2024mjp}.\\
 $G_0$ & $\TSInGzero$ & $Q^2=0$
 & Lattice axial charges~\cite{FlavourLatticeAveragingGroupFLAG:2021npn,Springmann:2024mjp}.\\
 $\Delta s$ & $\TSInDeltaS$ & $Q^2=0$
 & Strange axial charge~\cite{FlavourLatticeAveragingGroupFLAG:2021npn,Vonk:2020zfh}.\\
 $z$ & $\TSInZ$ & ---
 & Quark-mass ratio, adopted as in Ref.~\cite{Springmann:2024mjp}.\\
 $\delta_{\rm run}$ & $\TSInDeltaRun$ & ---
 & Singlet running~\cite{GrillidiCortona:2015jxo}.\\
 $F_\pi$ & $\TSInFpi\,\mathrm{MeV}$ & on shell
 & Pion decay constant~\cite{FlavourLatticeAveragingGroupFLAG:2021npn,Springmann:2024mjp}.\\
 $M_\pi,M_N$ & $\TSInMpi,\ \TSInMN\,\mathrm{MeV}$ & on shell
 & Mass inputs~\cite{ParticleDataGroup:2024cfk,Springmann:2024mjp}.\\
 $c_{u-d}$ & $\TSCvdUnc$ & ---
 & Derived, Eq.~\eqref{eq: cupmd cs IR UV relations}.\\
 $C_p,C_n$ & $\TSCpUnc,\ \TSCnUnc$ & $Q^2=0$
 & Derived, Eq.~\eqref{eq: cp cn in the IR}.\\
 $M_\Delta$ & $\TSInMDelta\,\mathrm{MeV}$ & ---
 & Breit--Wigner mass; range $\TSInMDeltaLo$--$\TSInMDeltaHi\,\mathrm{MeV}$~\cite{ParticleDataGroup:2024cfk}.\\
 $\Gamma^{\rm BW}$ & $\TSInGamma\,\mathrm{MeV}$ & $W=M_\Delta$
 & Breit--Wigner width; range $\TSInGammaLo$--$\TSInGammaHi\,\mathrm{MeV}$~\cite{ParticleDataGroup:2024cfk}.\\
 $\mathrm{BR}_{N\pi}$ & $\TSInBRNpi$ & $W=M_\Delta$
 & $\Delta\to N\pi$ branching fraction~\cite{ParticleDataGroup:2024cfk}.\\
 $G_{\pi N\Delta}$ & $\TSInGpi$ & $\Delta$ at rest, $W=M_\Delta$
 & Derived from the partial width, Eq.~\eqref{eq:width-vacuum-kernel}.\\
 $G_{aN\Delta}$ & $\TSInGax$ & $\Delta$ at rest, $Q^2=0$
 & Axial transition in pion production by neutrinos~\cite{Hernandez:2016yfb}, equal to $\Gpi$ at this order.\\
 $\hat c_2$ & $\TSInHatCTwo$ & $\nu=t=0$
 & Converted Roy--Steiner input, Eq.~\eqref{eq: hat ci coeff and residual}~\cite{Hoferichter:2015tha}.\\
 $\hat c_3$ & $\TSInCThree$ & $\nu=t=0$
 & Roy--Steiner input~\cite{Hoferichter:2015tha}.\\
 $\hat c_4$ & $\TSInHatCFour$ & $\nu=t=0$
 & Converted Roy--Steiner input, Eq.~\eqref{eq: hat ci coeff and residual}~\cite{Hoferichter:2015tha}.\\
 $\hat c_{2,3,4}^{\rm res}$ & $\CKHatCTwoRes,\ \TSResCThree,\ \TSHatCFour$ & $\nu=t=0$
 & Derived, Eqs.~\eqref{eq: hat ci coeff and residual}.\\
 $\hat c_5$ & $\DXcFiveHat$ & ---
 & Proton neutron mass difference~\cite{BMW:2014pzb}.\\
 $\hat c_{8,9}$ & $\TSInHatCEight,\ 0$ & ---
 & Unknown.\\
 $C_S$, $C_T$ & $-(40$--$70)$, $-(1$--$2)\GeV^{-2}$ &  $Q\sim0.2$--$0.4\GeV$
 & $s$-wave amplitudes~\cite{Stoks:1993tb}.\\
 $c_D$, $\tilde c_D$ & $0,\ 0$ & ---
 & Unknown.\\
 \end{tabular}
 \end{ruledtabular}
 \caption{Physical inputs and matched HB constants, with the kinematic point of each observable.
 The physical inputs are used at LO and NLO; the residual contacts enter at NLO.
 Contact coefficients are in $\mathrm{GeV}^{-1}$.
 Entries marked ``derived'' are obtained from the indicated matching equations and the cited inputs.
 $Q$ is the momentum transfer of the axial current, $W$ the invariant mass of the $\Delta$, and $\nu=t=0$ the subthreshold point of $\pi N$ scattering.}
 \label{tab:matched-constants}
\end{table}

\paragraph*{\textbf{Axion-nucleon couplings.}}
The couplings $c_q$ of the axion to the quarks $q=u,d,s$, Eq.~\eqref{eq:derivaive coupling axion and quarks}, induce its couplings to protons and neutrons,
\begin{equation}
    C_{p,n} = G_0\, c_{u+d} \pm G_A\, c_{u-d} + \Delta s\, c_s\,,\qquad c_{u\pm d} = \frac{c_u \pm c_d}{2}\,.
    \label{eq: cp cn in the IR}
\end{equation}
The coefficients are the axial charges of the proton, the matrix elements of the quark axial currents at zero momentum transfer,
\begin{equation}
    \braket{p(P)|\bar q \gamma^\mu \gamma_5 q|p(P)} = \Delta q\, \bar u_p(P)\gamma^\mu\gamma_5 u_p(P),
\end{equation}
which measure the contributions of the quark spins to the spin of the proton, in the combinations $G_A=\Delta u-\Delta d$ and $G_0=\Delta u+\Delta d$.
For the neutron $\Delta u$ and $\Delta d$ are interchanged, which reverses the sign of $G_A$.
The isovector charge $G_A$ is the axial coupling of neutron $\beta$ decay, and we take it from its measurement.
The isoscalar charge $G_0$ and the strange charge $\Delta s$ do not enter $\beta$ decays, and we take them from lattice evaluations of the matrix element in the $\overline{\rm MS}$ scheme at $\mu = 2\GeV$ (Table~\ref{tab:matched-constants}).
The couplings $c_q$ are obtained by running down to $\mu = 2\text{ GeV}$ the model dependent UV axion-quarks couplings, $c_q^{\rm UV} = c_q^0 - [Q_a]_q$, where
\begin{equation}
    Q_a = \frac{\text{diag}\left(1, z, w\right)}{1+z+w}\,,\quad z = \frac{m_u}{m_d}\,,\quad w =\frac{ m_u}{m_s},
\end{equation}
is an axial rotation matrix removing nonderivative axion-pion couplings up to NLO.
We approximate $w\simeq 0$ in the following.
For a KSVZ axion~\cite{Kim:1979if,Shifman:1979if}, which is our benchmark model, $c_q^0 = 0$.
The effect of the running is a shift of the isospin singlet coefficients, namely
\begin{equation}
    c_{u-d} = c_{u-d}^{\rm UV}\,,\quad c_{u+d} = c_{u+d}^{\rm UV} + \delta_{\rm run} \,,\quad c_s = \delta_{\rm run},
    \label{eq: cupmd cs IR UV relations}
\end{equation}
where $\delta_{\rm run}$ is the common shift of the light $c_q$ by the running from $f_a$ to $\mu=2\GeV$, with the top and bottom quarks integrated out at their masses~\cite{GrillidiCortona:2015jxo}.
Plugging \eqref{eq: cupmd cs IR UV relations} in \eqref{eq: cp cn in the IR} gives the numerical values reported in Table~\ref{tab:matched-constants}.

The uncertainties of $C_p,C_n,c_{u-d}$ are propagated from the same draws of $G_A,G_0,\Delta s,z,\delta_{\rm run}$; in particular the proton and neutron couplings have correlation coefficient approximately $\TSCpCnCorr$ and are not sampled independently.

\paragraph*{\textbf{Pion--nucleon--$\Delta$ coupling.}}
We determine $\Gpi$ in terms of the physical nucleon and pion mass, the BW $\Delta$ mass and the pion decay constant, by making use of the vacuum decay width of the $\Delta$ at rest into $N$ and $\pi$ \eqref{eq:width-vacuum-kernel}, the measured decay width of the $\Delta$ and its $\pi N$ branching ratio, collected in Table~\ref{tab:matched-constants}.
Since recoil corrections vanish at rest, the matching is valid both at LO and NLO.

\paragraph*{\textbf{Axial nucleon--$\Delta$ coupling.}}
The pion and the axion couple to the $N\to\Delta$ transition through the same isovector axial current, so at the order of our calculation they share one constant, $\Gax=\Gpi$, the off-diagonal Goldberger--Treiman relation~\cite{Procura:2008ze}.
Two observables determine it and agree within their uncertainties.
The width gives $\Gpi=\TSInGpi$, and the axial transition measured in pion production by neutrinos gives $\Gax=\TSInGax$~\cite{Hernandez:2016yfb}.
We use the first value at the pion vertex and the second at the axion vertex.

\paragraph*{\textbf{$\hat c_i$ and residual $\hat c_i^{\rm res}$ coefficients.}}
To obtain $\hat c_i$, we use the Roy--Steiner values~\cite{Hoferichter:2015tha} and sum the $1/m_N$ terms whenever necessary, see \eqref{eq: hat ci coeff and residual}.

The coefficients $\hat c_{2, 3, 4}$ receive corrections from the $\Delta$-mediated $N\pi \rightarrow N \pi$ interaction at low energy, when the $\Delta$ propagator can be integrated out.
In Ref.~\cite{Arcangeli:2026long} we show explicitly that these contributions are, at LO,
\begin{equation}
\begin{aligned}
 c_2^\Delta&=\frac{4G_{\pi N\Delta}G_{aN\Delta}}{9\delta},
 &c_3^\Delta&=-\frac{4G_{\pi N\Delta}G_{aN\Delta}}{9\delta},&
 c_4^\Delta&=\frac{2G_{\pi N\Delta}G_{aN\Delta}}{9\delta}.
\end{aligned}
\label{eq:matching-residual-contacts}
\end{equation}
Therefore, if we used $\hat c_i$ as Lagrangian parameters of the $\Delta$-full Lagrangian, we would double count.
To prevent this, we define \textit{residual} coefficients \eqref{eq: hat ci coeff and residual}, by subtracting \eqref{eq:matching-residual-contacts} from the numerical result for $\hat c_i$.

$c_8$ and $c_9$ are not yet available from lattice simulations.
With the same naturalness prescription already used in Ref.~\cite{Springmann:2024mjp}, we assign the independent contact contributions an equal-weight mixture of Gaussian priors centered at $\pm0.5\,\mathrm{GeV}^{-1}$, each with standard deviation $0.5\,\mathrm{GeV}^{-1}$.
The central prediction sets these contributions to zero, while retaining the fixed recoil term $\hat c_8=-G_AG_0/M_N$.

The strong part of the mass difference of neutron and proton, $(M_n-M_p)_{\rm QCD}=2.52(29)\MeV$ from lattice QCD~\cite{BMW:2014pzb}, fixes $\hat c_5=\DXcFiveHat\GeV^{-1}$~\cite{Springmann:2024mjp}.

\paragraph*{\textbf{Contact terms of two nucleons.}}
We match $C_S$ and $C_T$ at tree level to the ${}^1S_0$ and ${}^3S_1$ amplitudes of the partial-wave analysis of Ref.~\cite{Stoks:1993tb} at the relative momentum of the collision, which gives $C_S\approx-(40$ to $70)\GeV^{-2}$ and $C_T\approx-(1$ to $2)\GeV^{-2}$ (real parts) at $0.2$ to $0.4\GeV$.
We match at the momenta of the colliding nucleons, which absorbs the neglected iterations of the interaction into the constants at each scale.

The coefficient $c_D$ can be determined from the Gamow--Teller matrix element, however, with large uncertainty~\cite{Gnech:2023mvb}.
No determination of $\tilde c_D$ exists.
For both constants we choose a superposition of two Gaussians centered around $\pm 1$ with width $1$, which covers this value of $c_D$.

\section{Supernova profiles and in-medium self-energies}
\label{sec:medium-profiles}

\begin{figure}[tbp]
 \centering
 \includegraphics[width=\linewidth]{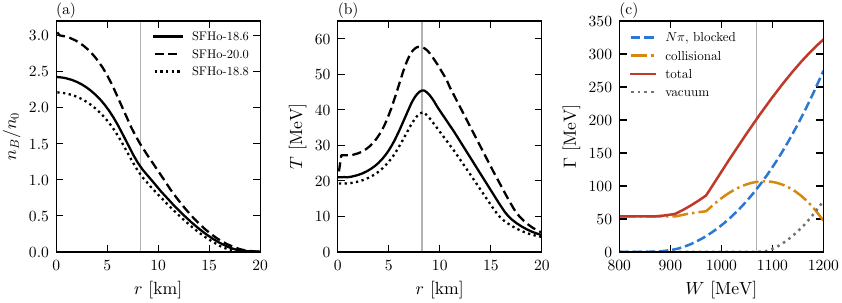}
 \caption{(a) Baryon density in units of $n_0=0.16\,\mathrm{fm}^{-3}$ and (b) temperature against radius at one second post-bounce for SFHo-18.6 (solid), SFHo-20.0 (dashed) and SFHo-18.8 (dotted).
 The vertical line shows the hottest cell of SFHo-18.6.
 (c) Reference in-medium $\Delta$ width in that cell, against the kinetic invariant mass $W$, for a momentum of $300\,\mathrm{MeV}$, namely the Pauli-blocked mesonic width, the collisional contribution of Ref.~\cite{Oset:1987re}, their sum, and the vacuum width.
 The vertical line marks the in-medium pole.}
 \label{fig:width-profiles}
\end{figure}

\paragraph*{\textbf{Supernova profiles.}}
We use the spherically symmetric Garching simulations SFHo-18.6, SFHo-18.8 and SFHo-20.0, with progenitor masses $18.6$, $18.8$ and $20.0\,M_\odot$, including muons~\cite{Bollig:2020xdr,Garching} and the SFHo equation of state~\cite{Steiner:2012rk}.
The radial profiles supply the temperature, proton and neutron densities and chemical potentials, nucleon effective masses and vector self-energies, together with the lapse and fluid velocity.
Figure~\ref{fig:width-profiles}(a),(b) shows the density and temperature at $t_{\rm pb}=1\,\mathrm{s}$.
We use this time as the benchmark for the cooling criterion $L_a(1\,\mathrm{s})\lesssim L_\nu(1\,\mathrm{s})$~\cite{Raffelt:2006cw,Springmann:2024mjp}; the stellar signal is integrated over the profiles from zero to nine seconds.
At one second, most emission comes from the hot shell near $r=8\,\mathrm{km}$.
For SFHo-18.6 the emissivity maximum is at $r=\TSPeakRadius\,\mathrm{km}$, with $T=\TSPeakT\,\mathrm{MeV}$ and $n=\TSPeakDensity\,n_0$, where $n_0=0.16\,\mathrm{fm}^{-3}$.
At $t_{\rm pb}=1\,\mathrm{s}$, SFHo-18.8 and SFHo-20.0 provide colder and hotter comparison models, with smaller and larger axion luminosities, respectively.

\paragraph*{\textbf{Self-energies and propagator poles.}}
In dense matter the surrounding nucleons give a nucleon a scalar and a vector self-energy.
In the relativistic mean-field models that provide the equation of state of the simulations~\cite{Hempel:2014ssa}, they are generated by the mean fields of scalar mesons, which lower the mass of the nucleon, and of vector mesons, which shift its energy like a potential.
We take both self-energies, $\Sigma_S^N$ and $\Sigma_V^N$, from the profiles as local, momentum-independent inputs, separately for protons and neutrons, $N=p,n$, and include them in the relativistic propagators before taking the HB projection.
In the fluid rest frame the nucleon self-energy is $\Sigma_S^N+\slashed v\,\Sigma_V^N$.
It shifts the mass and separates the canonical four-momentum $P$, which is conserved at the vertices, from the kinetic four-momentum $P-\Sigma_V^Nv$, which is on the mass shell.
A nucleon of momentum $\vp$ therefore has
\begin{equation}
 m_N^*=M_N+\Sigma_S^N,\qquad
 E=\Sigma_V^N+\epsilon,\qquad
 \epsilon=\sqrt{\vp^{\,2}+m_N^{*2}},
\label{eq:medium-nucleon-kinematics}
\end{equation}
with the vacuum mass $M_N$ of the proton or the neutron, as in the End Matter.
For the $\Delta$ we introduce scalar and vector self-energies $\Sigma_S^\Delta$ and $\Sigma_V^\Delta$ in the same way, separately for each charge state.
Then
\begin{equation}
 m_\Delta^*=M_\Delta+\Sigma_S^\Delta,\qquad
 W^2=(P^0-\Sigma_V^\Delta)^2-\vP^{\,2},\qquad
 E_\Delta=\Sigma_V^\Delta+\sqrt{\vl^{\,2}+m_\Delta^{*2}},
\label{eq:medium-delta-kinematics}
\end{equation}
where $W$ is the kinetic invariant mass of a $\Delta$ of canonical four-momentum $P$, and $E_\Delta$ the energy of a $\Delta$ of momentum $\vl$ on its mass shell.
Every internal $\Delta$ in the diagrams of $NN\to NNa$ and of the absorption is attached to an $aN\Delta$ vertex.
The axion is neutral, so this vertex connects only $p\leftrightarrow\Delta^+$ and $n\leftrightarrow\Delta^0$.
In the formulas for a $\Delta$ below, $N$ is the nucleon at this vertex, the proton for the $\Delta^+$ and the neutron for the $\Delta^0$.

The propagators of Eq.~\eqref{M-eq:props} then depend on the kinetic four-momenta and the effective masses.
In terms of the canonical momentum $P$ of the internal baryon their scalar factors become
\begin{equation}
\begin{aligned}
 S_N^{\rm med}(P)
 &=\frac{P^0-\Sigma_V^N+m_N^*}
 {(P^0-\Sigma_V^N)^2-\vP^{\,2}-m_N^{*2}+i0},\\[3pt]
 S_\Delta^{\rm med}(P)
 &=\frac{P^0-\Sigma_V^\Delta+m_\Delta^*}
 {(P^0-\Sigma_V^\Delta)^2-\vP^{\,2}-m_\Delta^{*2}+iW\Gamma},
\end{aligned}
\label{eq:medium-resummed-propagators}
\end{equation}
and the HB spin and isospin projectors remain as specified in \eqref{eq:hb-projectors}.
Before the width is included, the poles lie at $(P-\Sigma_V^Nv)^2=m_N^{*2}$ and $(P-\Sigma_V^\Delta v)^2=m_\Delta^{*2}$.
Keeping these denominators unexpanded preserves the relativistic quasiparticle pole positions, which is crucial to consistently estimate the production rate within the SN profiles of the Garching group, as the individual scalar and vector self-energies, provided in their tables, are of hundreds of MeV, even when their sum is much smaller.

\paragraph*{\textbf{$\Delta$ self-energy prescription and uncertainty.}}
To account for in-medium corrections to the $\Delta$ mass and energy consistently with our treatment of nucleons, we introduce scalar and vector self-energies, $\Sigma_S^\Delta$ and $\Sigma_V^\Delta$.
Since the $\Delta$ self-energies are not specified by the Garching simulations, we express in terms of the nucleon self energies using the following prescription.
Defining
\begin{equation}
 U_N=\Sigma_S^N+\Sigma_V^N,
 \qquad U_\Delta=\Sigma_S^\Delta+\Sigma_V^\Delta,
 \qquad D=U_\Delta-U_N,
\label{eq:medium-potential-difference}
\end{equation}
we adopt $U_\Delta/U_N=2/3$ as our reference prescription, following the transport-model treatment of a less attractive $\Delta$ potential~\cite{GiBUU,Mosel:2020zdw}.
This fixes the total potential but leaves its scalar/vector decomposition undetermined.
For simplicity, we assign the same factor to the scalar self-energy, $\Sigma_S^\Delta/\Sigma_S^N=2/3$, which also implies $\Sigma_V^\Delta/\Sigma_V^N=2/3$.
Our reference choice is therefore
\begin{equation}
 \Sigma_S^\Delta=\frac23\Sigma_S^N,
 \qquad \Sigma_V^\Delta=\frac23\Sigma_V^N.
\label{eq:medium-delta-reference}
\end{equation}
We assess this prescription by varying separately the potential difference $D$ and the scalar/vector split.
As shown below, the former test identifies the reference as a comparatively conservative benchmark, while the latter supports the use of the simple scalar prescription.

We first consider the uncertainty in $D$.
Published analyses favour both less attractive and more attractive $\Delta$ potentials relative to nucleons~\cite{Mosel:2020zdw,Oset:1987re,Kolomeitsev:2016ptu,Bodek:2020wbk,Cozma:2021tfu}.
The reference relation $D=-U_N/3$ should therefore be understood as a model prescription, without an associated statistical error.
To explore the spread of literature estimates, we adopt the model envelope $-30\leq D\leq25\,\mathrm{MeV}$~\cite{Kolomeitsev:2016ptu}, holding $\Sigma_S^\Delta=2\Sigma_S^N/3$ fixed and adjusting the vector self-energy.

Across the three supernova models, the endpoints $D=-30$ and $+25\,\mathrm{MeV}$ change the $\Delta$ luminosity at one second by factors $\RBOneSDmthirtyDMin$--$\RBOneSDmthirtyDMax$ and $\RBOneSDptwentyfiveDMin$--$\RBOneSDptwentyfiveDMax$, respectively; the corresponding factors for the total luminosity are $\RBOneSDmthirtyTMin$--$\RBOneSDmthirtyTMax$ and $\RBOneSDptwentyfiveTMin$--$\RBOneSDptwentyfiveTMax$.
Over zero to nine seconds, the number of axions above $200\,\mathrm{MeV}$ changes by at most $\RBDChangeHard\%$, while the $\Delta$ enhancement of the Hyper-Kamiokande and Fermi-LAT counts changes by at most $\RBDChangeHK\%$ and $\RBDChangeSE\%$, respectively.
Thus, relative to the reference prescription, the total luminosity at one second changes by at most $\fpeval{round(100*(\RBOneSDmthirtyTMax-1))}\%$, and the time-integrated hard-axion signals by at most $\fpeval{max(\RBDChangeHard,\RBDChangeHK,\RBDChangeSE)}\%$.
These variations affect the size of the $\Delta$ contribution while preserving its enhancement of the hard axion signal.
We quote this uncertainty separately from the bands associated with the matched constants and the chiral truncation.

Within the explored envelope, decreasing $D$ strengthens the $\Delta$ contribution.
In the emitting region at one second, the nucleon potentials are attractive, with $U_n\simeq-(20$--$37)\,\mathrm{MeV}$ and $U_p\simeq-(43$--$80)\,\mathrm{MeV}$.
The reference prescription therefore gives $D\simeq6$--$12\,\mathrm{MeV}$ for neutrons and $14$--$27\,\mathrm{MeV}$ for protons.
These values lie on the positive-$D$ side of the adopted envelope, with the largest proton values slightly above its upper endpoint.
Together with the observed increase of the $\Delta$ contribution at lower $D$, this makes our reference choice a comparatively conservative benchmark for the enhancement.

We next examine the scalar/vector decomposition.
At fixed $D$, we vary the scalar ratio $\alpha=\Sigma_S^\Delta/\Sigma_S^N$ while preserving the total potential by writing
\begin{equation}
 \Sigma_S^\Delta=\alpha\Sigma_S^N,
 \qquad
 \Sigma_V^\Delta=\Sigma_V^N+(1-\alpha)\Sigma_S^N+D.
\label{eq:medium-scalar-vector-split}
\end{equation}
The reference prescription is recovered for $\alpha=2/3$ and $D=-U_N/3$.
In a representative hot-shell sensitivity test, varying $-1\leq\alpha\leq1$  at fixed $U_\Delta/U_N=2/3$ gives maximum relative half-widths of about $16\%$ for the local $\Delta$ spectrum and $8\%$ for the total spectrum.
This weak dependence, smaller than the sensitivity to $D$ discussed above, supports our simplifying choice of the scalar/vector split.
We note that the estimates  of Ref.~\cite{Kolomeitsev:2016ptu} also give ratios near $0.7$.

\paragraph*{\textbf{Shift of the resonance.}}
Including the in-medium self energies has an impact on the axion energy giving the resonance.
For a final nucleon $N$ at rest and zero $\Delta$ width, the intermediate momentum is $P=(\Sigma_V^N+m_N^*,\bm0)+p_a$.
Its pole condition gives
\begin{equation}
 E_a^{\rm res}
 =\frac{m_\Delta^{*2}-(m_N^*+\Sigma_V^N-\Sigma_V^\Delta)^2}{2(m_N^*+\Sigma_V^N-\Sigma_V^\Delta)}
 =\delta+D+\frac{(\delta+D)^2}{2(m_N^*+\Sigma_V^N-\Sigma_V^\Delta)},
 \qquad \delta=M_\Delta-M_N .
\label{eq:medium-resonance-general}
\end{equation}
Using the reference choice \eqref{eq:medium-delta-reference}, we can rewrite it as
\begin{equation}
 E_a^{\rm res}
 =\left(\delta-\frac{U_N}{3}\right)
  +\frac{(\delta-U_N/3)^2}
         {2(m_N^*+\Sigma_V^N/3)} .
\label{eq:medium-resonance-reference}
\end{equation}
The leading potential shift is therefore $-U_N/3$, and it is independent of the scalar/vector split, which justifies why varying the parameter $\alpha$ in the previous paragraph has a small impact on the final result.
To develop some intuition about the order of the effect, we can consider that in the hottest cell of SFHo-18.6 at one second,  $\Sigma_S^n\simeq \PRLSMSigmaS\,\mathrm{MeV}$ and $\Sigma_V^n\simeq \PRLSMSigmaV\,\mathrm{MeV}$, so that $U_n\simeq\PRLSMU\,\mathrm{MeV}$ and the leading shift $-U_n/3\simeq+\PRLSMShift\,\mathrm{MeV}$ moves the resonance, and with it the peak of the axion emissivity, upwards; for protons $U_p\simeq\PRLSMUp\,\mathrm{MeV}$ and $-U_p/3\simeq+\PRLSMShiftp\,\mathrm{MeV}$.
However, moving nucleons add a direction-dependent Doppler shift, which broadens the $\Delta$ peak, and the Fermi-Dirac distributions penalize larger energies.

\paragraph*{\textbf{Consistent use in the emissivity.}}

The same self-energies enter the baryon propagators, the $\Delta$ width calculation in Sec.~\ref{sec:width}, external normalizations and phase space, and the Fermi--Dirac distributions when computing the axion emissivity \eqref{M-eq:emis} in the End Matter.

The normalization of the large-component of the nucleon Dirac spinors \eqref{eq:large component of spinor} becomes
\begin{equation}
    \xi_i = \chi_i\sqrt{\epsilon_i + m_N^*}\,,\quad \epsilon_i = \sqrt{m_N^{*2} + \mathbf p_i^2}
\end{equation}
and the relativistic normalization of the phase space measure, Eq.~\eqref{M-eq:dPhi} of the End Matter, $\prod_i1/(2\epsilon_i)$, is $m_N^*$ dependent as well.
Similarly, the Fermi-Dirac distributions can be written as
\begin{equation}
 f_N(\epsilon)
 =\left[1+\exp\!\left(\frac{\epsilon-\mu_N^*}{T}\right)\right]^{-1},
 \qquad \mu_N^*=\mu_N-\Sigma_V^N ,
\label{eq:medium-fermi-occupations}
\end{equation}
which equals $[1+\exp((E-\mu_N)/T)]^{-1}$ with the energy $E$ and the chemical potential $\mu_N$.
Finally, notice that the delta function $\delta(E_1+E_2-E_3-E_4-E_a)$ does not receive contributions by the in-medium vector self energies, but only by the scalar ones through the in-medium nucleon masses.
Notice also that exchanged pions carry the difference of canonical nucleon momenta, including the vector self-energy difference in charge-exchange channels.

\section{The \texorpdfstring{$\Delta$ width in vacuum and in medium}{Delta}}\label{sec:width}
\label{sec:delta-width}

The width in the $\Delta$ propagator \eqref{eq:hb-resummed-propagators} has a mesonic and a collisional part,
\begin{equation}
 \Gamma=\Gamma_{N\pi}+\Gamma_{\rm coll}.
\label{eq:width-split}
\end{equation}
The mesonic part describes the decay $\Delta\to N\pi$, with the final nucleon subject to Pauli blocking, and we compute it from the same $\pi N\Delta$ vertex as the axion amplitudes.
The collisional part describes collisions with the nucleons of the medium and vanishes in vacuum.
We take it from the many-body calculation of Ref.~\cite{Oset:1987re} and compare it with a different model at the end of this section.
Small electromagnetic decay channels are neglected throughout.
As in the End Matter we write the formulas for the $\Delta^+$.
Those for the $\Delta^0$ follow by interchanging $p$ and $n$.

\paragraph*{\textbf{Vacuum decay.}}
A $\Delta^+$ of invariant mass $W$ decays into $p\pi^0$ and $n\pi^+$ with the squared isospin Clebsch--Gordan coefficients $\tfrac23$ and $\tfrac13$.
With the vertex of Eq.~\eqref{eq:hb-lo-vertices}, averaged over the spin of the $\Delta$ and summed over that of the nucleon and over the two channels, the decay rate in the rest frame of the $\Delta$ is
\begin{equation}
\begin{aligned}
 \Gamma_{N\pi}&=\frac{\Gpi^2q^3(\epsilon_N+M_N)}{12\pi F_\pi^2W},\qquad
 q=\frac{\lambda^{1/2}(W^2,M_N^2,M_\pi^2)}{2W},\qquad
 \epsilon_N=\sqrt{M_N^2+q^2},\\
 \lambda(x,y,z)&=x^2+y^2+z^2-2xy-2yz-2zx,
\end{aligned}
\label{eq:width-vacuum-kernel}
\end{equation}
where $q$ is the momentum of the pion and $\epsilon_N$ the energy of the nucleon.
Below the threshold $W=M_N+M_\pi$ the width vanishes.
We use one nucleon mass $M_N$ and one pion mass $M_\pi$ (Table~\ref{tab:matched-constants}), so the two channels have the same kinematics, and the $\Delta^+$ and the $\Delta^0$ have the same width.
At $W=M_\Delta$ it is the $N\pi$ part of the width, $\mathrm{BR}_{N\pi}\Gamma^{\rm BW}\simeq116.3\,\mathrm{MeV}$, which fixes $\Gpi$ (Sec.~\ref{sec:matching}).

We use Eq.~\eqref{eq:width-vacuum-kernel} at LO and at NLO for the matching of the coupling.
The NLO recoil term of the transition vertex, Eq.~\eqref{eq:hb-delta-recoil-rule}, vanishes for a $\Delta$ at rest.

\paragraph*{\textbf{Decay in the medium.}}
In the medium the final nucleon of the decay, $N=p$ or $n$, has its own vector self-energy $\Sigma_V^N$, which differs from that of the $\Delta$.
Energy conservation with the canonical four-momentum $P$ of the $\Delta$ gives the $N\pi$ system the kinetic four-momentum $P-\Sigma_V^Nv$ and the invariant mass
\begin{equation}
 W_N^2=(P^0-\Sigma_V^N)^2-\vP^{\,2},
\label{eq:width-pair-invariant}
\end{equation}
as in Eq.~\eqref{M-eq:width} of the End Matter.
For the $\Delta^+$ of diagram (b) of Fig.~\ref{M-fig:diagrams} with $s=+$, for example, $P=p_3+p_a$ with $p_3^0=\Sigma_V^p+\epsilon_3$, so that $W_p^2=(\epsilon_3+E_a)^2-(\vp_3+\vp_a)^2$, while $W_n$ contains in addition $\Sigma_V^p-\Sigma_V^n$.

The final nucleon is Pauli blocked.\footnote{We neglect the density of pions.
With pions of occupation $n_\pi$ the factor $1-f_N$ would become $1-f_N+n_\pi=(1-f_N)(1+n_\pi)+f_Nn_\pi$ for the decay and its inverse, or $(1-f_N)(1+n_\pi)$ for the decay of a $\Delta$ alone.}
The factor $1-f_N$ is the statistical factor of the $N\pi$ cut of the retarded $\Delta$ self-energy.
The decay is isotropic in the rest frame of $P-\Sigma_V^Nv$, so the energy of the nucleon in the medium frame is distributed uniformly between
\begin{equation}
 \epsilon_N^\pm=\frac{(P^0-\Sigma_V^N)\,\epsilon_N\pm|\vP|\,q_N}{W_N},
\label{eq:width-daughter-energy-range}
\end{equation}
where $q_N$ and $\epsilon_N$ are $q$ and $\epsilon_N$ of Eq.~\eqref{eq:width-vacuum-kernel} with $W_N$ and $m_N^*$ in place of $W$ and $M_N$.
Averaged over the decay directions, the probability that the final state is empty is
\begin{equation}
 B_N=\frac{1}{\epsilon_N^+-\epsilon_N^-}\int_{\epsilon_N^-}^{\epsilon_N^+}\dd\epsilon\,\big[1-f_N(\epsilon)\big]
 =\frac{T}{\epsilon_N^+-\epsilon_N^-}\,
  \ln\frac{1+e^{(\epsilon_N^+-\mu_N^*)/T}}{1+e^{(\epsilon_N^--\mu_N^*)/T}} ,
\label{eq:width-finite-temperature-blocking}
\end{equation}
with $f_N$ of Eq.~\eqref{eq:medium-fermi-occupations}.
The mesonic width of the $\Delta^+$ is then $\Gamma_{N\pi}=\tfrac23\Gamma_p+\tfrac13\Gamma_n$, with
\begin{equation}
 \Gamma_N=\frac{\Gpi^2q_N^3(\epsilon_N+m_N^*)}{12\pi F_\pi^2W_N}\,B_N ,
\label{eq:width-medium-charges}
\end{equation}
Eq.~\eqref{M-eq:width} of the End Matter.
In vacuum, where $m_N^*=M_N$, $W_N=W$ and $B_N=1$, this coincides with Eq.~\eqref{eq:width-vacuum-kernel}.

For a $\Delta$ at rest $\epsilon_N^\pm=\epsilon_N$, and $B_N=1-f_N(\epsilon_N)$.
A moving $\Delta$ boosts its decay nucleon.
A nucleon emitted along $\vP$ gains energy and one emitted against it loses energy, so that the energies of Eq.~\eqref{eq:width-daughter-energy-range} fill an interval of width $2|\vP|q_N/W_N$ around $(P^0-\Sigma_V^N)\,\epsilon_N/W_N$.
At $T=0$, $B_N$ is the fraction of this interval that lies above $\mu_N^*$.
A moving $\Delta$ can therefore decay where a $\Delta$ at rest is blocked, and a finite temperature smooths the edge.
Protons and neutrons are blocked with their own occupations and thresholds.

\paragraph*{\textbf{Collisional broadening.}}
We use the absorptive collisional self-energy of Ref.~\cite{Oset:1987re} in the parametrization implemented in GiBUU~\cite{GiBUU}.
Its three parts describe quasi-elastic scattering $\Delta N\to\Delta N$, two-body absorption $\Delta N\to NN$ and three-body absorption $\Delta NN\to NNN$.
The parametrization depends on the density and on the energy $\omega$ of a pion that forms the $\Delta$ with a nucleon of the Fermi sea.\footnote{Notice that this energy $\omega$ only labels the parametrization of the GiBUU collisional width.
It is not related to the energy $q^0$ of the exchanged pion in the Bremsstrahlung amplitude.}
GiBUU obtains this energy from the mass of the $\Delta$ with vacuum kinematics, which no longer holds with the self-energies of the rate.
We therefore keep the fitted coefficient functions and recompute their argument.

For $\pi N\to\Delta$, with $\omega$ the energy of the pion, $W$ the invariant mass of the $\Delta$ and by averaging over the momentum of the nucleon in the Fermi sea, energy conservation gives
\begin{equation}
 W^2=\Big(\bar\epsilon+\frac{\Sigma_V^N}{3}+\omega\Big)^2-\omega^2+M_\pi^2-\frac35k_f^2,
 \qquad k_f=\Big(\frac{3\pi^2n}{2}\Big)^{1/3},
 \qquad \bar\epsilon=m_N^*+\frac{3k_f^2}{10m_N^*},
\label{eq:width-collision-fermi-average}
\end{equation}
with the baryon density $n=n_p+n_n$, the average kinetic energy $\bar\epsilon$ and squared momentum $3k_f^2/5$ of the Fermi sea, and $m_N^*$ and $\Sigma_V^N$ of the nucleon interacting with the $\Delta$ through an axion (in particular, the proton for the $\Delta^+$).
The term $\Sigma_V^N/3=\Sigma_V^N-\Sigma_V^\Delta$ is the difference of the vector self-energies \eqref{eq:medium-delta-reference}.
Up to cubic order in $k_f$,
\begin{equation}
 \omega=\frac{W^2-m_N^{*2}-M_\pi^2-(\Sigma_V^N/3)^2-2\bar\epsilon\,\Sigma_V^N/3}{2(\bar\epsilon+\Sigma_V^N/3)} ,
\label{eq:width-collision-energy-mapping}
\end{equation}
which without the self-energies is the relation used by GiBUU.
When we vary $D$ in Sec.~\ref{sec:medium-profiles}, $\Sigma_V^N-\Sigma_V^\Delta$ with the varied $\Sigma_V^\Delta$ replaces $\Sigma_V^N/3$ in Eqs.~\eqref{eq:width-collision-fermi-average} and \eqref{eq:width-collision-energy-mapping}.

We define $x=(\omega-M_\pi)/M_\pi$, the kinetic energy of the pion in units of its mass, and we use the pion mass of the GiBUU implementation~\cite{GiBUU}, $M_\pi=138\,\mathrm{MeV}$.
We evaluate the coefficient functions for $0\leq\omega-M_\pi\leq450\,\mathrm{MeV}$ and hold the argument at the nearest endpoint outside this interval.
With $y=n/(0.168\,\mathrm{fm}^{-3})$ the width is~\cite{Oset:1987re}
\begin{equation}
 \Gamma_{\rm coll}
 =2\left([C_Q]_+y^\alpha+[C_{A2}]_+y^\beta
                              +[C_{A3}]_+y^{2\beta}\right),
\label{eq:width-collision-fit}
\end{equation}
where $[C]_+=\max(C,0)$ sets negative fitted coefficients to zero, and the coefficients in MeV and the exponents are
\begin{equation}
\begin{aligned}
 C_Q&=-5.19x^2+15.35x+2.06\,,\quad
 C_{A2}=1.06x^2-6.64x+22.66\,,\quad
 C_{A3}=-13.46x^2+46.17x-20.34,\\
 \alpha&=0.382x^2-1.322x+1.466,
 \qquad \beta=-0.038x^2+0.205x+0.613 .
\end{aligned}
\label{eq:width-collision-coefficients}
\end{equation}

The coefficients of Ref.~\cite{Oset:1987re} describe cold symmetric matter, while the matter of the supernova profiles is hot and neutron rich.
We neglect the dependence of the coefficients on the temperature and on the asymmetry, and estimate the resulting error by the difference between the mesonic width at zero and at finite temperature, which is at most $10$--$20\,\mathrm{MeV}$.\footnote{The temperature opens the phase space that Pauli blocking closes, by smearing the Fermi--Dirac distribution around the Fermi momentum.
The momentum of the $\Delta$ already lifts much of the blocking, so the temperature changes the mesonic width by no more than $10$--$20\,\mathrm{MeV}$.
For the LO contribution of $\Delta N\to NN$ to the collisional width, computed with the $\Delta$-full chiral Lagrangian, we checked that this estimate of the temperature effect is conservative.}
In the emissivity and the luminosity this is small compared with the other sources of uncertainty.

Panel (c) of Fig.~\ref{fig:width-profiles} shows the contributions to the width in the hottest cell of SFHo-18.6 at one second, $r=\PRLSMPeakR\,\mathrm{km}$, $T=\PRLSMPeakT\,\mathrm{MeV}$, $n=\PRLSMPeakDens\,n_0$ and $Y_p=\PRLSMPeakYp$.
For a $\Delta$ momentum of $300\,\mathrm{MeV}$ the reference curve gives, at its in-medium pole $W=m_\Delta^*=\PRLSMPoleW\,\mathrm{MeV}$, $\Gamma_{N\pi}=\PRLSMGammaMes\,\mathrm{MeV}$ and $\Gamma_{\rm coll}=\PRLSMGammaColl\,\mathrm{MeV}$, hence $\Gamma=\PRLSMGammaTot\,\mathrm{MeV}$, about $\fpeval{\PRLSMGammaTot-\PRLSMGammaVacPole}\,\mathrm{MeV}$ larger than its vacuum and rest frame value.
Neglecting the collisional terms would reduce by almost a factor of 2 the total width \eqref{eq:width-split}, and therefore narrow the resonance, overestimating the role of the $\Delta$ in the axion Bremsstrahlung process.

\paragraph*{\textbf{Model dependence of the collisional width.}}
Unlike the mesonic width, the collisional width introduced in the previous paragraph does not follow from the chiral Lagrangian.
Its parametrization reproduces a many-body calculation for nuclear matter near saturation density, whereas the axions are emitted from denser and hotter matter.
We therefore test how much the emission depends on it.
We replace $\Gamma_{\rm coll}$ in Eq.~\eqref{eq:width-split} by three alternatives and keep everything else fixed, namely the vertices, the constants, the self-energies, the supernova profiles and the Monte Carlo events:
\begin{enumerate}
    \item no collisional width, $\Gamma=\Gamma_{N\pi}$, which gives the narrowest resonance;
    \item the two-body absorption width for $\Delta N\to NN$, computed consistently with the inverse production process $NN\to N\Delta$ entering the axion emission diagrams, as required by detailed balance. 
    We evaluate the direct and exchange one-pion-exchange diagrams at NLO using the vertices of Sec.~\ref{sec:lagrangian} and integrating over the in-medium phase space using the Fermi-Dirac distributions for every baryon in chemical and temperature equilibrium;
    \item the spreading-width of Ref.~\cite{VanDaele:2001cz}, which is obtained from the $N \pi$ contribution to the $\Delta$ self energy.
    Medium effects are included through mean-field self-energies and Pauli blocking in the nucleon propagator, while the pion propagator is dressed by its interactions with the surrounding nucleons, including short-range correlations.
    We adapt this calculation to the same effective masses, vector self-energies, temperature and proton fraction as the rate.
\end{enumerate}

\begin{figure}[t]
 \centering
 \includegraphics[width=\linewidth]{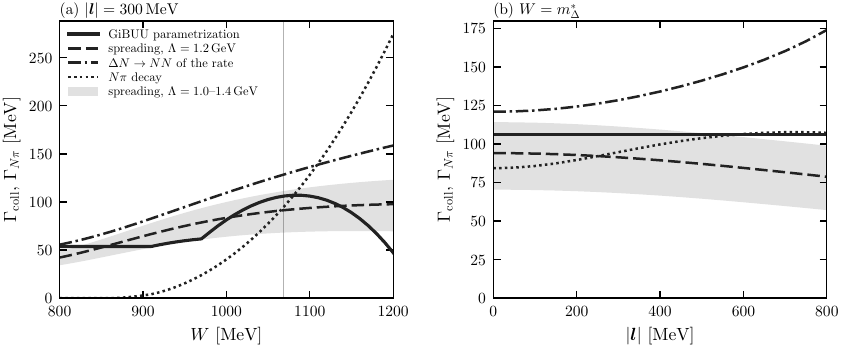}
 \caption{Collisional widths of the $\Delta^0$ in the hottest cell of SFHo-18.6 at one second.
 The solid line is the collisional width of Ref.~\cite{Oset:1987re} in its GiBUU parametrization, the dashed line the spreading width with the cutoff $\Lambda=1.2\,\mathrm{GeV}$ (the band covers $\Lambda=1.0$--$1.4\,\mathrm{GeV}$), and the dash-dotted line the two-body absorption $\Delta N\to NN$ with the one-pion exchange of the rate, the width required by detailed balance.
 The dotted line is the Pauli-blocked width $\Gamma_{N\pi}$ of the decay $\Delta\to N\pi$, the total width without collisional width.
 (a) Against the kinetic invariant mass $W$ for a momentum $|\vl|=300\,\mathrm{MeV}$.
 The vertical line marks the pole $W=m_\Delta^*$.
 The kinks of the parametrization lie at the lower end of its argument, $\omega=M_\pi$, and at the zero of $C_{A3}$, Eq.~\eqref{eq:width-collision-coefficients}.
 (b) At the pole, against the momentum $|\vl|$ of the $\Delta$.}
 \label{fig:width-compare}
\end{figure}

Figure~\ref{fig:width-compare} compares the three collisional widths in the hottest cell of SFHo-18.6 at one second, the cell of Fig.~\ref{fig:width-profiles}(c).
At the pole, for a $\Delta$ momentum of $300\,\mathrm{MeV}$, the GiBUU parametrization gives $\CWOSPole\,\mathrm{MeV}$, the spreading width $\CWSpPole\,\mathrm{MeV}$ ($\CWSpPoleLo$--$\CWSpPoleHi\,\mathrm{MeV}$ for cutoffs between $1.0$ and $1.4\,\mathrm{GeV}$) and the width from detailed balance $\CWAbsPole\,\mathrm{MeV}$, compared with $\Gamma_{N\pi}=\PRLSMGammaMes\,\mathrm{MeV}$.
The rate samples the width mostly below and around the pole.
For a neutron at rest, axion energies of $100$--$500\,\mathrm{MeV}$ give $W=860$--$1160\,\mathrm{MeV}$ in this cell.
In this range the parametrization lies within the band of the spreading width, and the width from detailed balance lies above both.
The GiBUU parametrization does not depend on the momentum of the $\Delta$, since its argument $\omega$ in Eq.~\eqref{eq:width-collision-energy-mapping} depends only on $W$, whereas the width from detailed balance rises with the momentum and the spreading width falls.

\begin{figure}[t]
 \centering
 \includegraphics[width=\linewidth]{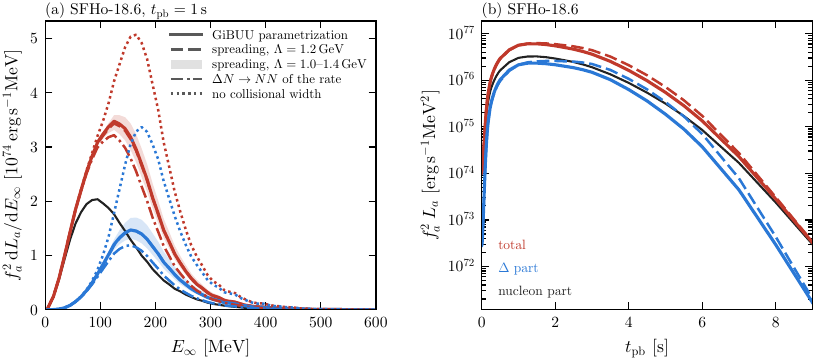}
 \caption{Axion emission with different collisional widths of the $\Delta$ and everything else fixed.
 The line style marks the width, namely the collisional width of Ref.~\cite{Oset:1987re} in its GiBUU parametrization, used for all other results (solid), the spreading width with the cutoff $\Lambda=1.2\,\mathrm{GeV}$ (dashed, the band covers $\Lambda=1.0$--$1.4\,\mathrm{GeV}$), the two-body absorption $\Delta N\to NN$ with the one-pion exchange of the rate, the width required by detailed balance (dash-dotted), and no collisional width (dotted).
 The color marks the total (red), the $\Delta$ part (blue) and the nucleon part (black), which does not depend on the width.
 (a) Spectrum at infinity at one second in SFHo-18.6.
 (b) Luminosity in SFHo-18.6 over the first nine seconds, with the GiBUU parametrization and the spreading width.}
 \label{fig:width-models}
\end{figure}

Figure~\ref{fig:width-models}(a) shows the spectrum at one second.
The luminosity calculated by the spreading width and GiBUU width agree well. The width from detailed balance lowers the luminosity slightly.
Without collisional width the luminosity instead would rise, since the narrower resonance emits more near its peak.

Panel (b) follows SFHo-18.6 over nine seconds.
At late times the $\Delta$ emission comes from matter at $\RBLateDensMidLo$--$\RBLateDensMidHi\,n_0$, beyond the densities for which both many-body models were constructed, and there the models differ more.
The spreading width gives a $\Delta$ part larger by factors $\RBVDLumDFiveMid$ and $\RBVDLumDNineMid$ at five and nine seconds, and it raises the $\Delta$ enhancement of the Hyper-Kamiokande count from $\RBHKrefMid$ to $\RBHKvdMid$.
We use the parametrization of Ref.~\cite{Oset:1987re}, the choice of the GiBUU transport code~\cite{GiBUU}, for all other results, and quote the dependence on the collisional width separately from the bands of the matched constants and the truncation.

\section{Integration over the supernova}
\label{sec:integration}

\paragraph*{\textbf{Phase space.}}
We evaluate the emissivity of Eq.~\eqref{M-eq:emis} of the End  Matter in each profile cell, a radial grid point of a snapshot of the simulation with its temperature, densities, chemical potentials and mean fields, with the self-energies and occupations specified in Sec.~\ref{sec:medium-profiles}.
For each channel, we introduce the total kinetic energy $\mathcal E$ and the total momentum $\bm P$ of the incoming nucleon pair, its invariant mass squared $s_i$, the cosine $\zeta$ of the angle between $\bm P$ and the axion momentum, and the invariant mass squared $s_f$ of the outgoing pair,
\begin{equation}
\begin{aligned}
 \mathcal E&=\epsilon_1+\epsilon_2,
 &\bm P&=\bm p_1+\bm p_2,
 &s_i&=\mathcal E^2-P^2,\\
 \zeta&=\widehat{\bm P}\cdot\widehat{\bm p_a},
 &s_f&=s_i-2E_a(\mathcal E-P\zeta),
 &P&=|\bm P|.
\end{aligned}
\label{eq:sn-pair-invariants}
\end{equation}
Here $p_a=(E_a,\bm p_a)$ and $|\bm p_a|=E_a$.
The sum of the external vector self-energies $\Sigma_V^N$ (see Sec.~\ref{sec:medium-profiles}) is the same before and after $nn$, $pp$ or $np$ Bremsstrahlung, so the outgoing kinetic pair four-momentum is $(\mathcal E-E_a,\bm P-\bm p_a)$.

The incoming and outgoing pairs have different rest frames.
In either frame the spatial delta function fixes the two momenta to be opposite, while the energy delta function fixes their magnitude,\footnote{Alternatively, the delta functions can be resolved in the fluid frame, with the total momentum $\bm P$ and the relative momenta $\bm p=(\bm p_1-\bm p_2)/2$ and $\bm p\,'=(\bm p_3-\bm p_4)/2$ of the two pairs.
The momentum delta function then fixes the total momentum $\bm P-\bm p_a$ of the outgoing pair, and the energy delta function fixes $|\bm p\,'|$ for each direction of $\bm p\,'$, with a Jacobian that depends on the angle between $\bm p\,'$ and $\bm P-\bm p_a$.
In the rest frames of the pairs this angle does not enter.}
\begin{equation}
\begin{aligned}
 q_i&=\frac{\sqrt{\lambda(s_i,m_1^{*2},m_2^{*2})}}{2\sqrt{s_i}}, &q_f&=\frac{\sqrt{\lambda(s_f,m_3^{*2},m_4^{*2})}}{2\sqrt{s_f}},& \dd\Phi_2(s;m_1^*,m_2^*)&=
 \frac{q(s;m_1^*,m_2^*)}{16\pi^2\sqrt{s}}\,\dd\Omega^*,
\end{aligned}
\label{eq:sn-two-body-measure}
\end{equation}
with $\lambda(x,y,z)=x^2+y^2+z^2-2xy-2xz-2yz$.
The angles $\Omega_i^*,\Omega_f^*$ specify the relative directions in these two rest frames.
Factoring the four-nucleon measure into these two-body measures and integrating the three common rotations gives $8\pi^2$ in an isotropic cell.
Equation~\eqref{M-eq:emis} then becomes
\begin{equation}
 \dot\varepsilon_a
 =\frac{1}{2^{13}\pi^9}
 \sum_{\rm ch}S_{\rm ch}
 \int P^2\dd P\,\dd\mathcal E\,E_a^2\dd E_a\,\dd\zeta\,
\dd\Omega_i^*\dd\Omega_f^*\,\frac{q_iq_f}{\sqrt{s_i s_f}}
 f_1f_2(1-f_3)(1-f_4)
 \sum_{\rm spins}|\mathcal M|^2 .
\label{eq:sn-reduced-emissivity}
\end{equation}
The symmetry factor $S_{\rm ch}$ and external-state normalization are those of the End Matter; the spin sum is not an initial-spin average.
The integration limits are
\begin{equation}
P\geq0\,,\quad\mathcal E\geq\sqrt{P^2+(m_1^*+m_2^*)^2}\,,\quad  -1\leq\zeta\leq1\,,\quad 0\leq E_a\leq \frac{s_i-(m_3^*+m_4^*)^2}{2(\mathcal E-P\zeta)} .
\label{eq:sn-phase-space-limits}
\end{equation}
Each pair is boosted back to the fluid rest frame before evaluating the occupations, canonical momentum transfers and amplitude.

We compute the eight-dimensional integral of Eq.~\eqref{eq:sn-reduced-emissivity} by quasi-Monte Carlo~\cite{LEcuyer:2002}.

\paragraph*{\textbf{Luminosity and spectra.}}
The luminosity seen at infinity is 
\begin{equation}
 L_a(t_{\rm pb})
 =4\pi\int
 \dd r\,r^2
   \alpha(r,t_{\rm pb})^2\left(1+2v_r(r,t_{\rm pb})\right)
   \dot\varepsilon_a(r,t_{\rm pb}) ,
\label{eq:sn-luminosity}
\end{equation}
The two powers of the lapse function $\alpha(r, t_{\rm pb})$ account for the redshift of the axion energy and for the conversion of the local emission time to the time at infinity, in the effective general-relativistic treatment of the simulations~\cite{Marek:2005if,Buras:2005rp,Rampp:2002bq}.
The simulations provide the thermodynamic state in the rest frame of the fluid~\cite{Bollig:2020xdr,Garching}, in which we evaluate the emissivity.
The factor $1+2v_r$ transforms the outward flux of the free-streaming axions from this frame to the frame of the star, to first order in the radial velocity $v_r$.
Both $\alpha$ and $v_r$ are taken from the tables of the simulations.

Figure~\ref{fig:sn-emission}(a) shows the radial emissivity and panel (b) its integral as a function of post-bounce time, separating the nucleon, $\Delta$ and interference contributions.

For the axion energy spectrum and number spectrum, we use the gravitational mapping $E_\infty=\alpha E_a$, keeping the same bulk-velocity factor as in Eq.~\eqref{eq:sn-luminosity}.
Its Jacobian removes one lapse factor:
\begin{equation}
\begin{aligned}
 \frac{\dd L_a}{\dd E_\infty}(t_{\rm pb})
 &=4\pi\int \dd r\,r^2\alpha(1+2v_r)
 \left.\frac{\dd\dot\varepsilon_a}{\dd E_a}\right|_{E_a=E_\infty/\alpha},\\
 \frac{\dd N_a}{\dd E_\infty}
 &=\int_0^{9\,\mathrm{s}}\dd t_{\rm pb}\,
        \frac{1}{E_\infty}\frac{\dd L_a}{\dd E_\infty}(t_{\rm pb}) .
\end{aligned}
\label{eq:sn-observer-spectra}
\end{equation}
Thus the observed number spectrum follows by dividing the energy spectrum by $E_\infty$ and integrating over the simulated times.
In Figure~\ref{M-fig:spectrum} of the main text, we show the source differential emissivity $E_a\,\dd N_a/\dd E_a$, without applying the redshift ($E_a$ is the local energy and not the energy at infinity).
All these rates scale as $f_a^{-2}$.
For the cooling criterion we compare $L_a(1\,\mathrm{s})$ with $L_\nu(1 \mathrm{s})$ for the same SN simulation. 
For SFHo-18.6, we have $L_\nu=5.5\times10^{52}\,\mathrm{erg\,s^{-1}}$ and $f_a^2L_a(1\,\mathrm{s})=\TSLumT\,{\rm erg\,s^{-1}MeV^2}$ (Table~\ref{tab:sn-model-results}), the criterion gives for the KSVZ axion $f_a\gtrsim\TSFaT$, or $m_a\lesssim\TSMaT$, compared with $f_a\gtrsim\TSFaN$ from the nucleon part alone.

\paragraph*{\textbf{Short-range nucleon interaction.}} 

Our amplitudes treat the $NN$ interaction through one-pion exchange in the Born approximation.
This introduces a potentially substantial uncertainty in the absolute emission rate: the soft-radiation analysis of Ref.~\cite{Hanhart:2000ae}, using measured vacuum $NN$ scattering amplitudes, gives rates roughly four times below the one-pion-exchange estimate in neutron matter.
This benchmark does not determine the correction under supernova conditions.
Even when combined with thermal occupations and Pauli blocking, vacuum amplitudes omit the modification of the interaction itself by the surrounding hot, dense nuclear medium.
We are leaving a dedicated systematic analysis of these effects for future work~\cite{Arcangeli:2026nuc}.

For the gamma-ray forecasts from axion conversion in the progenitor magnetic field, which constitute one of the applications of our results presented in the main text and discussed in detail in Sec.~\ref{sec:gamma}, we calibrate the nucleon spectrum to the phenomenological prescription $S_{\rm cal}$ adopted in Refs.~\cite{Fiorillo:2025gnd,Candon:2025sdm}.
We treat this choice as a common reference normalization against which to assess the impact of the $\Delta$, while retaining nuclear-model uncertainties.
Whether the corrections contained in this prescription reduce the $\Delta$ part in the same way is not known.
We therefore bracket it, with the $\Delta$ part reduced like the nucleon part or not reduced (Sec.~\ref{sec:gamma}).

\section{Uncertainty bands}
\label{sec:uncertainty}
We follow the prescription of Refs.~\cite{Springmann:2024mjp,Springmann:2024ret} where the dark band accouts for the the uncertainties of the constants in the chiral Lagrangian, whereas the light band include additionally the error from  truncating the chiral expansion at a given order.

\paragraph*{\textbf{Constants.}}
We draw the measured inputs from their distributions.
We assume all inputs to be Gaussian and uncorrelated. 
We draw $\TSDraws$ times and rematch the residual contact and recompute all rates on the same phase-space points; the band is the 16th--84th percentile relative to the central inputs, and ratios and fractions are formed draw by draw.
It is dominated by $z$, which enters the $\Delta$ part through $\cud$ and the nucleon part through $C_p$.
The self-energies, the collisional width and the profile are not varied.

\paragraph*{\textbf{Truncation.}}
The first omitted order is estimated as the last computed one, suppressed by one further power of the expansion parameter $Q$ of Eq.~(\ref{M-eq:Qtrunc})~\cite{Springmann:2024mjp}.
At LO the last computed order is the LO amplitude itself, so the omitted NLO changes the amplitude by a factor $1\pm Q$ and the emissivity by $(1\pm Q)^2$.
At NLO the omitted NNLO is $Q$ times the NLO correction, the difference of the emissivities at NLO and at LO, but not less than $Q^2$ times the LO value,
\begin{equation}
 \delta\Big(\frac{\dd\dot\varepsilon_a}{\dd E_a}\Big)=Q\max\bigg(\bigg|\frac{\dd\dot\varepsilon_a^{\rm NLO}}{\dd E_a}
 -\frac{\dd\dot\varepsilon_a^{\rm LO}}{\dd E_a}\bigg|,\ Q\,\bigg|\frac{\dd\dot\varepsilon_a^{\rm LO}}{\dd E_a}\bigg|\bigg) ,
 \label{eq:trunc}
\end{equation}
for the emissivity per unit axion energy of each part (nucleon, $\Delta$, interference, total) in each cell, with the local axion energy in $Q$ and, in its recoil term, the larger of the Fermi momentum of the cell (of protons or neutrons) and $\sqrt{M_NE_a}$.
The axion takes its energy from the kinetic energy of the two colliding nucleons: in their center-of-mass frame, where they have momenta $\pm\vp$ before the collision, $\vp^{\,2}/M_N\ge E_a$, and in the rest frame of the medium at least one of them moves at least as fast.
Above $E_a\simeq k_f^2/M_N$ the colliding nucleons therefore lie above the Fermi surface, and $|\vp|$, not $k_f$, sets the size of the recoil corrections: $|\vp|/M_N\ge0.56$ at $E_a=300\MeV$ and $0.69$ at $450\MeV$.
We integrate it like the emissivity itself, over the star for the luminosity and over the cells that contribute at $E_a=E_\infty/\alpha$ for the stellar spectrum.
Since Eq.~(\ref{eq:trunc}) uses the size of the correction, it applies also to the interference, which changes sign.
The second argument matters only where the NLO correction is smaller than $Q$ times the LO value.
Without it the estimate would vanish where the correction changes sign, although the omitted order need not vanish there.
In SFHo-18.6 at one second the emission peaks at $E_a\simeq165\MeV$, and the error is dominated by the recoil term, leading to a total uncertainty of around $30\%$.
The light band adds the truncation and the band of the constants in quadrature on each side.

\paragraph*{\textbf{Fitted quantities.}}
The temperatures and pion shares of Sec.~\ref{sec:inference} are fitted to the spectrum and do not depend linearly on it.
A pion share cannot be negative, for instance, while its shifts from the constants and the truncation, added in quadrature, can exceed the share itself.
For these quantities we fit samples of the spectrum instead.
Each sample is the central spectrum times the ratio of the spectra with one draw and with the central constants, both computed on the same phase-space points.
To it we add the truncation of the fitted spectrum, Eq.~\eqref{eq:trunc}, times a number from a standard Gaussian, one per sample and the same at all energies and times.
The Gaussian is the assumption behind adding the truncation in quadrature, and spectra that it would make negative are set to zero.
The dark band is the 16th--84th percentile of the fits without the truncation, the light band that of the fits with it.

\section{Results}
\label{sec:results}

\begin{figure}[tbp]
\centering
\includegraphics[width=.95\textwidth]{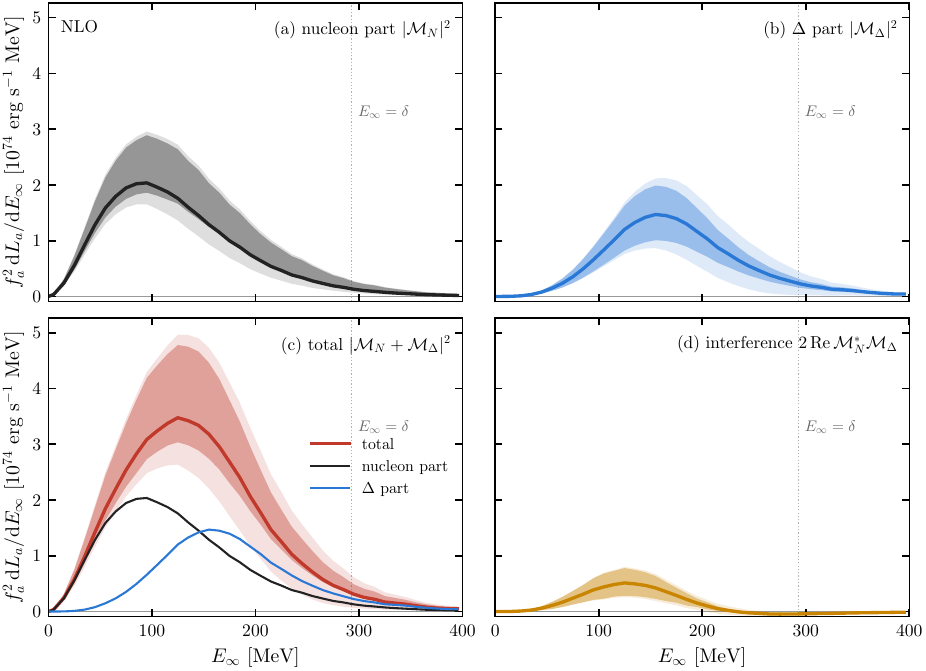}
\caption{Stellar axion spectrum $f_a^2\,\dd L_a/\dd E_\infty$ at NLO in SFHo-18.6 at $t_{\rm pb}=1\,$s.
(a) The nucleon part (nucleon diagrams and contact terms), (b) the $\Delta$ part, (c) the total, with the nucleon part (black) and the $\Delta$ part (blue) for comparison, and (d) the interference.
Dark bands show the uncertainty from the Lagrangian parameters (16--84\%), light bands that from the parameters and the truncation in quadrature (Sec.~\ref{sec:uncertainty}).
The dotted line marks $E_\infty=\delta$.}
\label{fig:spectrum1s}
\end{figure}

\begin{figure}[tbp]
 \centering
 \includegraphics[width=\linewidth]{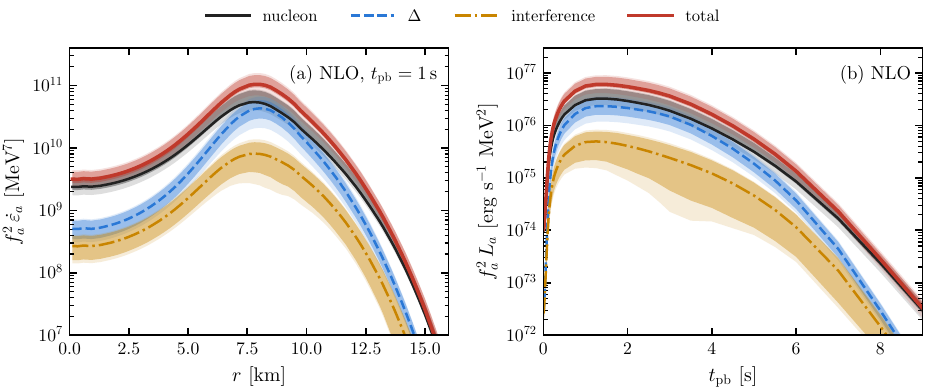}
 \caption{Emission at NLO in SFHo-18.6.
 (a) Local emissivity against radius at one second after bounce and (b) luminosity at infinity against time, for the nucleon part (solid black), the $\Delta$ part (dashed blue), their interference (dash-dotted amber) and the total (solid red).
 Dark bands show the uncertainty from the Lagrangian parameters (16--84\%), light bands that from the parameters and the truncation in quadrature.}
 \label{fig:sn-emission}
\end{figure}

\paragraph*{\textbf{NLO spectra and luminosity.}}
Figures~\ref{fig:spectrum1s} and \ref{fig:sn-emission} summarize the NLO results for SFHo-18.6.
Figure~\ref{fig:spectrum1s} shows the differential luminosity $\dd L_a/\dd E_\infty$ at $t_{\rm pb}=1\,\mathrm{s}$, while Fig.~\ref{fig:sn-emission} shows the radial emissivity at the same time and the time evolution of the integrated luminosity.
The $\Delta$ contribution peaks at $E_\infty\simeq\TSSpecPeakD\,\mathrm{MeV}$, farther below $\delta$ than in the estimate of Fig.~\ref{M-fig:analytic} of the main text, which peaks at $E_a=\PRLPeakDFortyFive\MeV$ for the temperature $T\simeq\EDHotT\MeV$ of the hottest shell.
Apart from the suppression of the high-energy side of the resonant enhancement by the Fermi--Dirac distributions, which the estimate contains, the full computation includes \textit{(i)} the kinematics of moving nucleons and $\Delta$ in the propagators and the phase space, with the in-medium masses and potentials and the Pauli-blocked mesonic width; \textit{(ii)} the collisional width; \textit{(iii)} the recoil of the NLO vertices; \textit{(iv)} the colder shells of the star; and \textit{(v)} the gravitational redshift, $E_\infty=\alpha E_a$ with $\alpha\simeq\TSPeakLapse$ at the emission peak.
Including the redshift has the largest impact, followed by the NLO recoil.

At one second, both the nucleon and $\Delta$ emissivities peak in the hot shell at $r\simeq8\,\mathrm{km}$.
Their luminosities reach a broad maximum around $t_{\rm pb}\sim1\,\mathrm{s}$ and subsequently decrease as the proto-neutron star cools.

\begin{figure}[t]
\centering
\includegraphics[width=.95\textwidth]{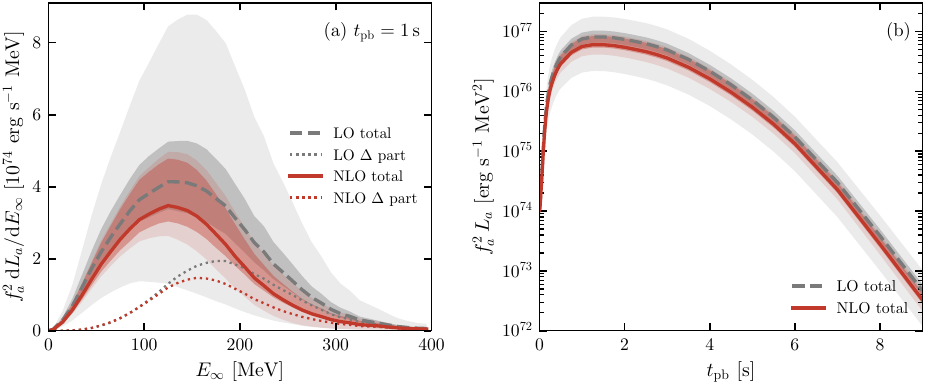}
\caption{LO (gray, dashed) and NLO (red, solid) in SFHo-18.6, with their bands.
(a) Stellar spectrum at $t_{\rm pb}=1\,$s, where the dotted lines are the $\Delta$ parts, and (b) luminosity against time.}
\label{fig:orders}
\end{figure}

\paragraph*{\textbf{Comparison with LO.}}
Figure~\ref{fig:orders} compares the differential and integrated luminosity at LO and NLO.
As discussed in the previous sections, both calculations use the same resummed propagators, in-medium widths, external-state normalization and phase space, and differ only in the inclusion of the NLO interaction vertices and contact terms.
Here and in the figures and tables, the NLO luminosity is the luminosity computed with the amplitude through NLO, which contains the LO amplitude, and the NLO correction is its difference from the LO luminosity.
As expected from power counting, the NLO central predictions remain within the LO chiral-truncation band over the ranges shown, with a reduced truncation uncertainty.
The sign of the NLO correction depends on the new LECs and on their interference with the LO amplitude.
For SFHo-18.6, the NLO luminosity at one second is approximately $25\%$ below LO, and the $\Delta$ contribution decreases by about $\fpeval{round(100*(1-\TSOrderRatioD))}\%$.
The WT term and the other contact terms, of one and of two nucleons, change the luminosity of the nucleon-exchange diagrams by $\TSContactConn\%$ and $\TSContactRes\%$, together $\TSContactAll\%$, indicating that vertex recoil drives the main reduction.
Importantly, the additional hard component from the $\Delta$ persists at both orders.
\paragraph*{\textbf{Recoil as a Doppler shift.}}\label{sec:doppler}
The NLO vertices of Eqs.~\eqref{eq:hb-nn-recoil-rule} and \eqref{eq:hb-delta-recoil-rule} have a simple kinematic meaning.
A baryon moving with velocity $\bm\beta$ sees the axion with the momentum of its own rest frame, which the inverse boost of $(E_a,\bm p_a)$ gives as
\begin{equation}
 \bm p_a^*=\bm p_a-E_a\bm\beta+\mathcal O(\beta^2E_a).
 \label{eq:doppler-boost}
\end{equation}
On a nucleon line the mean velocity at the vertex is $\bm\beta=(\bm p+\bm p')/2M_N$.
For the $\Delta$ it is $\bm\beta_\Delta=\bm l/\epsilon_\Delta$ with $\epsilon_\Delta=(m_\Delta^{*2}+\bm l^2)^{1/2}$, which the expansion replaces by $\bm l/M_N$ with its single heavy mass.
To first order in the velocity the NLO vertices are therefore the LO vertices evaluated with the rest-frame momentum of the axion, while its energy in the medium frame is Doppler shifted.

\begin{figure}[t]
\centering
\begin{minipage}[t]{0.48\textwidth}
\centering
\includegraphics[scale=1.12]{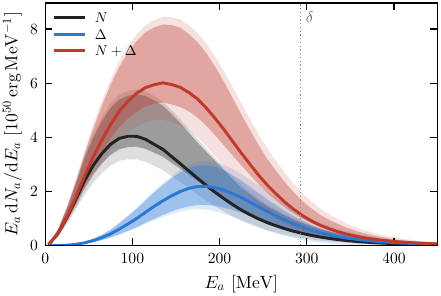}
\caption{As Fig.~\ref{M-fig:spectrum} of the main text, for SFHo-18.8.}
\label{fig:spectrum-s188}
\end{minipage}\hfill
\begin{minipage}[t]{0.48\textwidth}
\centering
\includegraphics[scale=1.12]{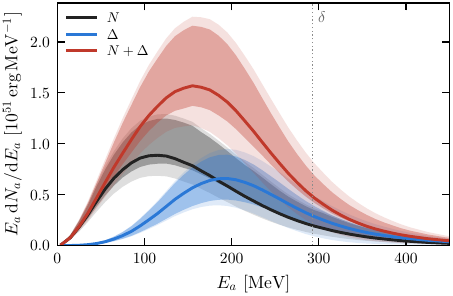}
\caption{As Fig.~\ref{M-fig:spectrum} of the main text, for SFHo-18.6.}
\label{fig:spectrum-s186}
\end{minipage}
\end{figure}

\paragraph*{\textbf{Dependence on the supernova model.}}
Table~\ref{tab:sn-model-results} compares the results of the numerics for the three reference supernova profiles considered.
Figures~\ref{fig:spectrum-s188} and~\ref{fig:spectrum-s186} show the spectrum at the source for SFHo-18.8 and SFHo-18.6, as Fig.~\ref{M-fig:spectrum} of the main text does for SFHo-20.0.
A hotter core increases the population of energetic nucleon pairs and hence the relative importance of the $\Delta$ channel.
In SFHo-20.0 at $1\,$s the $\Delta$ part carries $\TSHotFracDUnc\%$ of $L_a$, the nucleon part $\TSHotFracNUnc\%$ and their interference $\TSHotFracIUnc\%$, where the ranges come from the constants.
For the central values the $\Delta$ part thus exceeds the nucleon part.
Although the absolute luminosity varies by about an order of magnitude between the coldest and hottest models, the main conclusion is stable.
Including the $\Delta$ approximately doubles the luminosity and increases the number of axions above $E_\infty=200\,\mathrm{MeV}$ by a factor of about $2.4$.
The $\Delta$ therefore produces a substantial hardening of the spectrum across all three profiles.

\begin{table}[tbp]
\centering
\small
\setlength{\tabcolsep}{4pt}
\renewcommand{\arraystretch}{1.15}
\begin{tabular}{@{}lccc@{}}
\hline\hline
 & SFHo-18.8 & SFHo-18.6 & SFHo-20.0\\
\hline
$T(r_{\rm peak},1\,\mathrm{s})$ [MeV]
 & $\TSColdPeakT$ & $\TSPeakT$ & $\TSHotPeakT$\\
$f_a^2 L_a(1\,\mathrm{s})$
 & $\TSColdLumT$
 & $\TSLumT$
 & $\TSHotLumT$\\
$\Delta$ share at $1\,\mathrm{s}$ [\%]
 & $\TSColdFracD\ (\TSLOColdFracD)$ & $\TSFracD\ (\TSLOFracD)$ & $\TSHotFracD\ (\TSLOHotFracD)$\\
$L_a/L_N$ at $1\,\mathrm{s}$
 & $\TSColdRatioTN$
 & $\TSRatioTN$
 & $\TSHotRatioTN$\\
$L_a^{\rm NLO}/L_a^{\rm LO}$ at $1\,\mathrm{s}$
 & $\TSColdOrderRatioT$
 & $\TSOrderRatioT$
 & $\TSHotOrderRatioT$\\
$f_a^2 N_a$
 & $\TSColdNtotT$
 & $\TSNtotT$
 & $\TSHotNtotT$\\
$\langle E_\infty\rangle_{\rm total}/
 \langle E_\infty\rangle_N$ [MeV]
 & $\TSColdFluMeanET/\TSColdFluMeanEN$ & $\TSFluMeanET/\TSFluMeanEN$ & $\TSHotFluMeanET/\TSHotFluMeanEN$\\
$N_a/N_{a,N}$ above $200\,\mathrm{MeV}$
 & $\TSColdAboveTwoHundredRatio$
 & $\TSAboveTwoHundredRatio$
 & $\TSHotAboveTwoHundredRatio$\\
\hline\hline
\end{tabular}
\caption{Comparison between the three Garching models at NLO (LO in brackets where given): temperature at the emission peak at $1\,$s, luminosity at $1\,$s (constants and truncation), share of the $\Delta$ part, ratio of the total to the nucleon part, ratio of the NLO to the LO luminosity, and, over $0$--$9\,$s, the number of axions, the mean energy of the total and of the nucleon part, and the gain of the total over the nucleon part above $200\MeV$.
Luminosities in ${\rm erg\,s^{-1}MeV^2}$, $f_a^2N_a$ in ${\rm MeV^2}$.
Some bands lie entirely on one side of the central value (Sec.~\ref{sec:uncertainty}).}
\label{tab:sn-model-results}
\end{table}

\section{Gamma rays from stripped-envelope supernovae}
\label{sec:gamma}

\paragraph*{\textbf{Axion--photon conversion.}}
Axions produced in the supernova core can escape and convert into photons in the magnetic field of the progenitor, provided this field survives during the axion burst.
We consider conversion outside the photosphere, where photon absorption and plasma refraction can be neglected.
The interaction
\begin{equation}
 \mathcal L_{a\gamma}
 =-\frac{g_{a\gamma}}4 aF_{\mu\nu}\widetilde F^{\mu\nu}
 =g_{a\gamma}a\,\bm E\cdot\bm B
\end{equation}
mixes the axion with the photon polarization whose electric field is parallel to the transverse external magnetic field $B_T$.
For an axion wavelength short to the length scale of a coherent magnetic field, this coherent magnetic conversion is described by a two-state Schr\"odinger equation~\cite{Raffelt:1987im},
\begin{equation}
 i\frac{\dd}{\dd R}
 \begin{pmatrix}A_\parallel\\a\end{pmatrix}
 =-
 \begin{pmatrix}
  \Delta_\gamma&\Delta_{a\gamma}\\
  \Delta_{a\gamma}&\Delta_a
 \end{pmatrix}
 \begin{pmatrix}A_\parallel\\a\end{pmatrix},
\end{equation}
where
\begin{equation}
\begin{aligned}
 \Delta_{a\gamma}&=\frac{g_{a\gamma}B_T(R)}{2},
 &\Delta_a&=-\frac{m_a^2}{2E},&\Delta_\gamma&=-\frac{\omega_{\rm pl}^2}{2E}+\Delta_B,
 &\Delta_B&=\frac{7\alpha_{\rm em}}{90\pi}
             \left(\frac{B_T(R)}{B_{\rm crit}}\right)^2E,
 \qquad B_{\rm crit}=\frac{m_e^2}{e}.
\end{aligned}
\end{equation}
$m_a$ is the axion mass, $E$ is its energy, $\omega_{\rm pl}$ the plasma frequency, $\alpha_{\rm em}$ the fine structure constant; $m_e$, $e$ are the electron mass and charge.
The off-diagonal term produces conversion, while the diagonal difference determines the relative phase accumulated by axions and photons.
Besides their mass difference, this phase includes QED vacuum refraction: virtual electron--positron pairs make the magnetized vacuum behave as a medium with a polarization-dependent refractive index.
This gives the term $\Delta_B$.

We assume an equatorial trajectory through a dipole field, $B_T(R)=B_0(R_0/R)^3$, beginning at the photospheric radius $R_0$.
Since the conversion probability is small, we retain the photon amplitude to first order in $g_{a\gamma}$, neglecting axion depletion and reconversion.
Up to an irrelevant overall phase,
\begin{equation}
\begin{aligned}
 P_{a\gamma}(E)&=|\mathcal A_{a\gamma}(E)|^2\,,\quad \mathcal A_{a\gamma}(E)
 =\frac{g_{a\gamma}B_0R_0}{2}
   \int_1^\infty\frac{\dd x}{x^3}
   \exp\left[-i\chi_a x+\frac{i\chi_B}{5x^5}\right],\\
 \chi_a&=\frac{m_a^2R_0}{2E}\,,\quad \chi_B=\frac{7\alpha_{\rm em}}{90\pi}\left(\frac{B_0}{B_{\rm crit}}\right)^2ER_0.
\end{aligned}
\label{eq:gamma-conversion-probability}
\end{equation}

\begin{figure}[t]
\centering
\includegraphics[width=0.5\textwidth]{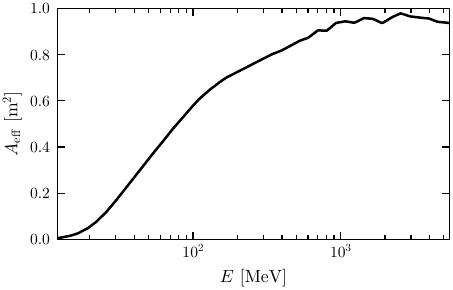}
\caption{On-axis effective area of the \emph{Fermi}-LAT for the transient event class, as used in Ref.~\cite{Candon:2025sdm}, against the photon energy.}
\label{fig:aeff}
\end{figure}

\paragraph*{\textbf{From the emitted spectrum to detector counts.}}
Let $\dd N_a/\dd E$ be the time-integrated axion number spectrum after gravitational redshift, with $E=E_\infty$.
The magnetic field of the progenitor is static, so the conversion conserves energy exactly and the photon carries the energy $E$ of the axion.
The field only takes up the difference of the momenta.
For a supernova at distance $d$, the expected number of detected photons is
\begin{equation}
\begin{aligned}
 N_\gamma
 &=\frac{1}{4\pi d^2}
   \int \dd E\,\frac{\dd N_a}{\dd E}P_{a\gamma}(E)A_{\rm eff}(E).
\end{aligned}
\label{eq:gamma-detector-counts}
\end{equation}
The effective area $A_{\rm eff}$, which includes the detector's collecting area, detection efficiency and event selection, rises strongly with energy over the relevant tens-to-hundreds of MeV range~\cite{Fermi-LAT:2009ihh}.
We take the on-axis effective area of the transient event class from Ref.~\cite{Candon:2025sdm}, Fig.~\ref{fig:aeff}.
Between $100\MeV$ and $1\GeV$ it grows from $0.6$ to $0.9\,\mathrm{m^2}$, so that a harder spectrum gives more photons per axion even where the conversion probability does not depend on the energy.

Following Ref.~\cite{Candon:2025sdm}, we adopt the detection criterion $N_\gamma\geq3$ during the short burst and use $d=2\,\mathrm{kpc}$ in the sensitivity plots.
The same calculation determines the maximum detectable distance through $N_\gamma\propto d^{-2}$.

\paragraph*{\textbf{Coherence, mass dependence and progenitor classes.}}
If $\chi_a\ll1$ and $\chi_B\ll1$, conversion amplitudes accumulated at different radii add coherently,
\begin{equation}
 P_{a\gamma}^{\rm coh}
 \simeq \frac{g_{a\gamma}^2B_0^2R_0^2}{16}.
\label{eq:gamma-coherent-limit}
\end{equation}
The probability is independent of the axion mass and energy, giving the low-mass plateaus in Fig.~\ref{M-fig:detection}(a).
    
When $\chi_a\gg\max(1,\chi_B)$, the relative axion--photon phase instead changes rapidly across the magnetic region.
In this case, the conversion probability grows with energy,
\begin{equation}
 P_{a\gamma}^{\rm mass}
 \simeq\frac{g_{a\gamma}^2B_0^2E^2}{m_a^4}.
\label{eq:gamma-mass-dominated-limit}
\end{equation}
Consequently, the threshold on $|g_{ap}g_{a\gamma}|$ grows as $m_a^2$.\footnote{We are using the KSVZ relation between the axion decay constant and the couplings with nucleons, $1/f_a = g_{ap}/(C_p m_p)$.}
It becomes parallel to the QCD axion lines in this plane, because both couplings scale as $f_a^{-1}$ and $m_a\propto f_a^{-1}$.

The progenitor bands span the assumed surface-field intervals: $B_0=1$--$100\,\mathrm G$ for blue and red supergiants, with $R_0=30R_\odot$ and $800R_\odot$, respectively, and $B_0=0.5$--$30\,\mathrm{kG}$ for Type Ibc progenitors with $R_0=R_\odot$.
Using \eqref{eq:gamma-coherent-limit} and \eqref{eq:gamma-mass-dominated-limit}, we then obtain the qualitative features of Fig.~\ref{M-fig:detection}(a): in the coherent limit, the number of received photons depends on $g_{a\gamma} B_0R_0$, placing the Type Ibc plateau between the supergiant plateaus for these choices.
In the mass-dominated limit, the radius drops out: the blue- and red-supergiant bands therefore overlap, while the stronger Type Ibc fields give a lower coupling threshold.

For the strongest Type Ibc fields, QED refraction modifies the low-mass plateau and the transition to the mass-dominated regime.
It suppresses conversion preferentially at high energy and produces the additional structure near the turnover that can be observed in the numerical result.

\paragraph*{\textbf{Implications of a harder spectrum.}}
As discussed and numerically verified in the previous sections, the $\Delta$ diagrams have two effects: they increase the integrated axion luminosity and shift the spectrum towards larger energies.
To understand analytically how spectral hardening affects the number of detected photons, we temporarily assume that the effect of the $\Delta$ is a rescaling of the axion number spectrum:
\begin{equation}
 \frac{\dd N_a^{(\lambda)}}{\dd E}
 =\frac1\lambda
   \left.\frac{\dd N_a}{\dd E'}\right|_{E'=E/\lambda},
 \qquad \lambda>1 .
\end{equation}
This preserves the total axion number and shifts spectral features to higher energies.
It enters equation~\eqref{eq:gamma-detector-counts} weighted by $A_{\rm eff}(E)$ in the coherent regime and by $E^2A_{\rm eff}(E)$ in the mass-dominated regime.
The $\Delta$ part raises the photon count in two ways.
It adds axions, and it hardens the spectrum, which in the mass-dominated regime raises the count per axion through the weight $E^2A_{\rm eff}(E)$.
For the hot one-zone source, the two together raise the count by a factor of about $\fpeval{round(\SEOZHotGainDelta,0)}$ if the $\Delta$ part is reduced like the nucleon part and of about $\fpeval{round(\SEOZHotGainDeltaFull,0)}$ if it is not, and the larger number of axions contributes somewhat more than the hardening.
In the coherent regime, only the effective area rewards the hardening, which raises the count per axion by about $\fpeval{round(100*(\SEOZHotLightShapeGainDelta-1),-1)}\%$ with the reduced and $\fpeval{round(100*(\SEOZHotLightShapeGainDeltaFull-1),-1)}\%$ with the unreduced $\Delta$ part.

\paragraph*{\textbf{Emission normalization and nuclear uncertainties.}}
We assess the $\Delta$ in two steps.
We first add our $\Delta$ part to the one-zone source of Ref.~\cite{Candon:2025sdm}, evaluated in the same cell, and then repeat the analysis on the full simulations, cell by cell and over time, which tests the one-zone description.
The source of Ref.~\cite{Candon:2025sdm} is a one-zone model of the core in which protons radiate through a parametric spin-fluctuation rate~\cite{Fiorillo:2025gnd},
\begin{equation}
 \frac{\dd\dot N_a}{\dd E_a\,\dd V}=\frac{\gap^2n_p}{8\pi^2M_p^2}\,\frac{E_a}{e^{E_a/T}+1}\,
 \frac{\Gamma_\sigma}{1+(\Gamma_\sigma/2E_a)^2},\qquad
 \Gamma_\sigma=40\MeV\,\frac{\rho}{4\times10^{14}\,{\rm g\,cm^{-3}}}\Big(\frac{T}{30\MeV}\Big)^{1/2},
 \label{eq:param}
\end{equation}
evaluated at rounded emission-weighted averages of $T$, $\rho$ and $Y_p$ of the Garching models SFHo-20.0 (hot) and SFHo-18.8 (cold).
As in Ref.~\cite{Fiorillo:2025gnd}, this rate per volume is multiplied by the volume of the emitting core and by the duration of the emission, and the axion energies are redshifted to the observer with an average lapse.
The factor $[1+(\Gamma_\sigma/2E_a)^2]^{-1}$ models the suppression from multiple scattering during emission,
while our computation models the NN interaction through one-pion-exchange in the Born approximation; see Sec.~\ref{sec:integration} for more details about the effects this approximation neglects.
A complete treatment would contain this suppression through the collisional width of the nucleons, as our rate contains the collisional width of the $\Delta$.
For the hard axions of the $\Delta$ part the factor is close to one.
In the hot cell, with $T=45\MeV$ and $\rho=6\times10^{14}\,{\rm g\,cm^{-3}}$, $\Gamma_\sigma\simeq74\MeV$, and the factor is $0.98$ at $E_a=\delta$.

\begin{figure}[t]
\centering
\includegraphics[width=\textwidth]{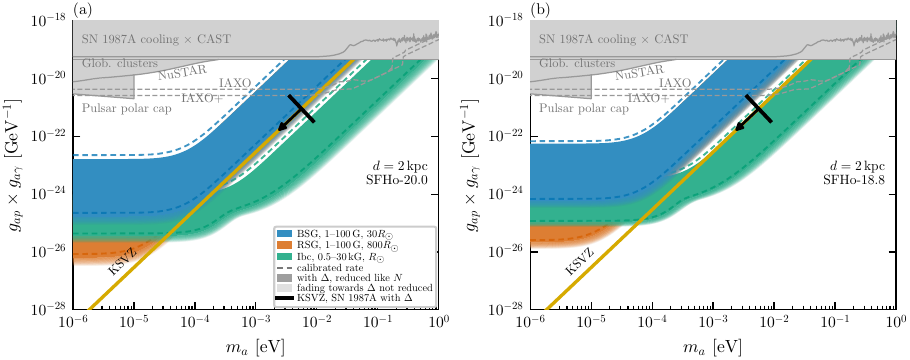}
\caption{As Fig.~\ref{M-fig:detection}(a), for the full simulations instead of the one-zone models, (a) SFHo-20.0 and (b) SFHo-18.8.
The dashed lines show the calibrated rate of Eq.~(\ref{eq:param}) on the cells and times of the simulation, for the weakest and the strongest field of each progenitor class.
The shaded regions add our NLO $\Delta$ contribution, combined cell by cell, reduced like the nucleon part and fading out towards an unreduced $\Delta$ contribution.}
\label{fig:smreach}
\end{figure}

To compare our result with this reference while isolating the impact of the $\Delta$, we calibrate our nucleon spectrum to the same prescription.
In each cell and at each local energy, our spectrum per volume is
\begin{equation}
 \mathcal S=\frac{\dd\dot N_a}{\dd E_a\,\dd V}=\mathcal S_N+\mathcal S_I+\mathcal S_\Delta ,
\end{equation}
with the nucleon part $\mathcal S_N$, the interference $\mathcal S_I$ and the $\Delta$ part $\mathcal S_\Delta$, and $\mathcal S_{\rm cal}$ is the spectrum of Eq.~\eqref{eq:param}.
We define
\begin{equation}
 r=\frac{\mathcal S_{\rm cal}}{\mathcal S_N},
 \qquad
 \mathcal M_N\longrightarrow\sqrt r\,\mathcal M_N .
\end{equation}
and multiply the nucleon component of the differential luminosity by $r$.
The corresponding correction to the $\Delta$ contribution is unknown.
The soft-radiation argument relating emission to elastic $NN$ scattering does not control the hard axions with $E_a\sim\delta$.
Nevertheless, short-range modifications of the tensor interaction can affect the exchanged pion common to the nucleon and $\Delta$ diagrams.
We therefore consider both equal reduction of the two amplitudes and an unchanged $\Delta$ amplitude:
\begin{equation}
\begin{aligned}
 \mathcal S_{\rm reduced}
 &=r(\mathcal S_N+\mathcal S_I+\mathcal S_\Delta),\\
 \mathcal S_{\rm unreduced}
 &=r\mathcal S_N+\sqrt r\,\mathcal S_I+\mathcal S_\Delta ,
\end{aligned}
\label{eq:gamma-calibration-prescriptions}
\end{equation}
These prescriptions define the additional shaded region in Fig.~\ref{M-fig:detection}(a) and Fig.~\ref{fig:smreach}, and illustrate the uncertain nuclear correction to the $\Delta$ contribution, which requires the in-medium nucleon interaction that we are developing~\cite{Arcangeli:2026nuc}.

\paragraph*{\textbf{Detection reach.}}
Figure~\ref{M-fig:detection}(a) compares our results with the hot one-zone reference, while Fig.~\ref{fig:smreach} shows the corresponding results for the SFHo-20.0 (hot) and SFHo-18.8 (cold) profiles.
For each progenitor class, the extremes of the corresponding band are obtained by using the appropriate maximum and minimum values of $B_0$, listed above.
The $\Delta$ contribution improves the reach through both a larger axion yield and a harder spectrum, with the latter particularly important in the mass-dominated regime.

On the KSVZ line, $|g_{ap}g_{a\gamma}|\propto m_a^2$ cancels the $m_a^{-4}$ dependence of the conversion probability in the photon count.
Detection is therefore characterized by a mass-independent threshold on $B_0/d$ within this regime: a smaller ratio implies better detection prospects, as photons from a more distant supernova could reach the detector.
The thresholds and their uncertainties are listed in Table~\ref{tab:gamma}.
For the hot one-zone source, including the equally reduced $\Delta$ contribution lowers the central threshold from $\SEBDHot$ to $\SEOZHotBDDelta\,\mathrm{G\,kpc^{-1}}$; for the cold source it decreases from $\SEBDCold$ to $\SEOZColdBDDelta\,\mathrm{G\,kpc^{-1}}$.
On the full hot and cold profiles, the corresponding changes are $\SEModHotBDParam\to\SEPFHotBDDelta$ and $\SEModColdBDParam\to\SEPFColdBDDelta\,\mathrm{G\,kpc^{-1}}$.
Keeping the $\Delta$ amplitude unreduced gives thresholds of $\SEPFHotBDDeltaFull$ and $\SEPFColdBDDeltaFull\,\mathrm{G\,kpc^{-1}}$ on those full profiles.

The thresholds on $B_0/d$ and their uncertainties are listed in Table~\ref{tab:gamma}, comparing three prescriptions for the emission spectrum.
First, we reproduce the one-zone sources of Refs.~\cite{Fiorillo:2025gnd,Candon:2025sdm}, by using the representative hot and cold one-zone parameters adopted therein, obtained through weighted averages of the numerical supernova profiles.

Second, we apply the local calibration $r=\mathcal S_{\rm cal}/\mathcal S_N$ to our nucleon spectrum.
This differs from the previous method in that we consider here the actual radial and temporal profiles, with the local temperature, density, proton fraction and lapse, rather than replacing them by representative averages.
Third, we use our microscopic NLO spectra on the full profiles without the phenomenological reduction.

For each of the first two prescriptions, we supplement the calibrated nucleon source with our $\Delta$ contribution, considering both an equally reduced $\Delta$ amplitude and an unreduced one, with the interference rescaled consistently.
The third prescription retains both microscopic amplitudes without additional reduction.

The comparison separates the effects of stellar averaging and the nuclear emission prescription.
Using the full profiles instead of the one-zone approximation changes the baseline nucleon detection thresholds by only about $7$--$10\%$.
In contrast, the uncalibrated OPE-based NLO nucleon spectrum lowers these thresholds by approximately a factor of three relative to the phenomenological prescription, illustrating the importance of the nuclear-interaction uncertainty.
Our numerical results show that the axion emission from the $\Delta$ diagrams increases the expected photon count and therefore lowers the required $B_0/d$, under both assumptions about its reduction.

\begin{table}[ht]
\begin{ruledtabular}
\begin{tabular}{llll}
source & hot (SFHo-20.0) & cold (SFHo-18.8) & SFHo-18.6\\
\hline
one-zone model~\cite{Candon:2025sdm} & $\SEBDHot$ & $\SEBDCold$ & ---\\
\quad with our $\Delta$ part, reduced & $\SEOZHotBDDeltaUnc$ & $\SEOZColdBDDeltaUnc$ & ---\\
\quad with our $\Delta$ part, not reduced & $\SEOZHotBDDeltaFullUnc$ & $\SEOZColdBDDeltaFullUnc$ & ---\\
parametric rate on the profiles & $\SEModHotBDParam$ & $\SEModColdBDParam$ & $\SEBDParam$\\
\quad with our $\Delta$ part, reduced & $\SEPFHotBDDeltaUnc$ & $\SEPFColdBDDeltaUnc$ & ---\\
\quad with our $\Delta$ part, not reduced & $\SEPFHotBDDeltaFullUnc$ & $\SEPFColdBDDeltaFullUnc$ & ---\\
NLO nucleon part & $\SEModHotBDNLONUnc$ & $\SEModColdBDNLONUnc$ & $\SEBDNLONUnc$\\
NLO total & $\SEModHotBDNLOTUnc$ & $\SEModColdBDNLOTUnc$ & $\SEBDNLOTUnc$\\
LO total & $\SEModHotBDLOTUnc$ & $\SEModColdBDLOTUnc$ & $\SEBDLOTUnc$\\
\end{tabular}
\end{ruledtabular}
\caption{Threshold on $B_0/d$, in G/kpc, for a detection of the KSVZ axion with \emph{Fermi}-LAT ($N_\gamma\ge3$, high-mass regime), with the uncertainty of the constants and the truncation.
The rows ``with our $\Delta$ part'' add our NLO $\Delta$ part to the calibrated nucleon rate of the row above, reduced like the nucleon part or not reduced, Eq.~\eqref{eq:gamma-calibration-prescriptions}.
The last three rows use our calculation alone, without the calibration, so that its nucleon part, one-pion exchange in the Born approximation, enters unreduced and is an upper estimate.
The LO row shows the size of the NLO correction.}
\label{tab:gamma}
\end{table}

\begin{figure}[t]
\centering
\includegraphics[width=0.5\textwidth]{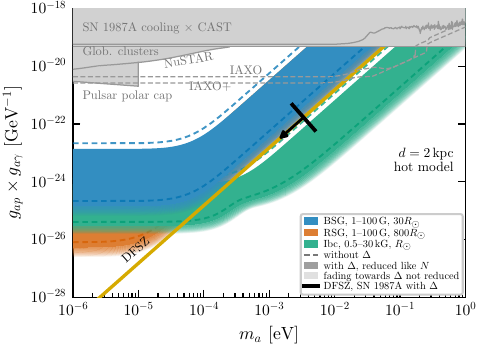}
\caption{As Fig.~\ref{M-fig:detection}(a), for the DFSZ axion with $E/N=8/3$ and $\tan\beta=\SEDFTanBeta$ (yellow line), with $\gap=|C_p|m_p/f_a$ of this model.
The dashed lines show the calibrated nucleon rate of the hot one-zone source with the DFSZ couplings, for the weakest and the strongest field of each progenitor class.
The shaded regions add our NLO $\Delta$ contribution with the DFSZ couplings in the same cell, reduced like the nucleon part and fading out towards an unreduced $\Delta$ contribution.
The black bar marks the SN~1987A bound on this axion with the $\Delta$, $m_a\lesssim\SEDFMaT$, with the arrow towards the allowed masses.}
\label{fig:dfszreach}
\end{figure}

\paragraph*{\textbf{DFSZ axion.}}
The size of the $\Delta$ part relative to the nucleon part depends on the axion model through $C_p$, $C_n$ and $\cud$ (Sec.~\ref{sec:fermigas}), so the reach of Fig.~\ref{M-fig:detection}(a) holds for the KSVZ axion only.
Figure~\ref{fig:dfszreach} shows the same analysis for the DFSZ axion~\cite{Dine:1981rt,Zhitnitsky:1980tq} with the ratio $E/N=8/3$ of the electromagnetic and colour anomalies and $\tan\beta=\SEDFTanBeta$.
At this value the ratio $\kappa_{\rm DFSZ}/\kappa_{\rm KSVZ}$ of Fig.~\ref{fig:dfsz} has reached its plateau.
With the constants of Table~\ref{M-tab:constants} and the running of Ref.~\cite{GrillidiCortona:2015jxo}, the couplings are $C_p=\SEDFCpUnc$, $C_n=\SEDFCnUnc$ and $\cud=\SEDFCvdUnc$, and the photon coupling is $\gag=\alpha_{\rm em}(E/N-1.92)/(2\pi f_a)$~\cite{GrillidiCortona:2015jxo}.
In the hot one-zone cell and per $\gap^2$, with $\gap=|C_p|m_p/f_a$ of each model, the $\Delta$ part of the DFSZ axion is $\SEDFRatioDelta$ times that of the KSVZ axion, and its nucleon part $\SEDFRatioN$ times, since the neutrons radiate as well.
The parametric rate of Eq.~\eqref{eq:param} contains only the proton coupling.
We therefore multiply it by the ratio of our DFSZ and KSVZ nucleon spectra in the same cell, which reduces the nucleon amplitudes of both axions by the same $\sqrt r$ at each energy, and combine the product with our DFSZ $\Delta$ part as in Eq.~\eqref{eq:gamma-calibration-prescriptions}.
In the high-mass regime the $\Delta$ part then raises the photon count by a factor $\SEDFGainDelta$ if it is reduced like the nucleon part and $\SEDFGainDeltaFull$ if it is not, compared with $\SEOZHotGainDelta$ and $\SEOZHotGainDeltaFull$ for the KSVZ axion.
On the DFSZ line the threshold on $B_0/d$ falls from $\SEDFBDParam$ to $\SEDFBDDeltaUnc\,\mathrm{G\,kpc^{-1}}$ with the $\Delta$ part reduced like the nucleon part, and to $\SEDFBDDeltaFullUnc\,\mathrm{G\,kpc^{-1}}$ with the $\Delta$ part not reduced.
These values lie above the KSVZ thresholds of Table~\ref{tab:gamma}, since the photon coupling of the DFSZ axion, proportional to $|E/N-1.92|=0.75$, is smaller than that of the KSVZ axion with $E/N=0$.
With the DFSZ couplings, the luminosity of SFHo-18.6 at one second is $\SEDFLumRatio$ times that of the KSVZ axion at equal $f_a$.
The cooling criterion of Sec.~\ref{sec:integration} then gives $f_a\gtrsim\SEDFFaT$, or $m_a\lesssim\SEDFMaT$, with the band of the constants and the truncation, the black bar in Fig.~\ref{fig:dfszreach}.

\section{Spectral inference from axion conversion in the Galactic magnetic field}
\label{sec:inference}

Light axion-like particles (ALPs) that escape the core of a Galactic supernova can convert into photons of the same energy in the magnetic field of the Galaxy.
The supernova then appears as a burst of gamma rays that lasts as long as the axion emission, several seconds, and whose spectrum follows the axion spectrum.
Ref.~\cite{Lella:2024hfk} studied what the Large Area Telescope (LAT) of the \emph{Fermi} satellite would learn from such a burst about the core of a supernova at $10\,\mathrm{kpc}$.
We first describe this analysis and, in the remaining paragraphs of the section, test it including the $\Delta$.

The analysis distinguishes two production processes with different spectra.
\textit{(i)} Nucleon Bremsstrahlung, $NN\to NNa$, which is quasi-thermal.
Its spectrum at Earth peaks near $40\,\mathrm{MeV}$ and falls exponentially at higher energies, and its mean energy grows with the temperature.
\textit{(ii)} Thermal pions, whose abundance in the core may be enhanced by strong interactions, relying on a large pion chemical potential,~\cite{Fore:2019wib}, convert into axions through $\pi^-p\to na$~\cite{Carenza:2020cis}.
In this process, the axion energy is at least the pion mass, such that this component begins near $100\,\mathrm{MeV}$ at Earth and extends to several hundred MeV.
Ref.~\cite{Lella:2024hfk} computes both spectra for the supernova model SFHo-18.8 and describes the Bremsstrahlung spectrum by a Gamma distribution in the energy $E$ at Earth,
\begin{equation}
 \frac{\dd^2N_{a,NN}}{\dd t_{\rm pb}\,\dd E}
 =A_{NN}\left(\frac{E}{E^0_{NN}}\right)^{\beta_{NN}}
 e^{-(\beta_{NN}+1)E/E^0_{NN}},
 \label{eq:sm_NN}
\end{equation}
with normalization $A_{NN}$, mean energy $E^0_{NN}$ and shape parameter $\beta_{NN}$. 
A thermal spectrum of massless particles has $\beta_{NN}=2$, and the Bremsstrahlung spectrum of SFHo-18.8 has $\beta_{NN}$ between $1.2$ and $1.5$.
Pion conversion is described by a Gamma distribution shifted by the threshold $\omega_c$,
\begin{equation}
 \frac{\dd^2N_{a,\pi}}{\dd t_{\rm pb}\,\dd E}
 =A_\pi\left(\frac{E-\omega_c}{E^0_\pi}\right)^{\beta_\pi}
 e^{-(\beta_\pi+1)(E-\omega_c)/E^0_\pi}
 \Theta(E-\omega_c),
 \label{eq:sm_pion}
\end{equation}
with the corresponding parameters $A_\pi$, $E^0_\pi$ and $\beta_\pi$.
The threshold is the pion mass redshifted to Earth, about $100$--$108\,\mathrm{MeV}$ for SFHo-18.8.
App.~A of Ref.~\cite{Lella:2024hfk} lists all parameters from $1$ to $8\,\mathrm{s}$, fitted to its computed spectra, together with smooth functions of time that interpolate them.
We refer to the resulting spectra as the \textit{published} Bremsstrahlung and pion spectra.

The burst is analyzed in windows of one second.
In each window, the photons are counted in $20$ energy bins between $60$ and $600\,\mathrm{MeV}$.
The counts are fitted with Eq.~\eqref{eq:sm_NN}, either alone or together with Eq.~\eqref{eq:sm_pion}, converted into expected counts with the response of the LAT, with all parameters free.
Postponing most of the technical details in later paragraphs of the section, we review now the two conclusions that Ref.~\cite{Lella:2024hfk} derives from these fits.

First, the fitted mean energy of the Bremsstrahlung component gives the temperature of the emitting region.
Ref.~\cite{Lella:2024hfk} calibrates this relation with its own Bremsstrahlung rate on a one-zone model, a core of constant temperature $T$ out to a radius of $13\,\mathrm{km}$, with a density of $3\times10^{14}\,\mathrm{g\,cm^{-3}}$ and an electron fraction of $0.3$, which gives a neutron star of about $1.4\,M_\odot$, and the remaining thermodynamic quantities from the SFHo equation of state.
Fitting Eq.~\eqref{eq:sm_NN} to the Bremsstrahlung spectrum of this model at temperatures between $10$ and $55\,\mathrm{MeV}$ gives the local mean energy $E^0_{NN}=15.4\,\mathrm{MeV}+2.2\,T$.
At Earth all energies are lower by the lapse, the factor by which the gravitational redshift lowers the energy of a particle that escapes from the emitting region to infinity.
Inverting the relation gives the reconstructed temperature
\begin{equation}
 \frac{T_{\rm rec}}{\mathrm{MeV}}
 =-6.93+0.45\,\frac{E^0_{NN}/\mathrm{MeV}}{0.78},
 \label{eq:sm_thermometer}
\end{equation}
where $0.78$ is the lapse that Ref.~\cite{Lella:2024hfk} takes for the emitting region of SFHo-18.8.
Ref.~\cite{Lella:2024hfk} concludes that such $T_{\rm rec}$ faithfully reproduces the emission-weighted temperature $\langle T\rangle$ of the profile, defined as the volume average of the local temperature weighted with the number of axions emitted per unit volume and time.

Second, if the fit with pion conversion describes the counts much better than the fit without it, the burst reveals thermal pions in the core.
Ref.~\cite{Lella:2024hfk} found such evidence in the first second, where in its model photons above $200\,\mathrm{MeV}$ come almost only from pion conversion.

Both conclusions assume that Bremsstrahlung produces no hard photons.
The $\Delta$ diagrams of this work produce axions of $100$--$400\,\mathrm{MeV}$ in $NN\to NNa$ itself, without thermal pions.
A fit whose only description of Bremsstrahlung is the soft Gamma distribution must then either raise its mean energy, and with it the inferred temperature, or assign the hard photons to pion conversion.
We repeat the analysis on spectra that contain the $\Delta$ but no thermal pions and measure both effects.

The following paragraphs describe the photon counts, the fit models, the Bremsstrahlung rate of Ref.~\cite{Lella:2024hfk}, how we add the $\Delta$, and the results.
At the end we repeat the analysis with our own nucleon part and discuss the photon counts published in Ref.~\cite{Lella:2024hfk}.

\paragraph*{\textbf{From axion spectrum to photon counts.}}
The differential axion and photon number fluxes at Earth, $d\Phi_a/dE$ and $d\Phi_\gamma/dE$ respectively, can be computed from the luminosity spectrum at infinity (Sec.~\ref{sec:integration}, we denote $E_\infty=E$ for simplicity throughout) as follows:
\begin{equation}
 \frac{\dd\Phi_a}{\dd E}(t_{\rm pb})
 =\frac{1}{4\pi d^2E}\frac{\dd L_a}{\dd E}(t_{\rm pb}),
 \qquad
 \frac{\dd\Phi_\gamma}{\dd E}(t_{\rm pb})
 =P_{a\gamma}\frac{\dd\Phi_a}{\dd E}(t_{\rm pb}),
 \label{eq:sm_flux}
\end{equation}
where $t_{\rm pb}$ is the time after bounce, $d$ the distance and $P_{a\gamma}$ the probability that an axion converts into a photon on its way through the Galactic magnetic field.
As in Ref.~\cite{Lella:2024hfk} we consider ultralight ALPs, for which the axion and the photon stay in phase along the path, the coherent regime \eqref{eq:gamma-coherent-limit} of Sec.~\ref{sec:gamma}.
The probability then does not depend on the energy, and the photon spectrum has the shape of the axion spectrum.
For the supernova direction and the model of the Galactic field used in Ref.~\cite{Lella:2024hfk} it is
\begin{equation}
 P_{a\gamma}
 =0.785\times10^{-10}
 \left(\frac{g_{a\gamma}}
 {10^{-15}\,\mathrm{GeV}^{-1}}\right)^2.
 \label{eq:sm_conversion_probability}
\end{equation}
The expected number of photons in the time window $b$ and the bin $i$ of reconstructed energy is
\begin{equation}
 n_{bi}=\int_{t_b^-}^{t_b^+}\dd t_{\rm pb}\int \dd E\,
 R_i(E,t_{\rm pb})\frac{\dd\Phi_\gamma}{\dd E}(t_{\rm pb}),
 \label{eq:sm_counts}
\end{equation}
where the detector response $R_i$ is the effective area times the probability that a photon of energy $E$ is reconstructed in bin $i$~\cite{Fermi-LAT:2009ihh}.
For the central results, we fit $n_{bi}$ directly, without statistical fluctuations (the Asimov data set~\cite{Cowan:2010js}).
To simulate an observation, we draw independent Poisson counts with means $n_{bi}$.
Repeating the fits to these simulated observations gives the statistical uncertainties.

Table~\ref{tab:sm_setup} lists all the inputs used for the numerical evaluation.

\begin{table}[tb]
\centering
\small
\begin{tabular}{lp{0.72\linewidth}}
\hline
Input & Value \\
\hline
Couplings &
$g_{a\gamma}=3\times10^{-15}\,\mathrm{GeV}^{-1}$,
$|g_{ap}|=5\times10^{-10}$, $g_{an}/g_{ap}=C_n/C_p$ \\
Conversion &
$P_{a\gamma}=0.785\times10^{-10}
\left(\frac{g_{a\gamma}}
{10^{-15}\,\mathrm{GeV}^{-1}}\right)^2$ \\
Source &
SFHo-18.8, $d=10\,\mathrm{kpc}$,
$(l,b)=(199.79^\circ,-8.96^\circ)$ \\
Nucleon part &
Bremsstrahlung spectrum of Ref.~\cite{Lella:2024hfk}, App.~A ($g_{an}=0$, fitted for $1$--$8\,$s) \\
$\Delta$ part &
this work (NLO), reduced like the nucleon part or not reduced, Eq.~\eqref{eq:sm_baseline} \\
Detector &
\texttt{P8R3\_TRANSIENT020\_V3}, FRONT+BACK,
$60$--$600\,\mathrm{MeV}$, $20$ logarithmic energy bins \\
Event selection &
all photons of the source\\
Exposure start & MET $510160000\,\mathrm{s}$ \\
Observation interval &
$1$--$6\,\mathrm{s}$ after bounce, one-second slices and $0.5\,$s windows \\
\hline
\end{tabular}
\caption{Benchmark inputs for the photon counts.
The conversion normalization, source distance and Galactic coordinates $(l,b)$, and LAT event class follow Ref.~\cite{Lella:2024hfk}.
$C_n/C_p$ is that of the KSVZ axion (Table~\ref{tab:matched-constants}), which differs negligibly from the choice $g_{an}=0$ of Ref.~\cite{Lella:2024hfk}.
The response consists of the effective area and the energy dispersion.
Since all photons of the source are counted, no point-spread-function containment enters.
MET denotes the Fermi mission elapsed time at the start of the burst, which fixes the orientation of the LAT and with it the angle between the source and the LAT axis.
Background is neglected.}
\label{tab:sm_setup}
\end{table}

\paragraph*{\textbf{Fit models.}}
We call the expected counts of a fit function its template.
Ref.~\cite{Lella:2024hfk} computes the templates of Eqs.~\eqref{eq:sm_NN} and \eqref{eq:sm_pion} with a simulation of the LAT on a grid of the shape parameters, $\beta_{NN}\in[0.5,2.5]$ and $E^0_{NN}\in[50,200]\,\mathrm{MeV}$ for Bremsstrahlung, and $\beta_\pi\in[0.2,4]$ and $E^0_\pi\in[100,400]\,\mathrm{MeV}$ for pion conversion, with $\omega_c$ fixed to its predicted value in each second.
The normalizations $A_{NN}$ and $A_{\pi}$ are free.

We compute the templates for any parameter values with Eq.~\eqref{eq:sm_counts} and fit the counts of each time window with three models.
\begin{itemize}
    \item \textbf{Model} $\mathbf{M_0}$ is Eq.~\eqref{eq:sm_NN} with $A_{NN}$, $E^0_{NN}$ and $\beta_{NN}$ free, the fit from which Ref.~\cite{Lella:2024hfk} obtains the temperature.
    It is also the hypothesis without pions of the flexible test below.
    \item \textbf{Model} $\mathbf{M_1}$ adds to the previous one Eq.~\eqref{eq:sm_pion} with $A_\pi$, $E^0_\pi$ and $\beta_\pi$ free and $\omega_c$ fixed, the two-component fit of the pion search of Ref.~\cite{Lella:2024hfk}.
    It is the hypothesis with pions of the flexible test.
    \item \textbf{Model} $\mathbf{M_2}$ adds to $M_0$ the published pion spectrum with only its normalization free.
    It measures how many photons a fit assigns to pion conversion if the pion spectrum has its predicted shape.
\end{itemize}

\paragraph*{\textbf{Bremsstrahlung beyond one-pion exchange.}}
Our nucleon part is the Born amplitude with one-pion exchange, as in most calculations of axion Bremsstrahlung.
The measured nucleon--nucleon amplitudes show that the Born rate overestimates soft Bremsstrahlung in neutron matter by a factor of about four~\cite{Hanhart:2000ae}.

The Bremsstrahlung spectrum of Ref.~\cite{Lella:2024hfk} uses the rate of Ref.~\cite{Carenza:2019pxu}, which corrects one-pion exchange in three ways.
\textit{(i)} It keeps the pion mass in the propagator and uses the effective nucleon masses of the medium, as our rate does.
\textit{(ii)} It adds the exchange of a $\rho$ meson, which models two-pion exchange~\cite{Ericson:1988wr}.
In the squared amplitude the pion factor $\bm k^2/(\bm k^2+m_\pi^2)$ becomes $\bm k^2/(\bm k^2+m_\pi^2)-C_\rho\,\bm k^2/(\bm k^2+m_\rho^2)$ with $C_\rho=1.67$, so the tensor force is reduced at the momentum transfers of a hot core.
\textit{(iii)} It finally includes the suppression of soft emission by multiple scattering~\cite{Raffelt:1991pw,Janka:1995ir}, as follows.
The axion couples to the nucleon spin, and the emission rate at energy $\omega$ is set by the spectrum of spin fluctuations.
A nucleon that collides many times during the emission cannot radiate coherently at energies below its collision rate, and the spectrum of spin fluctuations, $1/\omega^2$ in the Born approximation, becomes a Lorentzian of width $\Gamma$, $\Gamma_\sigma/(\omega^2+\Gamma^2)$, where $\Gamma_\sigma$ is the spin-fluctuation rate and $\Gamma$ is fixed by the sum rule of the spin structure function~\cite{Raffelt:1993ix}.

These corrections are phenomenological, not derived in chiral perturbation theory, where the same physics enters through the $NN$ contact terms, two-pion exchange and the resummation of collisions.
We include them in our analysis as well, in order to isolate the effect of the $\Delta$ in comparing our results with those of Ref.~\cite{Lella:2024hfk}.\footnote{Moreover, the relation \eqref{eq:sm_thermometer} between mean energy and temperature holds for the rate on which it was calibrated.}
At the end of this section, we assess the impact of this prescription by repeating the analysis with our unmodified emission rate.

\paragraph*{\textbf{Adding the $\Delta$ to the published Bremsstrahlung spectrum.}}
We take the published Bremsstrahlung spectrum as the nucleon part.
At every energy and time we scale the nucleon amplitude by $\sqrt r$, where $r$ is the ratio of the published spectrum to the nucleon part $\mathcal S_N$ of our rate.

How the corrections act on the $\Delta$ diagrams is not known.
The $\rho$ term changes the meson exchanged between the two nucleons, and the same exchange produces the $\Delta$, so the $\Delta$ amplitude may be reduced in a similar way.
However, the Lorentzian width accounting for multiple scattering suppresses the axion spectrum at energies lower than the collision rate, whereas the $\Delta$ component lies at energies far above that rate, and the collisions of the $\Delta$ are already contained in its width.
We therefore consider the two prescriptions, already introduced in Eq.~\eqref{eq:gamma-calibration-prescriptions},
\begin{equation}
\begin{aligned}
 \mathcal S_{\rm reduced}&=r\,(\mathcal S_N+\mathcal S_I+\mathcal S_\Delta),\\
 \mathcal S_{\rm unreduced}&=r\,\mathcal S_N+\sqrt r\,\mathcal S_I+\mathcal S_\Delta,
\end{aligned}
\label{eq:sm_baseline}
\end{equation}
with the interference $\mathcal S_I$ and the $\Delta$ part $\mathcal S_\Delta$ of our calculation.
In both prescriptions $r\,\mathcal S_N$ is the published spectrum.
In the reduced prescription the $\Delta$ contribution keeps its ratio to the nucleon part of our rate at every energy.
In the not-reduced prescription the $\Delta$ amplitude is left unchanged, and only the interference is scaled by $\sqrt r$.
Here $\mathcal S_N$, $\mathcal S_I$ and $\mathcal S_\Delta$ are the number spectra of the whole star at infinity, $(\dd L_a/\dd E)/E$ of Eq.~\eqref{eq:sn-observer-spectra} for each part, at each snapshot of SFHo-18.8, and $r$ is formed at each snapshot and energy.
The $\Delta$ part enters as computed, without a fitted function, and between snapshots the spectra are interpolated linearly in time.
The published spectrum is taken as exact, as in the control without the $\Delta$.
The truncation of the $\Delta$ contribution, the terms that Eq.~\eqref{eq:sm_baseline} adds to the published spectrum, is Eq.~\eqref{eq:trunc} applied to these terms in each cell, with $r$ formed from our nucleon part at LO and at NLO.

\begin{figure}[tb]
\centering
\includegraphics[width=\linewidth]{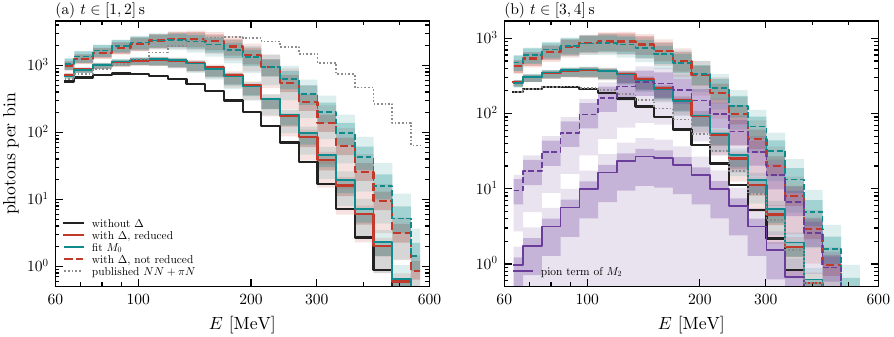}
\caption{Expected \emph{Fermi}-LAT photons per energy bin for SFHo-18.8 in (a) $t_{\rm pb}\in[1,2]\,$s and (b) $[3,4]\,$s at the benchmark couplings.
Black, the published Bremsstrahlung spectrum without the $\Delta$.
Red, with our $\Delta$ contribution reduced like the nucleon part (solid) or not reduced (dashed), Eq.~\eqref{eq:sm_baseline}.
Teal, the fit $M_0$ to the red counts.
Purple, the pion term that the fit $M_2$ assigns to the red counts, which contain no thermal pions.
Gray dotted, the published Bremsstrahlung and pion spectra together.
Dark bands show the 16--84\% range from the Lagrangian parameters, light bands that from the parameters and the truncation, both from samples of the spectrum (Sec.~\ref{sec:uncertainty}).
For the fitted curves every sample is fitted.}
\label{fig:sm_counts}
\end{figure}

\paragraph*{\textbf{Photon counts results.}}
We compute the photon counts at the detector in the time interval $1$--$6\,\mathrm{s}$.
Figure~\ref{fig:sm_counts} shows the counts for the first ($1$--$2\,\mathrm{s}$) time bin and an intermediate ($3$--$4\,\mathrm{s}$) one.
The published spectrum alone, without the $\Delta$, is the control, for which the analysis should return the temperature of the profile and no pions.
The uncertainty bands of all results contain the uncertainty on the parameters and the CHPT truncation propagated to our rates, for the fitted quantities through fits of sampled spectra (Sec.~\ref{sec:uncertainty}).

In $[1,2]\,$s the published spectrum gives $\LACountsLB$ photons, and the $\Delta$ raises them to $\LACountsRB$ in the reduced and to $\LACountsUB$ in the not-reduced prescription.
Between $1$ and $4\,$s the $\Delta$ contributes $\LAShareRB$--$\LAShareRD\%$ of the photons when reduced and $\LAShareUB$--$\LAShareUD\%$ when not reduced.
Importantly, the $\Delta$ contribution extends over $100$--$400\,\mathrm{MeV}$, the region that Ref.~\cite{Lella:2024hfk} attributes to pion conversion.

\paragraph*{\textbf{Temperature.}}
Figure~\ref{M-fig:detection}(b) of the main text shows $T_{\rm rec}$ as computed from \eqref{eq:sm_thermometer} from the fit model $M_0$ at each time step of the SFHo-18.8 profile.
Without the $\Delta$ the fit returns $\langle T\rangle$ within $\LATempCalLo$ to $+\LATempCalHi\,\mathrm{MeV}$, so the calibration works for the spectrum it was made for.
With the $\Delta$ the fit absorbs the hard photons into the Gamma distribution by raising its mean energy, and the reconstructed temperature lies above $\langle T\rangle$ at every time.

Table~\ref{tab:sm_temperature} lists in one-second slices the shift $\delta T_\Delta$, the difference between the temperatures reconstructed with and without the $\Delta$.
It is $\LAdTRMin$--$\LAdTRMax\,\mathrm{MeV}$ between $1$ and $6\,$s for the reduced $\Delta$ and $\LAdTUMin$--$\LAdTUMax\,\mathrm{MeV}$ for the not-reduced one.

\begin{table}[tb]
\centering
\small
\setlength{\tabcolsep}{9pt}
\renewcommand{\arraystretch}{1.3}

\begin{tabular}{lccccc}
\hline
\noalign{\vskip 0.7ex}
\shortstack{Time\\slice} &
$\langle T\rangle$ &
\shortstack{$T_{\rm rec}$\\without $\Delta$} &
\shortstack{$\delta T_\Delta$\\reduced} &
\shortstack{$\delta T_\Delta$\\not reduced} &
\shortstack{$\delta T_\Delta$\\our nucleon part}
\\[0.6ex]
\hline
$[1,2]\,\mathrm{s}$ &
$\LAMeanTB$ & $\LATLB$ &
$\LAdTRBBand$ & $\LAdTUBBand$ & $\SIdTMzeroBBand$ \\

$[2,3]\,\mathrm{s}$ &
$\LAMeanTC$ & $\LATLC$ &
$\LAdTRCBand$ & $\LAdTUCBand$ & $\SIdTMzeroCBand$ \\

$[3,4]\,\mathrm{s}$ &
$\LAMeanTD$ & $\LATLD$ &
$\LAdTRDBand$ & $\LAdTUDBand$ & $\SIdTMzeroDBand$ \\

$[4,5]\,\mathrm{s}$ &
$\LAMeanTE$ & $\LATLE$ &
$\LAdTREBand$ & $\LAdTUEBand$ & $\SIdTMzeroEBand$ \\

$[5,6]\,\mathrm{s}$ &
$\LAMeanTF$ & $\LATLF$ &
$\LAdTRFBand$ & $\LAdTUFBand$ & $\SIdTMzeroFBand$ \\
\hline
\end{tabular}
\caption{Emission-weighted temperature $\langle T\rangle$ of SFHo-18.8, temperature $T_{\rm rec}$ reconstructed with the fit $M_0$ and Eq.~\eqref{eq:sm_thermometer} from the published Bremsstrahlung spectrum without the $\Delta$, and the shift $\delta T_\Delta$ when our $\Delta$ contribution is added, reduced like the nucleon part or not reduced, all in MeV.
The last column gives the shift with our own nucleon part instead of the published spectrum.
Bands (16--84\%) from the Lagrangian parameters and the truncation, from fits of sampled spectra (Sec.~\ref{sec:uncertainty}).}
\label{tab:sm_temperature}
\end{table}

At the largest photon number at which the Gamma distribution still describes the counts,
$\delta T_\Delta$ exceeds three standard deviations of $T_{\rm rec}$ in \LATempZthreeAllR{} windows of $\LATempWin\,$s for both the reduced and the non-reduced $\Delta$.
The not-reduced $\Delta$, which carries three quarters of the photons, deforms the spectrum so much that the Gamma distribution is rejected with ten times fewer photons, $N_\gamma=\LATempNtotU$ instead of $\LATempNtotR$ in $1$--$6\,$s.

\paragraph*{\textbf{Thermal pions.}}
Ref.~\cite{Lella:2024hfk} searches for pions by comparing the fit with Bremsstrahlung alone with the fit that adds pion conversion.
The comparison uses the Bayesian evidence of each model $M$, its likelihood averaged over the prior distribution of its parameters,
\begin{equation}
 Z_M=\int \dd\boldsymbol p\,
 \mathcal L(\{n_{bi}\}\mid\boldsymbol p,M)\,
 \pi(\boldsymbol p\mid M),
 \qquad
 \ln B=\ln Z_{NN+\pi N}-\ln Z_{NN},
 \label{eq:sm_evidence}
\end{equation}
with the Poisson likelihood $\mathcal L$ of the counts in all bins and the prior $\pi$ of the parameters $\boldsymbol p$ of $M$.
Positive $\ln B$ therefore favors pion conversion.
The reason why the Bayesian evidence is used instead of directly comparing likelihoods is that the model with pion conversion has more parameters, therefore its evidence is higher only if the improvement of the fit outweighs the larger parameter space over which its likelihood is averaged.

We generate the photon count from the published Bremsstrahlung spectrum without the $\Delta$ and with the $\Delta$ in both prescriptions, each without and with the published pion spectrum, and perform two tests.
\begin{itemize}
    \item \textbf{The flexible test}, which compares $M_0$ with $M_1$, both with all shape parameters free, the procedure of Ref.~\cite{Lella:2024hfk}.\footnote{Ref.~\cite{Lella:2024hfk} does not state its priors.
    We take the amplitudes log-uniform over $10^{45}$--$10^{57}\,\mathrm{s^{-1}\,MeV^{-1}}$, $E^0_{NN}$ and $E^0_\pi$ uniform over $20$--$250$ and $20$--$400\,\mathrm{MeV}$, and $\beta_{NN}$ and $\beta_\pi$ uniform over $0.2$--$6$ and $0.1$--$6$.}
    \item \textbf{The fixed test}, which assumes that the shape of the Bremsstrahlung spectrum is known and asks only whether the predicted pions are present.
    Its Bremsstrahlung template is that of the input spectrum itself, without or with the $\Delta$, and its pion template is that of the published pion spectrum.
    Only the two normalizations are free, each with a uniform prior from zero to three times its value at the benchmark couplings.
    The nucleon and $\Delta$ parts are not varied separately, since their ratio is given by the calculation and its uncertainty by the two prescriptions.
\end{itemize}

\begin{table}[tb]
\centering
\small
\renewcommand{\arraystretch}{1.15}
\begin{tabular}{lccc}
\hline
Bremsstrahlung input, $[1,2]\,$s & Thermal pions & Flexible test & Fixed test\\
\hline
without $\Delta$ & absent & $\LAEvFlexLB$ & $\LAEvFixedLB$ \\
without $\Delta$ & added & $\LAEvFlexLPiB$ & $\LAEvFixedLPiB$ \\
with reduced $\Delta$ & absent & $\LAEvFlexRB$ & $\LAEvFixedRB$ \\
with reduced $\Delta$ & added & $\LAEvFlexRPiB$ & $\LAEvFixedRPiB$ \\
with not-reduced $\Delta$ & absent & $\LAEvFlexUB$ & $\LAEvFixedUB$ \\
with not-reduced $\Delta$ & added & $\LAEvFlexUPiB$ & $\LAEvFixedUPiB$ \\
\hline
\end{tabular}
\caption{Logarithm of the Bayes factor of Bremsstrahlung plus pion conversion over Bremsstrahlung alone, Eq.~\eqref{eq:sm_evidence}, for counts of SFHo-18.8 in $[1,2]\,$s at the benchmark couplings.
The Bremsstrahlung input is the published Bremsstrahlung spectrum without or with our $\Delta$ contribution, the thermal pions those of the published pion spectrum.
Flexible test, Gamma distributions with all shape parameters free (models $M_0$ and $M_1$).
Fixed test, the input Bremsstrahlung shape and the published pion shape with free normalizations.
Positive (negative) $\ln B$ favors (disfavors) pion conversion.}
\label{tab:sm_evidence}
\end{table}

Table~\ref{tab:sm_evidence} lists the results.
Without pions and without the $\Delta$, neither test favors a second component, as the analysis of Ref.~\cite{Lella:2024hfk} assumes.
With the not-reduced $\Delta$ the flexible test favors a second component, $\ln B=\LAEvFlexUB$, although no pions are present, whereas with the reduced $\Delta$ it gives $\ln B=\LAEvFlexRB$.
The answer of the flexible test thus depends on how strongly the $\Delta$ is reduced.
The fixed test, whose hypothesis without pions contains the hard emission of the $\Delta$, disfavors pions in all pion-free inputs and finds the predicted pions in all inputs that contain them.
The published pion spectrum is large compared with the published Bremsstrahlung spectrum, $\SILellaPiB$ against $\LACountsLB$ photons in $[1,2]\,$s, which explains the large values with pions.

The two tests ask whether pions are present.
How many there are follows from the normalization of the pion spectrum, which we determine with the fit $M_2$.
It keeps the predicted pion shape, like the fixed test, and leaves the Bremsstrahlung shape free, like the flexible test.
In the pion-free inputs $M_2$ assigns no photons to pion conversion in $[1,2]\,$s and, for the central constants, none in $[2,3]\,$s, where the band of the not-reduced $\Delta$ reaches $\LAfpiUCHi\%$.
After about three seconds the published pion spectrum becomes soft.
The temperature drops, and in the hot core of SFHo-18.8 the difference between the neutron and proton chemical potentials exceeds the pion mass, so that pions would form a Bose condensate there.
Ref.~\cite{Lella:2024hfk} sets pion conversion to zero in these regions, and the remaining emission comes from the cooler outer layers.
The published pion spectrum then resembles the $\Delta$ contribution.
In $[3,4]\,$s $M_2$ assigns $\LAfpiRDBand\%$ of the photons to pion conversion for the reduced and $\LAfpiUDBand\%$ for the not-reduced $\Delta$, against $\LAfpiLD\%$ without the $\Delta$ (Fig.~\ref{fig:sm_counts}(b)), and in $[4,5]\,$s $\LAfpiREBand\%$ and $\LAfpiUEBand\%$.

In conclusion, without the $\Delta$ in the templates, the spectrum of the converted axions thus determines neither the temperature nor the pion content of the core.

\paragraph*{\textbf{Comparison with our nucleon part.}}
With our own nucleon part, the one-pion-exchange Born rate with no suppression factor applied (equivalently, setting $r=1$ in \eqref{eq:sm_baseline}) instead of the published spectrum, the conclusions are qualitatively the same.

The shift of the reconstructed temperature is $\SIdTMzeroF$--$\SIdTMzeroB\,\mathrm{MeV}$ between $1$ and $6\,$s (Table~\ref{tab:sm_temperature}) and $\SIdTMzeroABand\,\mathrm{MeV}$ in the first second,
Since our nucleon part gives $\LANucRatioMin$--$\LANucRatioMax$ times more photons, the flexible test favors a second component more strongly, $\ln B=\SIEvFlexND$ in the first second against $\SIEvFlexN$ for the nucleon part alone.
The temperature calibration was not made for this spectrum, and indeed for the nucleon part alone the fit lies $\SICalMzeroLo$--$\SICalMzeroHi\,\mathrm{MeV}$ above $\langle T\rangle$.
This offset cancels in $\delta T_\Delta$.

\paragraph*{\textbf{The published photon counts.}}
We conclude the section with a technical remark.
Fig.~7 of Ref.~\cite{Lella:2024hfk} shows the expected LAT counts of the two components in $[0,1]\,$s and $[2,3]\,$s at the benchmark couplings.
We have not been able to reproduce these counts from the published spectra of its App.~A.

\begin{figure}[tb]
\centering
\includegraphics[width=\linewidth]{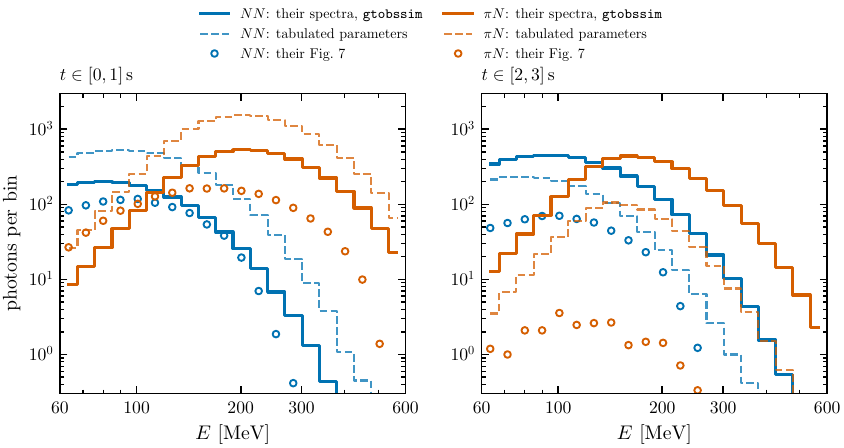}
\caption{Expected photons per bin for the Bremsstrahlung (blue) and pion-conversion (orange) spectra stated in App.~A of Ref.~\cite{Lella:2024hfk} for SFHo-18.8 at the benchmark couplings, in $[0,1]\,$s (left) and $[2,3]\,$s (right), simulated with \texttt{gtobssim} with their time dependence (solid) and with their tabulated parameters of the slice end (dashed), against the counts of their Fig.~7 (circles).}
\label{fig:sm_lella_check}
\end{figure}

We simulated these spectra with \texttt{gtobssim} of the \emph{Fermi} Science Tools, with the same event class, direction, explosion time and selection, and with the livetime and the dead time of the LAT.
Without dead-time losses the simulation agrees with our folding, Eq.~\eqref{eq:sm_counts}, within $\SIGtAgree\%$.
In $[0,1]\,$s it gives $\SIGtNNA$ Bremsstrahlung and $\SIGtPiA$ pion-conversion photons, against $\SIFigNNA$ and $\SIFigPiA$ in their Fig.~7.
In $[2,3]\,$s it gives $\SIGtNNB$ and $\SIGtPiB$, against $\SIFigNNB$ and $\SIFigPiB$ (Fig.~\ref{fig:sm_lella_check}).

The largest difference is the pion component after two seconds, $\SIGtPiFacB$ times smaller in their figure than in their stated spectra.
The ratio of the two components does not depend on the detector, and integrated over $60$--$600\,\mathrm{MeV}$ without any detector their spectra give a pion-to-Bremsstrahlung ratio of $\SINoDetRatioB$ in $[2,3]\,$s, against $\SIFigRatioB$ in their figure.
The conclusion of Ref.~\cite{Lella:2024hfk} that the burst is pure Bremsstrahlung after two seconds therefore does not follow from its stated spectra.
We could not identify the origin of the difference.

In any case, this discrepancy does not enter the results presented above, which use the stated spectra and fold all inputs and templates with the same response, nor would it qualitatively change the results of this section.

\section{Water Cherenkov detectors}
\label{sec:cherenkov}
\paragraph*{\textbf{Event yield.}}
Ref.~\cite{Cavan-Piton:2025nsj} proposed to detect supernova axions through absorption on hydrogen in water, followed by $\pi^0\to\gamma\gamma$.
Their forecast used axions produced by thermal-pion conversion, arguing that axions from nucleon Bremsstrahlung are not energetic enough.
Here we apply the same detection channel to the Bremsstrahlung spectrum calculated in this work.
Including the $\Delta$ enhances its high-energy tail and increases the flux above the pion-production threshold, without requiring thermal pions.
We compute the expected number of $ap\to p\pi^0$ interactions in one Hyper-Kamiokande tank, taking $M_{\rm det}=187\,\mathrm{kt}$~\cite{Hyper-Kamiokande:2018ofw} as the mass of the water in the tank.
Treating the two hydrogen protons per water molecule as free targets gives the total number of protons $N_p=2N_A M_{\rm det}/\mu_{\mathrm{H_2O}}$, where $N_A$ is the Avogadro constant and $\mu_{\mathrm{H_2O}}=18.015\,\mathrm{g\,mol^{-1}}$.

For a supernova at distance $d$, the time-integrated number spectrum $dN_a/dE_\infty$ of Eq.~\eqref{eq:sn-reduced-emissivity}, divided by $4\pi d^2$, gives the number of axions crossing a unit area over the burst.
The differential cross section $d\sigma(E_\infty)/dq^0$ is the effective interaction area per target proton and per unit energy of the outgoing pion, whose total energy is $q^0$.
We discuss in details how to compute it in the next paragraph.
Multiplying this cross section by the fluence and by $N_t$, and summing over incoming axion energies, gives the number of neutral pions produced by the axion,
\begin{equation}
 \frac{dN_{\pi^0}}{dq^0}
 =N_t\int dE_\infty\,
 \left[\frac{1}{4\pi d^2}\frac{dN_a}{dE_\infty}\right]
 \frac{d\sigma(E_\infty)}{dq^0},
 \qquad
 N_{\pi^0}=\int_{M_{\pi^0}}^\infty dq^0\,
 \frac{dN_{\pi^0}}{dq^0}.
 \label{eq:cherenkov-yield}
\end{equation}

\begin{figure}[t]
\centering
\begin{tikzpicture}[scale=0.85]
\begin{scope}[shift={(0,0)}]\FDabs{N}{s}{(a)}\end{scope}
\begin{scope}[shift={(3.4,0)}]\FDabs{D}{s}{(b)}\end{scope}
\begin{scope}[shift={(6.8,0)}]\FDabscontact{(c)}\end{scope}
\end{tikzpicture}\\[4pt]
\includegraphics[width=0.6\linewidth]{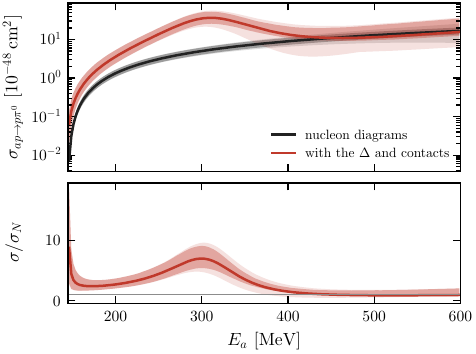}
\caption{Absorption $ap\to N\pi$ at NLO.
The upper row shows the diagrams with an intermediate nucleon (a) and $\Delta$ (b), each with its crossed diagram, and the contact terms (c) of the chiral connection (charged pions only) and of the LECs.
The lower panels show the cross section of $ap\to p\pi^0$ on a proton at rest for the KSVZ axion with $f_a=10^9\GeV$, with the nucleon diagrams (black) and with all diagrams (red), and their ratio.
The nucleon diagrams include the fixed $1/M_N$ parts of $\hat c_2$ and $\hat c_8$ in Eq.~\eqref{eq: hat ci coeff and residual}, which come from integrating out the lower components of the nucleon field.
Bands as in Fig.~\ref{M-fig:spectrum} of the main text.}
\label{fig:absorption}
\end{figure}

\paragraph*{\textbf{Absorption at LO and NLO.}}
For $a(p_a)+p(P_i)\to p(P_f)+\pi^0(q)$ on a proton at rest, neglecting the axion mass, the differential cross section is
\begin{equation}
 \frac{d\sigma}{dq^0}
 =\frac{\overline{|\mathcal M|^2}}{32\pi M_p E_a^2},
 \qquad
 \overline{|\mathcal M|^2}
 =\frac12\sum_{s_i,s_f}|\mathcal M_{s_fs_i}|^2,
 \label{eq:cherenkov-cross-section}
\end{equation}
where $E_a$ and $q^0$ are the incoming axion and outgoing pion energies; the bar denotes an initial-spin average and a final-spin sum.
The amplitude $\mathcal M$ is constructed from the LO and NLO Feynman rules of Sec.~\ref{sec:lagrangian}.
At LO, the contributing diagrams are (a) and (b) at the top of Fig.~\ref{fig:absorption}, each with its crossed ordering.
Using the resummed propagators of Eq.~\eqref{eq:hb-resummed-propagators}, the amplitudes with LO interaction vertices are
\begin{align}
 \mathcal M_N^s
 &=\frac{G_A C_p}{4F_\pi f_a}
 \,S_N(P)\,
 \xi_f^\dagger
 \left[
 \mathbf p_a\cdot\mathbf q
 +is\,\boldsymbol\sigma\cdot(\mathbf p_a\times\mathbf q)
 \right]\xi_i,
 \notag\\
 \mathcal M_\Delta^s
 &=\frac23\,
 \frac{G_{\pi N\Delta}G_{aN\Delta}c_{u-d}}{F_\pi f_a}
 \,S_\Delta(P)\,
 \xi_f^\dagger
 \left[
 \frac23\mathbf p_a\cdot\mathbf q
 -\frac{is}{3}\boldsymbol\sigma\cdot(\mathbf p_a\times\mathbf q)
 \right]\xi_i .
 \label{eq:cherenkov-LO}
\end{align}
Here $s=+$ labels the axion vertex adjacent to the outgoing proton, with $P=P_f-p_a=P_i-q$, and $s=-$ labels the axion vertex adjacent to the incoming proton, with $P=P_i+p_a=P_f+q$.
The amplitudes here and below are written with two-component spinors $\xi_{i,f}$ normalized to one.
As in the rate, each external proton of energy $E$ carries the normalization $(E+M_p)^{1/2}$ of its Dirac spinor, so that $\mathcal M$ in Eq.~\eqref{eq:cherenkov-cross-section} contains the factor $[2M_p(P_f^0+M_p)]^{1/2}$.
The kinematics use the physical masses $M_p$ and $M_{\pi^0}$ of the proton and the neutral pion, and the constants are those of Table~\ref{tab:matched-constants}.

At NLO there are two additional contributions.
First, the LO diagrams receive the vertex corrections of Eqs.~\eqref{eq:hb-nn-recoil-rule} and~\eqref{eq:hb-delta-recoil-rule}, with a single recoil insertion at either the axion or the pion vertex.
Second, the $a\pi NN$ vertex contributes, giving diagram (c) at the top of Fig.~\ref{fig:absorption}.
The WT interaction \eqref{eq:hb-wt-vertex}, including its recoil correction, and the $\hat c_4^{\rm res}$ contact both contain the isospin factor $\epsilon^{3bj}\tau^j$, which vanishes for a neutral pion, $b=3$.
The surviving contact amplitude is
\begin{equation}
 \mathcal M_c =\frac{\xi_f^\dagger\xi_i}{F_\pi f_a}\Big[\,(2c_{u-d}\hat c_3^{\rm res}+c_{u+d}\hat c_9)
       \mathbf p_a\cdot\mathbf q -E_a q^0\left(
       2c_{u-d}(\hat c_2^{\rm res}+\hat c_3^{\rm res})
       +c_{u+d}\left(\frac{\hat c_8}4+\hat c_9\right)
                       \right)\Big].
\label{eq:nlo-abs-neutral-contact}
\end{equation}
Both the spatial and energy-dependent terms are retained.
In particular, $\hat c_2^{\rm res}$ and $\hat c_8$ enter at NLO here because the outgoing real pion has an unsuppressed energy, whereas the energy of the exchanged pion in Bremsstrahlung is a recoil correction.

\paragraph*{\textbf{Relation to the main-text results.}}
We evaluate Eq.~\eqref{eq:cherenkov-yield} using the axion spectra computed in this work as presented in Sec.~\eqref{sec:results}, integrated over the first nine seconds.
For SFHo-18.6, $f_a=10^9\,\mathrm{GeV}$ and $d=0.2\,\mathrm{kpc}$ and with the nucleon-only cross section for absorption, including the $\Delta$ in the computation of the Bremsstrahlung axion spectrum raises the expected number of detected pions from $0.2$ to $0.6$, a factor of three.
Using the full NLO absorption cross section gives a further factor $\PRLCkGainDetector$.
The combined enhancement is therefore $\PRLCkGainUnc$.

Each curve in Fig.~\ref{M-fig:cherenkov} is obtained by varying the distance and finding the axion mass such that two neutral pions are produced in the Cherenkov detector.
Since production and absorption each scale as $f_a^{-2}$, \eqref{eq:cherenkov-yield} gives $N_{\pi^0} = N_{\pi^0}(d, f_a) \propto f_a^{-4}d^{-2}$.
We then obtain a relation between the minimum mass of the axion making
\begin{equation}
 m_{a,\min}(d)=5.7\,\mathrm{meV}
 \left[
 \frac{2}{N_{\pi^0}(0.2\,\mathrm{kpc},10^9\,\mathrm{GeV})}
 \right]^{1/4}
 \sqrt{\frac{d}{0.2\,\mathrm{kpc}}}.
 \label{eq:cherenkov-reach}
\end{equation}
All contours in Fig.~\ref{M-fig:cherenkov} rise as $d^{1/2}$, while a larger yield lowers the mass threshold.
Reconstruction efficiency and event overlap are not included in this analysis.

\end{document}